\documentclass[a4paper,fleqn,usenatbib,useAMS]{mnras}

\usepackage{graphicx}	% Including figure files
\usepackage{amsmath}	% Advanced maths commands
\usepackage{amssymb}	% Extra maths symbols
\usepackage{multicol}        % Multi-column entries in tables
\usepackage{bm}		% Bold maths symbols, including upright Greek
\usepackage{pdflscape}	% Landscape pages
\usepackage[T1]{fontenc}
\usepackage{ae,aecompl}
\usepackage{newtxtext,newtxmath}
\usepackage{xcolor} %AJB
\title[Interactions between tidal flows and convection.]{Tidal flows with convection: frequency-dependence of the effective viscosity and evidence for anti-dissipation}
\author[C. D. Duguid et al.]{
	Craig. D. Duguid,$^{1}$\thanks{E-mail: sccd@leeds.ac.uk}
	Adrian. J. Barker,$^{2}$
	and C. A. Jones $^{2}$
	\\
	$^{1}$EPSRC Centre for Doctoral Training in Fluid Dynamics, University of Leeds, Leeds LS2 9JT, UK\\
	$^{2}$School of Mathematics, University of Leeds, Leeds LS2 9JT, UK\\
}

\date{Accepted XXX. Received YYY; in original form ZZZ}
\pubyear{2019}

\begin{document}
\label{firstpage}
\pagerange{\pageref{firstpage}--\pageref{lastpage}}
\maketitle

% Abstract of the paper
\begin{abstract} 
 Tidal interactions are important in driving spin and orbital evolution in planetary and stellar binary systems, but the fluid dynamical mechanisms responsible remain incompletely understood. One key mechanism is the interaction between tidal flows and convection. Turbulent convection is thought to act as an effective viscosity in damping large-scale tidal flows, but there is a long-standing controversy over the efficiency of this mechanism when the tidal frequency exceeds the turnover frequency of the dominant convective eddies. This high frequency regime is relevant for many applications, such as for tides in stars hosting hot Jupiters. We explore the interaction between tidal flows and convection using hydrodynamical simulations within a local Cartesian model of a small patch of a convection zone of a star or planet. We adopt the Boussinesq approximation and simulate Rayleigh-B\'{e}nard convection, modelling the tidal flow as a background oscillatory shear flow. We demonstrate that the effective viscosity of both laminar and turbulent convection is approximately frequency-independent for low frequencies. When the forcing frequency exceeds the dominant convective frequency, the effective viscosity scales inversely with the square of the tidal frequency. We also show that negative effective viscosities are possible, particularly for high frequency tidal forcing, suggesting the surprising possibility of tidal anti-dissipation. These results are supported by a complementary high-frequency asymptotic analysis that extends prior work by Ogilvie \& Lesur. We discuss the implications of these results for interpreting the orbital decay of hot Jupiters, and for several other astrophysical problems.

\end{abstract}


%%%%%%%%%%%%%%%%% BODY OF PAPER %%%%%%%%%%%%%%%%%%

\section{Introduction}
Tidal interactions are important for the spin and orbital evolution of short-period planets and close binary stars. In particular, the prevalence of circular orbits amongst the closest binary stars is thought to be produced by the dissipation of tidal flows inside the stars, which can drive systems in initially eccentric orbits towards circularity (e.g.~\citealt{ZahnBouchet1989,VerbuntPhinney1995,Meibom2005,Meibom2006,mazeh_observational_2008,PriceWhelan2018,Beck2019}). There are also recent exciting early indications of tidally-driven orbital decay for two hot Jupiters (WASP-12 b and WASP-4 b; \citealt{Maciejewski2016,Patra2017,Maciejewski2018,Bouma2019}) based on transit timing variations over decadal timescales (though future observations are required to rule out alternative explanations; see also \citealt{Birkby2014,wilkins_searching_2017}). These observations motivate theoretical studies to understand the mechanisms of tidal dissipation in stars.

The fluid dynamical mechanisms responsible for tidal dissipation in stars and giant planets remain incompletely understood (e.g.~\citealt{Mathis2013,ogilvie_tidal_2014}). One key mechanism is believed to be the interaction between tidal flows and turbulent convection in the envelopes of giant planets and solar-type stars \citep{zahn_les_1966,zahn_tidal_1977,ZahnBouchet1989}. Convection is thought to act as an effective viscosity ($\nu_E$) in damping the large-scale tidal flows, but this process remains theoretically uncertain, particularly in the regime of fast tides. Indeed, there is a long-standing theoretical controversy over the efficiency of this process when the tidal frequency ($\omega$) exceeds the dominant convective frequency $(\omega_c)$ \citep{zahn_les_1966,goldreich_turbulent_1977,Zahn1989,GoodmanOh1997}. It has been realised that convection should become less efficient at dissipating the tidal flow when $\omega>\omega_c$, but to what extent is still a matter of debate. Based on ideas from mixing-length theory, \cite{zahn_les_1966} (see also \citealt{Zahn1989}) proposed $\nu_E\propto \omega^{-1}$ when $\omega\gg \omega_c$, taking the mean free path to be the distance the convective eddies have moved in one tidal period. \cite{goldreich_turbulent_1977} suggested that this reduction is insufficient, and that it is the resonant eddies, those with turnover frequencies comparable with $\omega$, that should primarily contribute to this interaction. Assuming a Kolmogorov turbulent cascade, this phenomenological argument implies $\nu_E\propto \omega^{-2}$, which would imply much less efficient tidal dissipation at high frequencies. 

Why does it matter which of these prescriptions, if any, are correct? To give just one example, a crude estimate for the inspiral time of a hot Jupiter orbiting a Sun-like star in a one day orbit suggests that (we discuss this further in \S~\ref{implications}) if there is no frequency-reduction in the effective viscosity the planet would decay in tens of Myr. On the other hand, if $\nu_E\propto \omega^{-1}$, the planet would decay in 1 Gyr, and if $\nu_E\propto \omega^{-2}$, the planet would decay in tens to a hundred Gyr. It is clearly essential to determine which of these (if any) are correct before we can predict the orbital decay (or otherwise) of hot Jupiters, and so that we can interpret existing observations.

This problem is relevant for a number of other astrophysical applications. In particular, the interaction between tidal flows and convection is thought to be the dominant mechanism for producing circularisation and synchronisation of late-type binaries with convective envelopes (e.g.~\citealt{ZahnBouchet1989,Meibom2005,Meibom2006,mazeh_observational_2008}), the enhanced rates of orbital circularisation in cool-cool and hot-cool binaries over hot-hot binaries \citep{VanEylen2016}, and in low-mass binary systems containing fully convective stars (e.g.~\citealt{Triaud2017,VonBoetticherTriaud2019}). The clearest observational example indicating the efficacy of this mechanism is in binary systems containing giant stars (e.g.~\citealt{VerbuntPhinney1995,PriceWhelan2018,Sun2018,Beck2019}). This last example is the only one for which the uncertainty in the high frequency reduction in $\nu_E$ is probably unimportant.

It is now possible to tackle this problem directly using hydrodynamic simulations of convection. Pioneering work in this direction was undertaken by \cite{penev_dissipation_2007}; \cite{penev_direct_2009} and \cite{penev_dissipation_2009}. \cite{penev_direct_2009} directly simulated the interaction between convection (in a deep layer, adopting the anelastic approximation) and a large-scale flow driven by an oscillatory body force in a Cartesian domain. Their simulations (and the associated perturbative calculations in \citealt{penev_dissipation_2007} based on \citealt{GoodmanOh1997}) measured the effective viscosity and found support for a frequency-reduction that is more consistent with the linear scaling of \cite{zahn_les_1966}, with some evidence of a weak anisotropy in the components of the eddy viscosity tensor. A further important contribution was subsequently made by \cite{ogilvie_interaction_2012}, who performed a high-frequency asymptotic analysis in a local Cartesian model to understand the fluid response to an imposed oscillatory tidal (shear) flow. They elucidated the visco-elastic nature of the response, and performed complementary simulations to probe the interaction between this flow and convection in a triply-periodic Cartesian box (so-called ``homogeneous convection"). Their analysis and simulations were both consistent with a quadratic reduction in the effective viscosity for high-frequency tides. They also obtained tentative evidence suggesting the surprising result that $\nu_E$ can become negative at high frequencies, indicating the possibility of tidal anti-dissipation (where energy is transferred from the convective flow). However, their negative values of $\nu_E$ contained substantial error bars, partly as a result of the bursty nature of homogeneous convection and the computational expense of running long-duration simulations. \cite{braviner_stellar_2015} continued this work further by simulating the interaction between an oscillatory tidal shear flow and a convective-like flow (ABC flow). He found support for both the asymptotic analysis of \cite{ogilvie_interaction_2012} and a quadratic reduction in the effective viscosity at high frequencies for this flow.

The independent studies of \cite{penev_direct_2009} and \cite{ogilvie_interaction_2012} obtained apparently contradictory results regarding the nature of the effective viscosity in the regime of high frequency tidal forcing. The former work simulated multiple pressure scale heights, whereas the latter assumed a Boussinesq domain but explored a wider range of tidal frequencies. The two sets of simulations exhibit different turbulent temporal power spectra, which may be responsible for the different frequency scalings in the effective viscosity, but these differences have not yet been explained. In our opinion, this motivates further work to understand the interaction between tidal flows and convection. Our approach follows \cite{penev_direct_2009} and \cite{ogilvie_interaction_2012} in simulating the dynamics of convection in a local Cartesian model that represents a small patch of a convection zone of a star or planet. We differ from these works by adopting the well-studied Rayleigh-B\'{e}nard setup to model (Boussinesq) convection, which allows us to overcome some of the peculiar properties of ``homogeneous convection" (as studied by \citealt{ogilvie_interaction_2012}). To drive convection in this system, we impose different temperatures on two boundaries in the vertical direction, with a hot plate underlying a cool one. We wish to understand the fundamental fluid dynamical interactions between convection and tidal flows in this model, using a combination of a hydrodynamical simulations and asymptotic theory.

The advantages of adopting a local Cartesian model that represents a small patch of the convection zone of a star or planet are that these simulations are much less computationally expensive, and this model is much simpler to set up and to analyse than a corresponding global model. The former allows us to simulate more turbulent convection (at higher Rayleigh numbers), and for these simulations to be run for much longer, allowing us to reduce turbulent noise. This is likely to be essential to accurately determine the effective viscosity, particularly in the regime of high frequency tides. The disadvantage of a local model is that global aspects, including the structure of the convection zone and spatial variations in the stellar or planetary properties are not considered. In this study, we have chosen to focus on Boussinesq convection, since it is simpler to analyse and more efficient to simulate than compressible (or anelastic) convection. This allows us to undertake a wider parameter survey, and to run our simulations for longer to reduce noise.

The plan of this paper is as follows. In \S~\ref{localmodel}, we describe our model, the governing equations and the numerical method used to simulate the convection and to analyse its interaction with the tidal flow. We also describe how the most important quantities are evaluated from our simulations. In \S~\ref{LaminarR2}, we describe the results of simulations of laminar convection and its interaction with the tidal flow, comparing our results with a complementary asymptotic analysis which is described in detail in Appendix~\ref{appendix_extension_of_ol2012}. We move on to explore more turbulent convection in \S~\ref{turbulent}. We then discuss the astrophysical implications of this work in \S~\ref{discussion} and conclude in \S~\ref{conclusions}.

%%%%%%%%%%%%%%%%%%%%%%%%%%%%%%%%%%%%%%%%%%%%%%%%%%%%%%%%%%%%%%%%%%%%%%%%%%%%%%%%%%%%%%%%%%
\section{Local Cartesian model: small patch of a convection zone} 
\label{localmodel}

\begin{figure*}
	\includegraphics[width=0.75\textwidth]{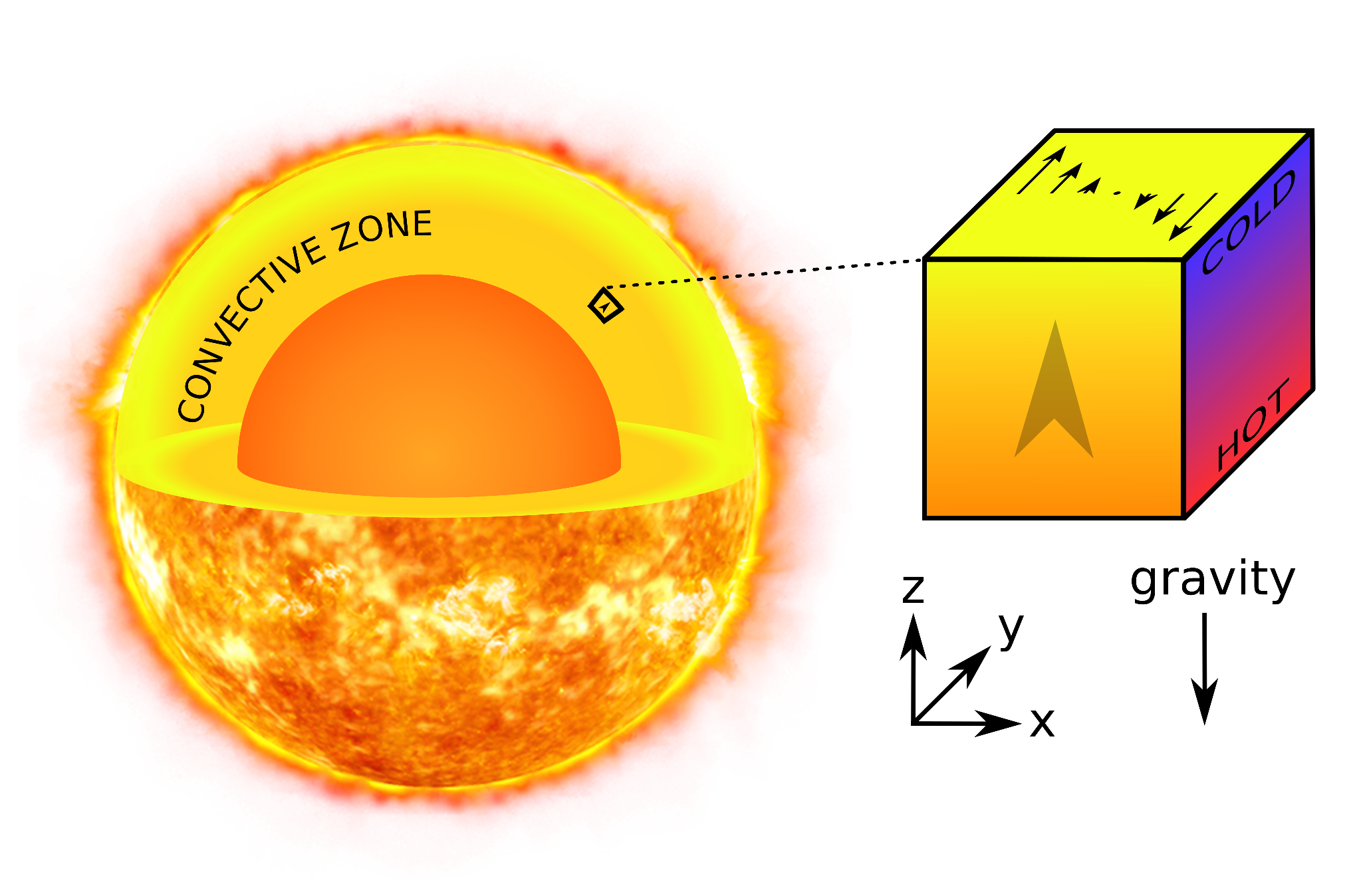}
	\caption{Local Cartesian model to study the interaction between tidal flows and convection. The tidal flow is modelled as an oscillatory shear flow in $y$ that varies linearly in $x$, which represents one component of the large-scale non-wave-like tidal flow in a star (or planet). The local vertical direction is $z$.}
	\label{fig_box_model}
\end{figure*}

We consider a local Cartesian representation of a small patch of a convection zone of a star or giant planet subjected to gravitational tidal forcing from a companion (see Fig.~\ref{fig_box_model}). This is the simplest model in which to explore the interaction between tidal flows and convection. We use Cartesian coordinates $(x,y,z)$, and simulate a Cartesian domain with $x\in [0,L_x d]$, $y\in [0,L_y d]$ and $z\in[0,d]$. A local model has the significant advantage that it is much simpler to set up and to analyse, and the numerical methods are also computationally more efficient, than in a corresponding global model. Our approach is similar to, and builds upon, the pioneering works of \cite{penev_dissipation_2007}, \cite{penev_direct_2009} and \cite{ogilvie_interaction_2012}.

We model convection using the Boussinesq approximation \citep{spiegel_boussinesq_1960} and adopt the well-studied Rayleigh-B\'{e}nard setup (e.g.~\citealt{chandrasekhar_hydrodynamic_2013}). The Boussinesq approximation is appropriate for studying small-scale convection with short length-scales that are much smaller than a pressure or density scale height, and with flows that are slow compared with the sound speed. Our approach is similar to \cite{ogilvie_interaction_2012}, except that they studied convection with periodic boundary conditions in the vertical (so-called ``homogeneous convection"). That model is peculiar due to the existence of exact convection solutions (``elevator modes"), which lead to bursty temporal dynamics. We have chosen to instead adopt impenetrable, stress-free and fixed temperature walls at the top and bottom of our domain in $z$. These walls are strictly artificial for stellar convection, but they prevent the occurrence of elevator modes and allow a statistically steady turbulent state to be attained more readily.

Since tidal deformations of stars are typically small, it is sensible to explore the regime of linear tides, in which the fluid response to each component of the tidal potential can be studied separately. We will follow \cite{ogilvie_interaction_2012} in modelling the large-scale tidal flow as an oscillatory shear flow that is linear in the local Cartesian coordinates. In particular, we model the tidal flow as a ``background flow" of the form
\begin{equation}
	\boldsymbol{u}_0 = 
	S x\cos(\omega t) \boldsymbol{e}_y = \frac{\dot{a}x}{d} \boldsymbol{e}_y \,,
\end{equation}	
 where $\omega$ is the tidal frequency, $S=a_0\omega/d$ is the amplitude of the tidal shear, $a_0$ is the amplitude of the tidal displacement, and $\dot{a}$ is the time-derivative of the displacement $a(t)=a_0\sin\omega t$. Note that this is only one component of the non-wave-like tidal flow even in the simplest case of a circular aligned orbit (e.g.\citealt{ogilvie_tidal_2014}). If we consider the tide in a star due to a planet on a circular orbit in the equatorial plane, the appropriate tidal frequency is $\omega=2(\Omega_o-\Omega_\star)$, where $\Omega_o$ is the orbital frequency of the planet and $\Omega_\star$ is the spin frequency of the star. There is no tidal forcing when $\omega=0$, which corresponds with spin-orbit synchronisation. If we instead consider the tide in a spin-synchronised (and spin-orbit aligned) star or planet on a weakly eccentric orbit, the relevant tidal frequency would be $\omega=\Omega_o$. 

In our simulations we will typically assume $a_0\ll d$ (with a typical value $a_0=0.05d$), so that the tidal displacement is much smaller than a density or pressure scale height. This is reasonable for studying tides in solar-type stars interacting with the convective eddies in the majority of the convection zone (and in particular near the base), for example (the tidal amplitude is typically smaller than $10^{-4}$ stellar radii, whereas the pressure scale height is of the order of $0.1$ stellar radii except near the surface). Since planet-hosting stars are usually slow rotators, we ignore rotation in this initial study, but we note that sufficiently rapid rotation is expected to affect the resulting effective viscosity \citep{mathis_impact_2016}. This is because rotation modifies convective velocities and length-scales (e.g.~\citealt{S1979,Jones2007,King2013,BDL14}), but we relegate the incorporation of rotation to future work.

Within the Boussinesq approximation, perturbations to a background flow $\mathbf{u}_0$ are governed by
\begin{gather}
	\dfrac{\partial \mathbf{u}}{\partial t} + \mathbf{u}\cdot\nabla\mathbf{u} 
	+\mathbf{u}_0 \cdot\nabla\mathbf{u} +\mathbf{u}\cdot\nabla\mathbf{u}_0= - \nabla P + \theta \mathbf{e}_z + \nu \nabla^2 \mathbf{u},\\
	\dfrac{\partial \theta}{\partial t} + \mathbf{u} \cdot \nabla \theta + \mathbf{u}_0 \cdot \nabla \theta = N^2 u_z + \kappa \nabla^2 \theta,  \\
	\nabla \cdot \mathbf{u} = 0  \,,
\end{gather}
where $\mathbf{u}$ is a velocity perturbation (related to the total velocity $\mathbf{u}^\ast$ by $\mathbf{u}^\ast=\mathbf{u}+\mathbf{u}_0$), $P$ is a pressure variable, $\nu$ is the (constant) kinematic viscosity and $\kappa$ is the (constant) thermal diffusivity. The background reference density has been set to unity. We define our ``temperature perturbation" by $\theta=\alpha g T$, where $\alpha$ is the thermal expansion coefficient, $g$ is the acceleration due to gravity and $T$ is the usual temperature perturbation, so that $\theta$ has the units of an acceleration. The above equations describe perturbations to a linear background temperature profile $T(z)$, with uniform gradient $\alpha g \partial_z T = N^2$, where $N^2$ is the square of the buoyancy (Brunt--V\"ais\"al\"a) frequency, which is negative in a convection zone. This describes fluid that is hotter at the bottom of the domain than at the top.

We choose units such that $d$ is our unit of length and the thermal timescale $d^2/\kappa$ is our unit of time (so that velocities are measured in units of $\kappa/d$). The latter is chosen to enable  a direct comparison with the linear theory of convection (e.g.~\citealt{chandrasekhar_hydrodynamic_2013}), though we will later interpret our results in terms of the convective turnover timescale of the dominant eddies (or ``free-fall" timescale), since this is the most relevant physical timescale in this problem. The resulting non-dimensional equations are
\begin{gather}
	\dfrac{\partial \mathbf{u}}{\partial t} + \mathbf{u}\cdot\nabla\mathbf{u} 
	+\mathbf{u}_0 \cdot\nabla\mathbf{u} +\mathbf{u}\cdot\nabla\mathbf{u}_0= - \nabla P + \mathrm{Ra Pr}\,\theta \mathbf{e_z} + \mathrm{Pr} \nabla^2 \mathbf{u},\\
	\dfrac{\partial \theta}{\partial t} + \mathbf{u} \cdot \nabla \theta + \mathbf{u}_0 \cdot \nabla \theta = u_z + \nabla^2 \theta,  \\
	\nabla \cdot \mathbf{u} = 0  \,,
\end{gather}
where we have avoided introducing new ``hatted" dimensionless variables to avoid confusing the presentation. The key dimensionless parameters describing the convection are the Rayleigh number
\begin{gather}
\mathrm{Ra}=\frac{-N^2 d^4}{\nu \kappa},
\end{gather}
and the Prandtl number
\begin{gather}
\mathrm{Pr}=\frac{\nu}{\kappa}.
\end{gather}
We also define a scaled Rayleigh number
\begin{gather}
R=\frac{\mathrm{Ra}}{\mathrm{Ra}_c},
\end{gather}
where $\mathrm{Ra}_c=27\pi^4/4$ is the critical Rayleigh number for the onset of convection (in the absence of shear) with our adopted boundary conditions in the vertical
\citep{chandrasekhar_hydrodynamic_2013}. In principle, the shear could affect the onset of convection, but we find no evidence that this occurs for the adopted values of $a_0$.

With this non-dimensionalisation, according to mixing-length theory, the convective velocity in the limit of large $\mathrm{Ra}$ should be independent of the microscopic diffusivities, meaning that it should scale as $\sqrt{\mathrm{Ra \, Pr}}$, and therefore the dominant convective frequency ($\omega_c$) should similarly scale as $\sqrt{\mathrm{Ra \, Pr}}$. In the simulations below, we will compute a convective frequency $\omega_c = \bar{u}_z^{\text{rms}}/d$, using the time-averaged rms vertical velocity defined using
\begin{gather}
\label{convvel}
   u_z^{\text{rms}} = \sqrt{\langle u_z^2\rangle},
\end{gather}
where angled brackets denote a volume average and an overbar denotes a time average. This is observed to scale as $\sqrt{\mathrm{Ra \, Pr}}$, as expected.

In this work we will first explore laminar cases close to the onset of convection, with $R \leq 10$ and $\mathrm{Pr}=1$, in order to understand the physics behind the interaction between tides and convection. An advantage of this laminar regime is that we can compare our results with a complementary asymptotic analysis. We then move on to explore more turbulent cases with $R \geq 100$, which are more relevant for convection in stars. However, in a Sun-like star we expect $\mathrm{Ra} \in [10^{21}, 10^{24}]$ and $\mathrm{Pr} \in[10^{-7}, 10^{-3}]$ in the convection zone \citep{hanasoge2016}. Reaching such extreme parameter regimes with simulations (even those adopting the Boussinesq approximation) is unfeasible, and we are inevitably restricted to modest values of these parameters, such that $R\leq 10^4$ and $\mathrm{Pr}\gtrsim 10^{-2}$. We hope that our simulations can be used to understand the interaction between tides and convection in stars, but this inevitably requires us to extrapolate our results to the astrophysical parameter regime.

In the vertical we adopt stress-free, impenetrable, fixed temperature boundary conditions such that
\begin{equation}\label{BCs}
  \left.\begin{aligned}
  u_z&=0\\
  \theta&=0\\
  \partial_z u_x=\partial_z u_y&=0
\end{aligned}\right\} \quad\mathrm{on} \;\;\;\; z=0 \;\;\; \& \;\;\; z=1.
\end{equation}
We use a modified version of the Cartesian pseudo-spectral code SNOOPY for our simulations \citep{Lesur2005,lesur_angular_2010}.
This uses a basis of shearing waves with time-dependent horizontal wavevector components to deal with the linear spatial variation of $\mathbf{u}_0$. This is equivalent to using shearing-periodic boundary conditions in $x$. In the $y$-direction we assume periodic boundary conditions, and in the vertical ($z$), variables are expanded as either sines or cosines so that they satisfy the boundary conditions given by Eq.~\ref{BCs} (e.g.~\citealt{Cattaneo2003,lesur_angular_2010}). Flow variables are expanded such that 
\begin{gather}
u_x(\mathbf{x} ,t) = \Re \Big(\sum_{k_x,k_y,n} \hat{u}_x(t)\mathrm{e}^{i \mathbf{k}_\perp (t)\cdot\mathbf{x}}\cos n \pi z  \Big),
\end{gather}
where $\mathbf{k}_\perp(t)=(k_x(t),k_y)$, and the sum is over
\begin{equation}
    k_i = \left\{ \frac{2\pi k_i^{\ast}}{L_i} : k_i^{\ast} \in \left[\frac{-N_i}{2}, \frac{N_i}{2} -1\right] \cap \mathbb{Z} \right\} 
\end{equation}
and
\begin{equation}
    n = \{n \in \mathbb{Z} : n = [0, N_z -1] \},
\end{equation}
where $N_i$ is the number of modes in each dimension and $i\in\{x,y\}$ denotes the dimension.
A similar expansion is used for $u_y(\mathbf{x} ,t)$ and $p(\mathbf{x} ,t)$, but we also have
\begin{gather}
 [u_z(\mathbf{x} ,t),\theta(\mathbf{x} ,t)] = \Re \Big( \sum_{k_x,k_y,n} [\hat{u}_z(t),\hat{\theta}(t)]\mathrm{e}^{ i \mathbf{k}_\perp (t)\cdot\mathbf{x}}\sin n \pi z  \Big).
\end{gather}
The horizontal wavevector evolves according to
\begin{gather}
	\dot{\mathbf{k}}_\perp =  -S k_y \cos(\omega t ) \mathbf{e}_x .
\end{gather}
so that 
\begin{gather}
    k_x(t) = k_{x,0}-a_0 k_{y,0} \sin(\omega t ) \,, \\
    k_y(t)=k_{y,0} \,, 
\end{gather}
where
\begin{equation}
    \mathbf{k}_\perp (t=0)=(k_{x,0},k_{y,0}).
\end{equation}
The code uses a 3rd order Runga-Kutta time-stepping scheme and deals with the diffusion terms using an integrating factor. To accurately integrate (and analyse) high-frequency shear, we impose an additional time-step constraint such that the timestep $\delta t \ll 1/\omega$ (in addition to the usual time-step constraint due to the CFL condition). Further details regarding the code can be found in e.g.~\citet{Lesur2005} or \citet{lesur_angular_2010}. The nonlinear terms are fully de-aliased using the 3/2 rule.

\subsection{Quantities of interest}

We wish to explore the interaction between the background tidal flow $\mathbf{u}_0$ and convectively-driven flows. The corresponding volume-averaged energy transfer is described by the Reynolds stress component
\begin{gather}
R_{xy}(t) = \frac{1}{V} \int_V u_x u_y \, \text{d}V,
\end{gather}
where $V=L_x L_y$ is the volume of our simulated domain. We are interested in the response of this Reynolds stress component at the frequency of the tidal forcing. 

We are most interested in exploring the effective viscosity $\nu_E$, which quantifies the rate of energy transfer between the tidal and convective flows. The value of $\nu_E$ represents the viscosity that is required to produce the same energy transfer rate from/to the convective flow to/from the tidal flow as the simulated flow, which may be turbulent. Following \cite{GoodmanOh1997} and \cite{ogilvie_interaction_2012}, we may define $\nu_E$ by equating the mean rate at which the shear does work on the flow 
\begin{equation}
    - \frac{1}{V} \int_V \mathbf{u} \cdot (\mathbf{u} \cdot \nabla ) \mathbf{u}_0 \, \text{d} V= -S\cos\omega t R_{xy}(t),
\end{equation}
with the mean rate at which energy is dissipated by a viscosity $\nu_E$ acting on the background tidal flow
\begin{equation}
    \frac{2\nu_E}{V} \int_V e^0_{ij}e^0_{ij} \, \text{d} V=\nu_E S^2 \cos^2\omega t,
\end{equation}
where $e^0_{ij}=\frac{1}{2}(\partial_i u_{0,j}+\partial_j u_{0,i})$ is the rate-of-strain tensor for the background flow. The effective viscosity at the frequency of the shear is therefore
\begin{align}\label{maths_effective_viscosity}
	\nu_E(\omega) =  \frac{-2}{a_0\omega (T-T_0)}\int_{T_0}^{T} R_{xy}(t) \cos(\omega t) \, \text{d}t  \,. 
\end{align}
where $T_0$ is an appropriate initial time and $T-T_0$ is (strictly) an integer number of periods of the shear flow $2\pi/\omega$. This gives the response that is out of phase with the tidal displacement (and in phase with the tidal shear). In our more turbulent simulations we will typically integrate for hundreds or thousands of tidal periods to get well-converged values for $\nu_E$, and $T-T_0$ will not necessarily be taken to be an integer number of tidal periods. 

In the limit of low frequencies (and small $a_0$) the tidal flow becomes quasi-steady as the tidal period becomes very long compared with the convective timescale, and so $\nu_E$ represents the ``eddy viscosity" of convection, and we therefore expect it to scale as a convective velocity multiplied by a lengthscale, as predicted by mixing-length theory. This would give the prediction $\nu_E\propto \sqrt{\mathrm{RaPr}}$ for $\omega\ll\omega_c$. We will explore the frequency dependence of $\nu_E$, as a function of $\omega/\omega_c$. Note that $\nu_E$ can be related to the potential Love number or (modified) tidal quality factor $Q'$, and is therefore the most relevant quantity for tidal dissipation. For example, the orbital period derivative $\dot{P}/P$ due to tidal dissipation is proportional to $\nu_E$. The effective viscosity is related to the misalignment between the tidal bulges and the line of centres of the star and planet.

It can be shown (e.g.~\citealt{ogilvie_interaction_2012}; see also Appendix \ref{appendix_extension_of_ol2012}) that the fluid responds viscoelastically to high-frequency shear, and that the dominant response is elastic (with a weaker viscous component). To explore this component, and to compare with asymptotic theories (in Appendix \ref{appendix_extension_of_ol2012}), we will also compute the effective elasticity of the flow. This is less directly relevant for tidal dissipation than the effective viscosity, but is important for quantifying the amplitude of the tidal response, and could be important for e.g.~modifying the rates of non-dissipative tidally-driven apsidal precession. We can obtain an effective elasticity $S_E$ by considering the ratio of tidal shear stress to shear strain, which is a measure of the deformation (e.g.~\citealt{braviner_stellar_2015}). This gives the quantity
\begin{align}\label{maths_effective_elasticity}
	S_E(\omega) = - \frac{2   \int_{T_0}^{T}  R_{xy} \sin(\omega t)  \, \text{d}\,t}{a_0 (T-T_0)}    \,,
\end{align}	
which we will also compute in our simulations.
This measures the response that is in phase with the tidal displacement (out of phase with the tidal shear). 

In our simulations we will evaluate both $\nu_E$ and $S_E$ by using a large time interval $T-T_0$ such that these quantities are adequately converged. An equivalent way of evaluating $\nu_E$ and $S_E$ is by using the Fourier transform of $R_{xy}(t)$ \citep{ogilvie_interaction_2012} as
\begin{equation}
	\nu_E(\omega) = \Re \left(\frac{\hat{R}_{xy}(\omega)}{S \pi} \right)\,, \\
	S_E (\omega)= \Im \left(\frac{\hat{R}_{xy}(\omega)}{a_0 \pi} \right)\,,
\end{equation}
where we select the value of the Fourier transformed quantity at the shear frequency. We prefer to use the integral forms given above, since we found this alternative approach to be very sensitive to the precise frequencies chosen, and the numerical errors obtained were considerably larger.

\subsection{Parameter survey}

In our simulations we will vary the scaled Rayleigh number $R$, which measures the strength of the convection, the tidal frequency $\omega$ and amplitude $a_0$. In a few simulations we will also vary $\mathrm{Pr}$ (which is set to unity otherwise). Unless otherwise stated we will take the dimensions of the box to be $(L_x,L_y,d) = (2,2,1)$, though $L_x$ and $L_y$ will be varied in a number of cases below. The initial conditions will be small amplitude, solenoidal, homogeneous random noise for the velocity field. This is initialized using the system clock so that each simulation has unique initial conditions (to high probability).

These simulations are much more computationally demanding than those that only aim to explore the dynamics of the convection, since we must integrate them for multiple tidal periods and for many convective timescales to accurately probe the interaction between tides and convection. This is the reason that we have limited our study to modest values of $R\leq 10^3$ in this paper.

\section{Laminar convection with $R=2, 5$ and 10}
\label{LaminarR2}

We begin our investigation by exploring the interaction between tidal flows and laminar convection with $R=2, 5$ and 10. By considering the critical Rayleigh number for each mode (e.g.~\citealt{chandrasekhar_hydrodynamic_2013}), we can show that the modes that are first unstable with $L_x=L_y=2$ have the vertical wavenumber $n=1$ and are 2D $y$-aligned rolls with $n_x=\pm 1,n_y=0$ or 2D $x$-aligned rolls with $n_x=0,n_y=\pm1$ (these modes first onset when $R\gtrsim 1.05$), where we have defined integers $n_x$ and $n_y$ by taking $k_x=2\pi n_x/L_x$ and $k_y=2\pi n_y/L_y$. With the larger box $L_x=L_y=4$, convection occurs when $R>1$ by exciting 3D modes with $n_x=\pm1$ and $n_y=\pm1$ (which have a horizontal wavenumber magnitude of $\pi/\sqrt{2}$). The advantage of simulating laminar convection when only a small number of modes are unstable is that it allows us to explore the dependence of the effective viscosity on the nature of the flow most easily. We can also compare our results with a complementary asymptotic analysis (see Appendix~\ref{appendix_extension_of_ol2012}). As $R$ is increased, additional modes become linearly unstable.

\subsection{Rolls aligned with $y$ with $L_x=L_y=2$}
\label{yaligned}

\begin{figure}
    \includegraphics[width=\columnwidth]{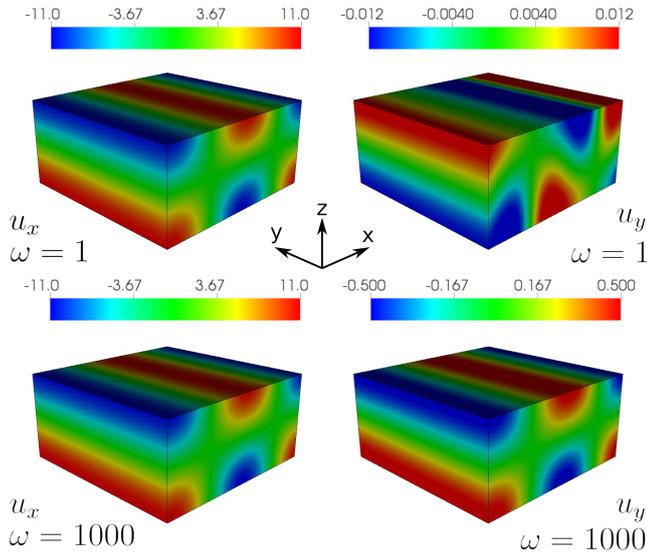}
	\caption{Flow structure for $y$-aligned rolls in a snapshot with $R=2$, with $\omega = 1$ and $\omega=1000$ (top and bottom respectively), both with $a_0 = 0.05$. The plotted quantities are $u_x$ (left) and $u_y$ (right). Note that only $u_y$ differs noticeably between low and high frequency cases.}
	\label{fig_flow_structure_y_aligned}
\end{figure}

We first select simulations with initial conditions such that 2D $y$-aligned convection rolls are preferentially excited, which saturate nonlinearly with an approximately steady amplitude. These simulations were performed with a resolution of at least ($N_x,N_y,N_z)=(64,64,16)$, which was found to be adequate. In the absence of shear, such a linear mode would have nonzero $u_x$ and $u_z$ velocity components (but $u_y=0$) and be independent of $y$. The shear (plus weak nonlinearity) induces an additional oscillatory $u_y$ component of the velocity, which is typically much smaller than that of the unperturbed convection roll. We present the horizontal flow structure for $u_x$ and $u_y$ in a snapshot from two simulations with $R=2$, with $\omega=1$ (top panels) and $\omega=1000$ (bottom panels) in Fig.~\ref{fig_flow_structure_y_aligned} at times $t=100$ and $29.9$, respectively. In these simulations $\omega_c\approx 5.5$, which means that $\omega=1$ is in the low-frequency regime ($\omega/\omega_c < 1$), and $\omega=1000$ is in the high-frequency regime ($\omega/\omega_c\gg 1$). The shear does not strongly modify the convection in these simulations, and $u_x$ is similar in both cases. On the other hand, the spatial structure of $u_y$ differs between the low and high-frequency regimes, as shown in the right panels of Fig.~\ref{fig_flow_structure_y_aligned}, indicating that we might expect the Reynolds stresses to differ. The flow is qualitatively similar in simulations with $R=5$ and 10.

\begin{figure}
	\includegraphics[width=\columnwidth]{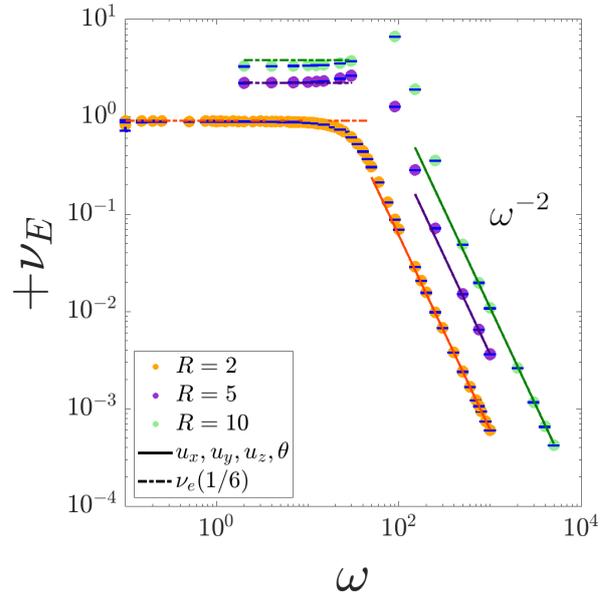}
	\caption{Frequency-dependence of $\nu_E(\omega)$ for laminar $y$-aligned convection rolls with $R\in\{2,5,10\}$, indicating that $\nu_E$ is reduced at high frequencies such that $\nu_E\propto \omega^{-2}$. The mixing-length theory prediction for $\nu_E$, with a constant of proportionality of $1/6$, such that $\nu_E=\frac{1}{6}\nu_e$, is plotted as the horizontal line, which matches the data for low frequencies. We also plot the prediction from an independent asymptotic analysis (Appendix \ref{appendix_extension_of_ol2012}) as the solid lines at high frequency. These are all proportional to $\omega^{-2}$ and are in excellent agreement with the data. Error bars for $\nu_E$ are plotted, but these are very small since the flow is laminar.}
	\label{figure_nu_E_vs_frequency_y_aligned}
\end{figure}

We compute $\nu_E$ using Eq.~\ref{maths_effective_viscosity}, and this is plotted as a function of $\omega$ for these simulations with $R=2, 5$ and 10 in Fig.~\ref{figure_nu_E_vs_frequency_y_aligned}. Each value of $R$ exhibits two distinct behaviours. For $\omega<20$ i.e. ``low frequencies", we find $\nu_E$ to be independent of $\omega$. For $\omega > \{30,90,200\}$ (for $R=2,5,10$, respectively), we find that the effective viscosity is reduced and decays with frequency such that $\nu_E \propto \omega^{-2}$. This matches the results of \cite{ogilvie_interaction_2012} for homogeneous convection and \cite{braviner_stellar_2015} for ABC flow. In Appendix \ref{appendix_extension_of_ol2012} we provide an asymptotic analysis which explains why laminar convection gives $\nu_E\propto \omega^{-2}$. We also provide a simpler mathematical derivation of this result for steady $y$-aligned rolls in Appendix \ref{appendix_simple_model}. Such a high-frequency scaling is in accordance with \cite{goldreich_turbulent_1977}, but their model was based on Kolmogorov turbulence rather than laminar flow, and it disagrees with \cite{zahn_les_1966}.  Note that the amplitude of the $u_y$ component increases with $\omega$ (e.g.~Fig.~\ref{fig_flow_structure_y_aligned}), following the behaviour of the oscillatory shear flow $\mathbf{u}_0$, but this component becomes increasingly in phase with the tidal displacement at high frequencies, thereby reducing $\nu_E$. On this figure, we have included simulations with several different values of $a_0 = \{0.01, 0.02, \dots , 0.09, 0.1 \}$, but these are not highlighted in the figure since our results were observed to be independent of $a_0$ for these values. This is consistent with us probing the regime of linear tides.

To get converged results for $\nu_E(\omega)$, it is important to integrate these simulations for long enough. In order to check this we plot the integral that arises in the expression for $\nu_E$ (Eq.~\ref{maths_effective_viscosity}) for various end times $T$, excluding terms outside of the integral, and plot its variation with time $T$. We then determine the linear regression line for this quantity over a time interval, $[T_0,T]$, such that the exponential convective growth phase is omitted (thus $T_0 > 0$). This linear fit provides the value of the integral in Eq.~\ref{maths_effective_viscosity}, allowing $\nu_E$ to be calculated. We evaluate error bars conservatively at $2\sigma$ (where $\sigma$ is the standard deviation) from the regression line. In laminar simulations with small $R$, we find that calculating the integral in Eq.~\ref{maths_effective_viscosity} directly gives similar results. But in more turbulent runs with larger $R$, the above approach was found to give cleaner results. A similar approach is carried out for $S_E$ (Eq.~\ref{maths_effective_elasticity}).

For low frequencies, and in particular as $\omega \rightarrow 0$ such that the tidal shear becomes quasi-steady, we expect $\nu_E$ to be approximated well by the mixing-length theory of convection if the flow is turbulent, i.e. by $\nu_E\propto \nu_e = u_{\text{mlt}} \ell$. We define $\nu_e$ to be an effective viscosity from MLT (without any constant multiplicative factors), $u_{\text{mlt}}$ to be the mixing-length velocity, which we can approximate by $u_{\text{mlt}} \approx \bar{u}_z^{\text{rms}}$, defined in equation~\ref{convvel}, after a suitable time averaging, and $\ell\sim d$ is a length-scale. We also expect $\nu_e\propto u_{\text{mlt}} \propto \sqrt{\mathrm{RaPr}}$ when $R\gg 1$. Even in laminar convection, where we might not expect mixing-length theory to apply, we find the simulations to be very well by $\nu_E\approx\frac{1}{6}\nu_e$ for $\omega\lesssim \omega_c$, as is shown in Fig.~\ref{figure_nu_E_vs_frequency_y_aligned}. Note that a frequency-independent $\nu_E$ is consistent with a constant tidal lag-time for this tidal component (e.g.~\citealt{Alexander1973,Mignard1980,Hut1981}).

\begin{figure}
	\includegraphics[width=\columnwidth]{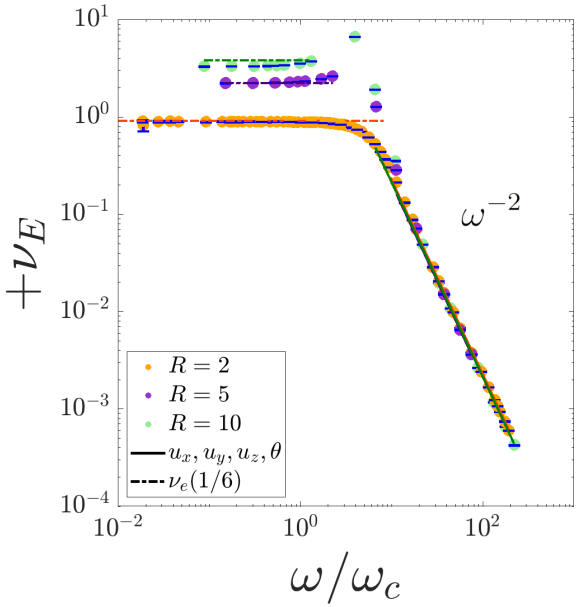}
	\caption{Same as Fig~\ref{figure_nu_E_vs_frequency_y_aligned}, showing $\nu_E(\omega)$ for laminar convection with $R=2,5,10$, but with frequencies scaled by the convective frequency, $\omega_c$. This clearly highlights the transition from frequency-independent $\nu_E$ to $\nu_E\propto \omega^{-2}$ occurs when $\omega\approx5\omega_c$ in these cases.}
	\label{figure_shearfreq_vs_VE_2sigma_mathematica_convectivescaled_yaligned}
\end{figure}

In Fig.~\ref{figure_nu_E_vs_frequency_y_aligned}, to conveniently separate results with different $R$, the unit of time is kept as the thermal timescale for both the $x$ and $y$ axes, therefore cases with larger $R$ values have larger convective velocities and frequencies in these units, and we expect the $\nu_E$ to increase with $R$. In Fig.~\ref{figure_shearfreq_vs_VE_2sigma_mathematica_convectivescaled_yaligned}, we instead scale frequencies with a convective frequency $\omega_c$ (without re-scaling the $y$-axis), where $\omega_c = \bar{u}^{\text{rms}}_z /d$ where $\bar{u}^{\text{rms}}_z$ is the time-averaged rms value of $u_z$. This figure shows clearly that the transition from constant $\nu_E$ to $\nu_E \propto \omega^{-2}$ occurs when $\omega\approx 5\omega_c$. Note that there is also an enhancement in $\nu_E$ for $\omega\approx \omega_c$ in the simulations with $R=5$ and 10.

 For high frequencies such that $\omega\gtrsim \omega_c$, we expect the rapid oscillatory nature of the tidal shear to inhibit the effective viscosity. In Appendix \ref{appendix_extension_of_ol2012}, we present an asymptotic linear analysis using the method of multiple scales to calculate $\nu_E$ (and $S_E$) in the high frequency regime in which $\omega\gg \omega_c$. This calculation builds upon prior work by \cite{ogilvie_interaction_2012} by extending their formalism so that it applies to Rayleigh-B\'{e}nard convection. 
 
 Since we are here simulating laminar convection, the convective flow primarily consists of only a single convective mode, with an approximately steady spatial structure $u_z=\Re (\hat{u}_z \mathrm{e}^{i k_x + i k_y}\sin \pi z)$, and similarly for other variables. We use our simulation snapshots (e.g.~Fig.~\ref{fig_flow_structure_y_aligned}) to fit the amplitude of the flow (such as $\hat{u}_z$) and input this into a Mathematica script that calculates the asymptotic predictions for $\nu_E$ and $S_E$ following Appendix \ref{appendix_extension_of_ol2012}. For $y$-aligned rolls, $k_y=0$, and since we are just above onset, we can use the linearised equations to approximate the relations between $u_x$, $u_y$, $u_z$ and $\theta$, so that only a single value is needed for this fit (we use the amplitude of $u_z$). Alternatively, we can fit the amplitudes of each of $u_x$, $u_y$, $u_z$ and $\theta$ separately, but still assuming a single linear mode for each variable, by taking $\max(u_x)$ (and similarly for other variables) within the domain. We use the latter method throughout this paper. Further details regarding this calculation are relegated to Appendix \ref{appendix_extension_of_ol2012}. 
 
 The asymptotic analytical prediction is shown as the solid lines at high $\omega$ in Figs.~\ref{figure_nu_E_vs_frequency_y_aligned} and \ref{figure_shearfreq_vs_VE_2sigma_mathematica_convectivescaled_yaligned}. The theory is in excellent agreement with our simulations for each $R$ value considered. For $y$-aligned rolls, we also obtain good agreement between each of the methods of evaluating the magnitudes of the velocity components. This provides an independent check that our simulations are correctly probing the high frequency regime.

\begin{figure}
	\includegraphics[width=\columnwidth]{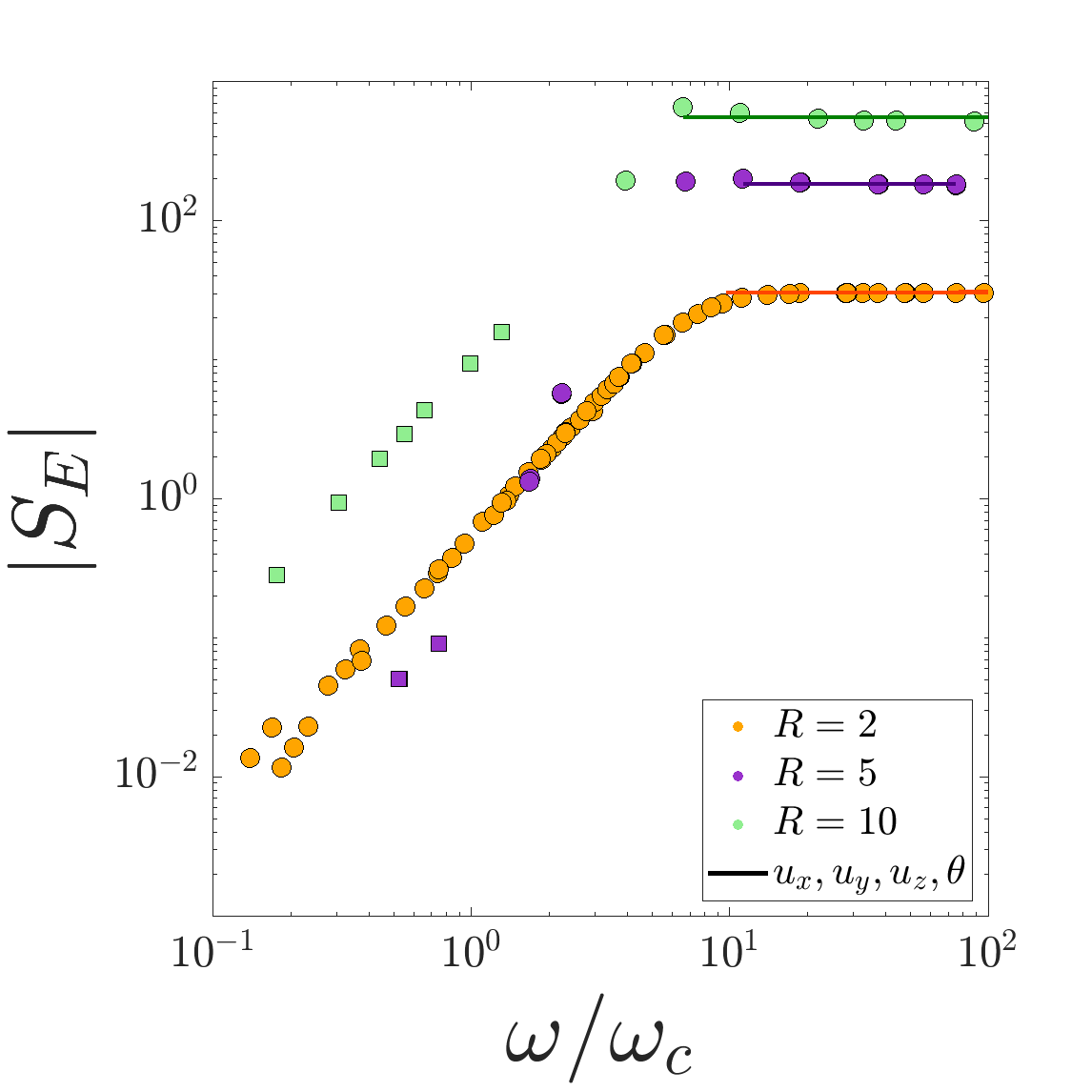}
	\caption{Frequency-dependence of $S_E(\omega)$ for laminar $y$-aligned convection rolls with $R\in\{2,5,10\}$, where frequencies have been scaled by the convective frequency, $\omega_c$. Theoretical predictions of $S_E$ for $\omega\gg \omega_c$ are computed using the theory presented in Appendix \ref{appendix_extension_of_ol2012} and are plotted here as solid lines. Square symbols denote negative values of $S_E$ while circles denote positive values. We have omitted error bars since they are small, but they typically fit within the symbol plotted except for small $\omega$ ($\omega/\omega_c \lesssim 0.1$).}
	\label{figure_shearfreq_vs_SE_2sigma_mathematica_convectivescaled_yaligned}
\end{figure}

We show the corresponding results for the effective elasticity, $S_E$, in Fig.~\ref{figure_shearfreq_vs_SE_2sigma_mathematica_convectivescaled_yaligned}, where we have scaled frequencies by the convective frequency. We find that $S_E \propto \omega^{1.77}$ for $\omega\lesssim 5\omega_c$, which is an empirical scaling, with a transition to a frequency-independent $S_E$ for $\omega\gtrsim 5\omega_c$. This frequency independence is in qualitative agreement with the findings of \cite{braviner_stellar_2015} for the related problem of ABC flow. The value of $S_E$ in the high frequency regime from the theoretical prediction in Appendix \ref{appendix_extension_of_ol2012} is shown as the solid lines.
Error bars are omitted from this figure for clarity. As with $\nu_E$ we observe that $S_E$ is independent of $a_0$ for the values of $a_0$, $R$ and $\omega$ simulated here (so we do not explicitly show this). One interesting feature is that for $R\in\{5,10\}$ we observe a statistically significant sign change in $S_E$, going from negative at low $\omega$ to positive at high $\omega$, at $\omega\approx \omega_c$, which does not occur for $R=2$. We have also observed that $S_E$ appears to transition from increasing with $\omega$ to being flat at approximately the same frequency (as opposed to scaled frequency $\omega / \omega_c$) in each case. The case with $R=5$ is intermediate between the case with $R=2$, where $S_E$ is always positive, and the case with $R=10$, which exhibits a change in sign. As a consequence we observe a more pronounced transition between positive and negative values for $R=5$.

Note that $S_E\gg \nu_E$ at high frequencies, emphasizing that the response is primarily elastic, with a weaker viscous component.

\subsection{Rolls aligned with $x$ with $L_x=L_y=2$}
\label{xaligned}

In this section we analyse similar simulations to \S~\ref{yaligned} except that the initial conditions select $x$-aligned convection rolls. For these cases we use the same box size and set of $R$ values but the resolution is chosen to be $(N_x,N_y,N_z)=(64,64,32)$, which was found to be sufficient in all cases. In the absence of shear such linear convection solutions would have nonzero $u_y$ and $u_z$ velocity components, with $u_x=0$, and they would saturate nonlinearly with a steady amplitude. The flow is similar to that shown in Fig.~\ref{fig_flow_structure_y_aligned} except that the roll is aligned with $x$ rather than $y$. The $u_x$ component behaves differently in the high and low frequency cases however.

\begin{figure}
	\includegraphics[width=\columnwidth]{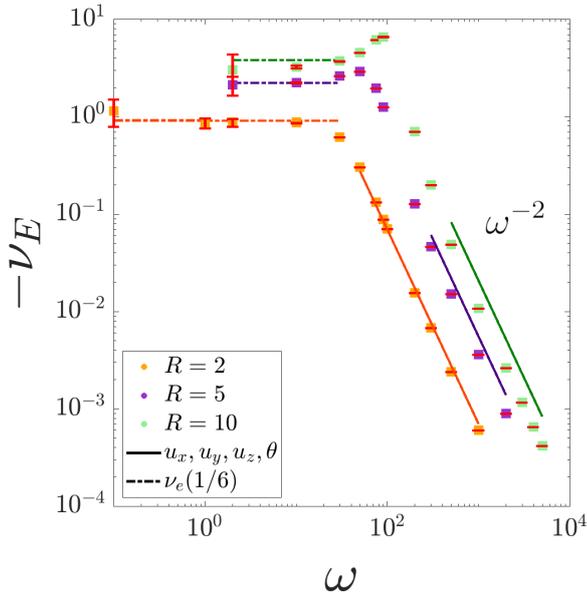}
	\caption{Frequency-dependence of $\nu_E(\omega)$ for laminar $x$-aligned convection rolls with $R\in\{2,5,10\}$. This demonstrates the surprising result that $\nu_E$ is negative in these cases. $\nu_E$ is also reduced at high frequencies such that $\nu_E\propto \omega^{-2}$. The mixing-length theory prediction for $|\nu_E|$, with a constant of proportionality of $1/6$, such that $|\nu_E|=\frac{1}{6}\nu_e$, is plotted as the horizontal line, which matches the magnitude of the data for low frequencies. We also plot the prediction from an independent asymptotic analysis (Appendix \ref{appendix_extension_of_ol2012}) as the solid lines at high frequency, which confirms our observation of negative $\nu_E$. These predictions are proportional to $\omega^{-2}$ and are in excellent agreement with the data. Error bars for $\nu_E$ are plotted, but these are very small in most cases as the flow is laminar.}
	\label{figure_shearfreq_vs_VE_2sigma_mathematica}
\end{figure}

We show $\nu_E$ as a function of $\omega$ (not scaled by $\omega_c$ so as to separate data with different $R$) in Fig.~\ref{figure_shearfreq_vs_VE_2sigma_mathematica}, using the same method as for the $y$-aligned simulations in the previous section. The most surprising feature is that $\nu_E$ is negative for all $\omega$. This indicates that the convective flow is transferring energy to the tidal flow. Negative values were previously obtained in the (more turbulent) simulations of \cite{ogilvie_interaction_2012}, though only at high frequency and with error bars that could not conclusively rule out positive values. Our results in Fig.~\ref{figure_shearfreq_vs_VE_2sigma_mathematica} conclusively demonstrate that statistically significant negative values\footnote{We have also performed preliminary simulations to explore whether $\nu_E$ remains negative for very small $\omega$, and we find some evidence of a possible transition to positive values for $\omega \lesssim 10^{-4}$, although exceptionally long run times were required and the error bars were large in these cases.} are attained in laminar convection consisting of rolls aligned with $x$.

As in the previous section, the value of $|\nu_E|$ is approximately independent of frequency until $\omega\approx 5\omega_c$ (this is most clearly shown by scaling frequencies with $\omega_c$, but we have omitted this figure to save space), above which it falls off with frequency such that $|\nu_E|\propto\omega^{-2}$. We have also analysed simulations with $R=2$ in which we vary the tidal amplitude such that $a_0\in\{0.0005, 0.005, 0.05 \}$, finding that our results for $\nu_E$ (and $S_E$) are independent of $a_0$ for the simulated values. 

As in the previous section, we have extended the theory of \cite{ogilvie_interaction_2012} to also apply to $x$-aligned convection rolls, as explained in Appendix~\ref{appendix_extension_of_ol2012}. This allows us to independently predict $\nu_E$ (and $S_E$) in the high frequency regime by fitting the amplitude of the convection roll. Our results are shown as the solid lines in Fig.~\ref{figure_shearfreq_vs_VE_2sigma_mathematica}. These results are in good agreement with our simulation data, which provides independent confirmation of the negative values for $\nu_E$, as well as the quadratic fall-off with $\omega$. Note that the asymptotic prediction departs most strongly from the simulation results for $R=10$, presumably because we have adopted a single mode in the analysis, which is no longer strictly applicable. 

\begin{figure}
	\includegraphics[width=\columnwidth]{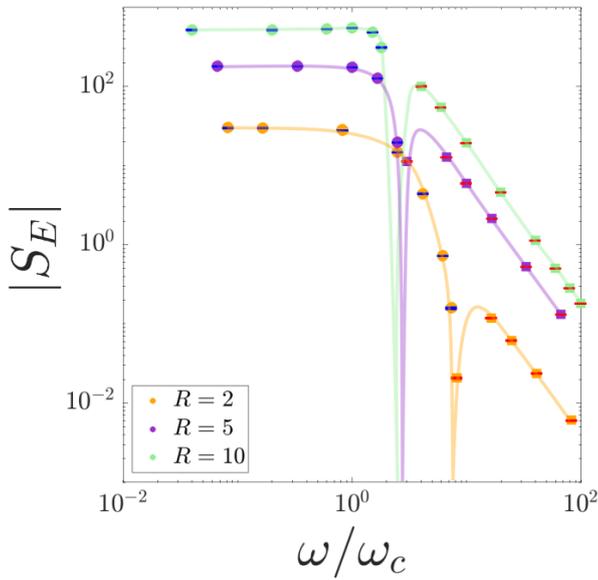}
	\caption{Frequency-dependence of $S_E(\omega)$ for laminar $x$-aligned convection rolls with $R\in\{2,5,10\}$, where frequencies have been scaled by the convective frequency, $\omega_c$. The lines connect the data points with each value of $R$ and highlight the sign change in $S_E$ as $\omega$ is varied. 
	Note that the high-frequency behaviour is not predicted according to the theory in Appendix~\ref{appendix_extension_of_ol2012}, since it requires the computation of higher-order terms. In this plot negative (positive) values of $S_E$ are denoted by red error bars on square symbols (blue error bars on round symbols).
	}
	\label{figure_shearfreq_vs_SE_2sigma_mathematica_convectivescaled}
\end{figure}

We show $S_E$ as a function of $\omega$ in Fig.~\ref{figure_shearfreq_vs_SE_2sigma_mathematica_convectivescaled}, where we have scaled the frequencies with $\omega_c$. For $y$-aligned rolls, we previously found that $S_E$ increased for small $\omega$ and was independent of frequency for large $\omega$, similar to the results of \cite{braviner_stellar_2015} for ABC flow. However, for $x$-aligned rolls, we instead find $S_E$ to be independent of frequency for small $\omega$ and to fall off rapidly with frequency such that $S_E\propto \omega^{-2}$ at high frequencies. For each of the $R$ values considered, $S_E$ also changes sign at a certain frequency, transitioning from positive at low $\omega$ to negative at high $\omega$. This again differs from $y$-aligned rolls, where $S_E$ was instead negative for small $\omega$. In both $x$ and $y$-aligned cases, $S_E$ is negative when it is small and varying with $\omega$, and is positive when it is larger in magnitude and independent of $\omega$. 

The asymptotic theory presented in Appendix~\ref{appendix_extension_of_ol2012} predicts $S_E=0$, and so is unable to explain the high frequency behaviour of $S_E$ for $x$-aligned rolls shown in Fig.~\ref{figure_shearfreq_vs_SE_2sigma_mathematica_convectivescaled}. However, this is consistent with our observation that $S_E\sim \omega^{-2}$ at high frequency, since this implies we must consider higher order terms (in $\omega^{-1}$) to explain this with theory.

%%%%%%%%%%%%%%%%%%%%%%%%%%%%%%%%%%%%%%%%%%%%%%%%%%%%%%%%%%%%%%%%%%%
\subsection{$R=2$ with larger boxes such that $L_x=L_y=4$}
\label{R24x4x1}

%omega = 1000, folder cases 014 and 021. R= 2. a0=0.05. case 014 is positive nuE and case 021 is negative nuE. 014 is at 97 time units and 021 is at  290 time units
\begin{figure}
    \includegraphics[width=\columnwidth]{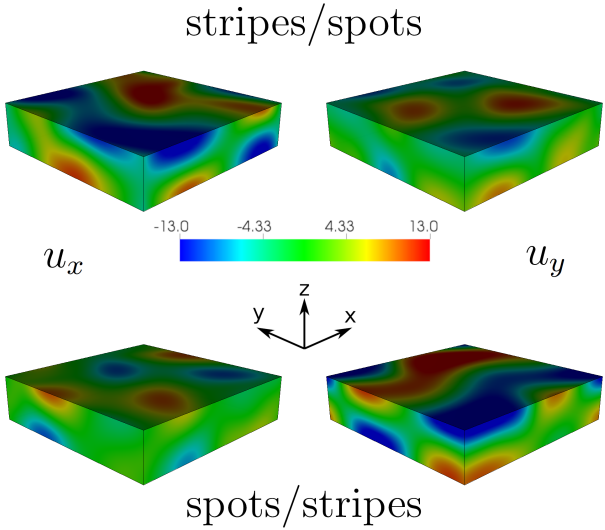}
	\caption{Snapshots of the flow for two cases with different initial conditions that we refer to as stripes/spots and spots/stripes (top and bottom, respectively) with $\omega = 1000$ at $t=97$ and $290$ (respectively). These have $R=2$ and $(L_x, L_y, L_z) = (4,4,1)$. The plotted quantities are $u_x$ (left) and $u_y$ (right).}
	\label{fig_flow_structure_4x4x1_R2}
\end{figure}

Our final set of simulations to explore the interaction between tidal shear and laminar convection have $R=2$ but with a larger box such that $L_x=L_y=4$ using a resolution of $(N_x, N_y, N_z) = (64, 64, 16)$. In this case, we expect multiple convective modes to be excited, including 3D modes with $n_x=\pm 1$, $n_y=\pm1$ as well as the quasi-2D $x$ and $y$-aligned convection rolls that we have studied in the previous two sections. Each mode is expected to provide its own contribution to the effective viscosity (and elasticity), which we can quantify and compare with theory. Since 3D modes are preferentially excited in these cases, unlike those with $L_x=L_y=2$, we can use this set of simulations to probe the contribution of 3D modes to $\nu_E$.

We show a snapshot of $u_x$ and $u_y$ in two simulations with $\omega=1000$ in Fig.~\ref{fig_flow_structure_4x4x1_R2}, which indicates that the flow consists of a superposition of a number of convective modes. We explored different random initial conditions and determined that there were two main flow configurations, as shown in the top and bottom panels of this figure. We will refer to these as stripes/spots (corresponding to the patterns in $u_x$ and $u_y$) and spots/stripes, respectively. These cases provide a more robust test of the asymptotic theory in Appendix.~\ref{appendix_extension_of_ol2012}, since many of the terms in the theory are identically zero when only an $x$ or $y$-aligned convection roll is considered.

\begin{figure}
	\includegraphics[width=\columnwidth]{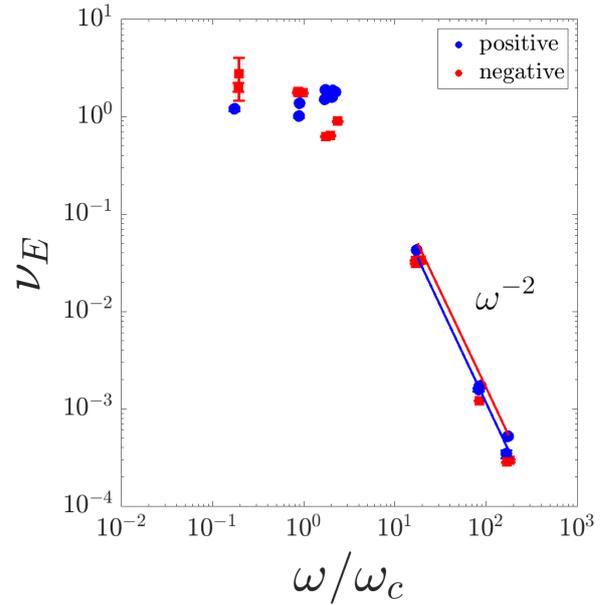}
	\caption{Frequency-dependence of $\nu_E(\omega)$ for laminar convection with $R=2$ in a box with $L_x=L_y=4$ for two different flow configurations. The asymptotic high-frequency prediction (shown as the solid lines), is computed by incorporating multiple modes in the theory in Appendix.~\ref{appendix_extension_of_ol2012}. We obtain very good agreement with the values and sign of $\nu_E$ in each case, corresponding with the two flows in Fig.~\ref{fig_flow_structure_4x4x1_R2}, where the stripes/spots flow has $\nu_E>0$ and spots/stripes flow has $\nu_E<0$.}
	\label{figure_shearfreq_vs_VE_2sigma_mathematica_convectivescaled_R2}
\end{figure}

We show the frequency-dependence of $\nu_E$ for each of these cases in Fig.~\ref{figure_shearfreq_vs_VE_2sigma_mathematica_convectivescaled_R2}. As in \S~\ref{yaligned} and \ref{xaligned}, both cases exhibit an approximately frequency-independent $\nu_E$ for $\omega\lesssim \omega_c$, and for $\omega\gtrsim \omega_c$, $|\nu_E|\propto \omega^{-2}$. Interestingly, the different flow configurations shown in Fig.~\ref{fig_flow_structure_4x4x1_R2} exhibit different signs for $\nu_E$, which remain the same for all $\omega$. The case with stripes/spots (top panels of Fig.~\ref{fig_flow_structure_4x4x1_R2}) has $\nu_E>0$ for all $\omega$ and the case with spots/stripes (bottom panels of Fig.~\ref{fig_flow_structure_4x4x1_R2}) has $\nu_E<0$. This agrees with what we might expect based on \S~\ref{yaligned} and \ref{xaligned}, since the stripes/spots flow contains a larger amplitude $y$-aligned roll component, and the spots/stripes flow contains a larger amplitude $x$-aligned roll component. These examples illustrate once more that for laminar convection, the effective viscosity can be negative.

Computing the asymptotic predictions using the approach outlined in Appendix.~\ref{appendix_extension_of_ol2012} is more difficult in this case, due to the presence of multiple modes that could each contribute to $\nu_E$. We can calculate the contribution from the most important modes by taking a horizontal Fourier transform of a set of flow snapshots to determine the amplitude of each Fourier $k_x,k_y$ mode in the flow (selecting the maximum value over all $z$). The amplitude of each mode may vary in time and so we use a suitable time-average to obtain these values. Once we have the amplitude of $u_x, u_y, u_z$ and $\theta$ for each mode, we compute their contributions to $\nu_E$ separately using the approach outlined in Appendix.~\ref{appendix_extension_of_ol2012}, and then sum up their contributions to obtain a prediction for $\nu_E$. We consider all modes which have a contribution within 3 orders of magnitude of the dominant (typically the largest scale) mode, and we find large wavenumber modes contribute less to $\nu_E$.

We show our prediction for the high-frequency behaviour of $\nu_E$ as the solid lines in Fig.~\ref{figure_shearfreq_vs_VE_2sigma_mathematica_convectivescaled_R2}, where each of the two flow configurations have been treated separately. These agree well with the simulation data, and are found to correctly predict the magnitude and sign of $\nu_E$ in each case, such that the spots/stripes case is negative and the stripes/spots case is positive. The case with negative values is slightly less well predicted by theory, probably because this depends more strongly on the $\theta$ component of the flow, which is more strongly modified by nonlinearity.

These simulations agree with the asymptotic theory in Appendix.~\ref{appendix_extension_of_ol2012} in cases where multiple modes contribute, and we have also shown that 3D laminar convection can exhibit a negative effective viscosity. Next, it is essential to determine how robust these results are to increasing $R$, allowing us to explore more turbulent convective flows.

\section{More turbulent convection with $R=100$ and $R=1000$}
\label{turbulent}

The simulations of laminar convection described in \S~\ref{LaminarR2} provide a starting point to explore the interaction between tidal flows and convection. Those cases had the significant advantage that the flow was sufficiently simple that we could compare our results with an independent asymptotic analysis. Since convection in stars is highly turbulent, it is important to study more astrophysically relevant cases with much larger values of $R$. We begin by studying cases with $R=100$, both with $\mathrm{Pr}=1$ and then with different $\mathrm{Pr}$, before moving on to cases with $R=1000$.

\subsection{$R=100$ with $L_x=L_y=2$}
\label{R100}

%case 040. a0 = 0.05. omega = 5000. vtk 90
\begin{figure}
	\includegraphics[width=\columnwidth]{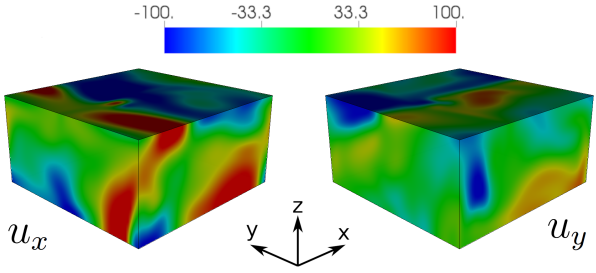}
	\caption{Flow in weakly turbulent simulations with $R=100$, with $\omega = 5000$ and $a_0=0.05$, where $u_x$ is plotted in the left panel and $u_y$ in the right panel at $t=90$. The flow is three-dimensional but is dominated by $x$ and $y$-aligned convection rolls, which can be seen by the tendency for $u_x$ to align with $y$ and $u_y$ to align with $x$.}
	\label{fig_flow_structure_R100}
\end{figure}

We now present results with $R=100$, which corresponds with a weakly turbulent regime. These simulations have $(L_x, L_y, L_z) = (2, 2, 1)$, are run with $(N_x,N_y,N_z)=(64,64,64)$, and result in a fully three-dimensional time-dependent convective flow. We show this in a snapshot with $\omega=5000$ at $t=90$ in Fig.~\ref{fig_flow_structure_R100}. The flow is qualitatively similar for all $\omega$ considered, and consists of several modes, including both $x$ and $y$-aligned rolls. This can be seen in Fig.~\ref{fig_flow_structure_R100} by observing that $u_x$ preferentially aligns with $y$ and $u_y$ aligns with $x$, indicating the dominance of these rolls in the flow, even if other components are also present.

\begin{figure}
	\includegraphics[width=\columnwidth]{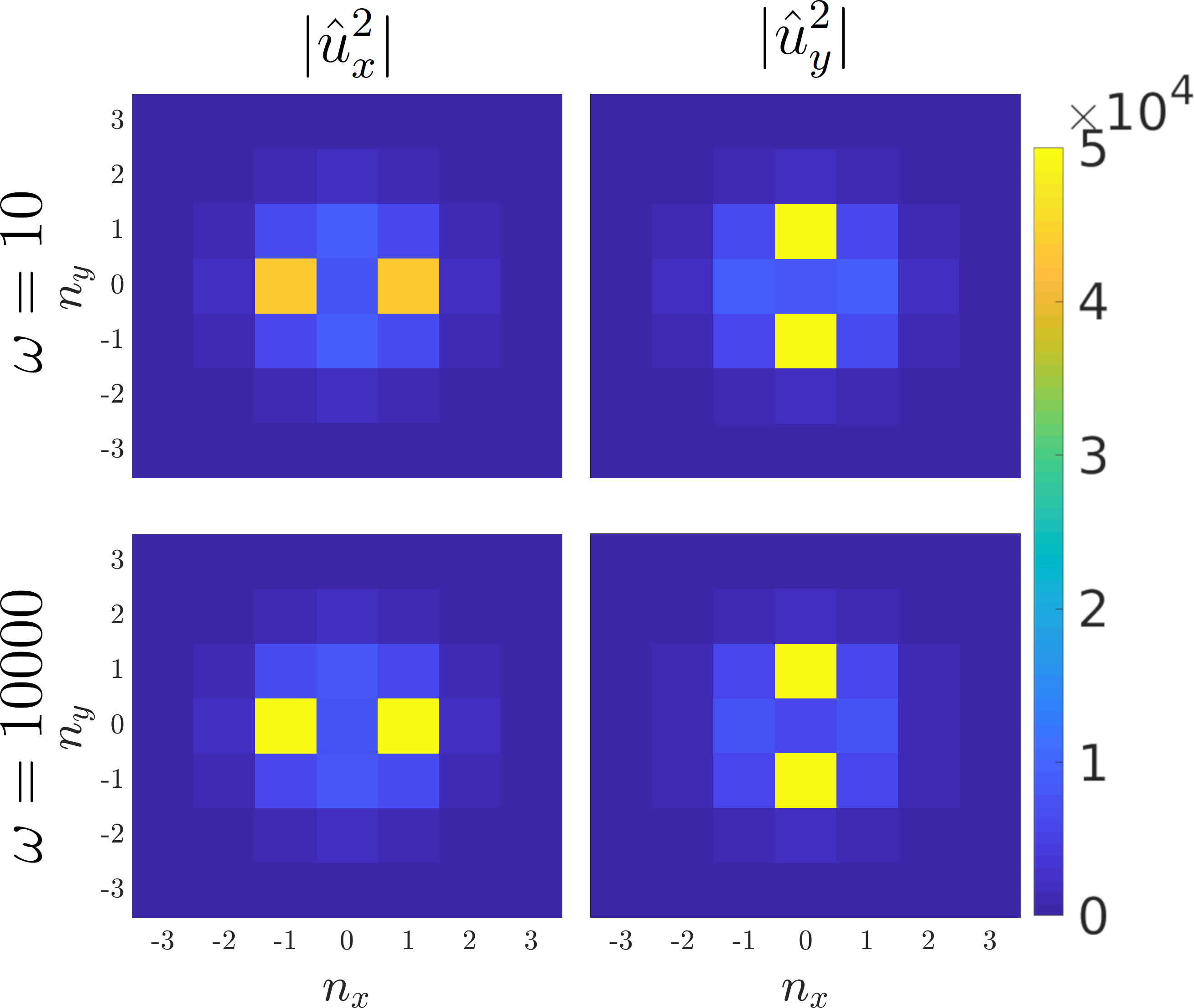}
	\caption{Vertically integrated and temporally averaged horizontal power spectrum of $|\hat{u}_x|^2$ (left) and $|\hat{u}_y|^2$ (right) as a function of $(k_x,k_y)$ in simulations with $R=100$ and $a_0= 0.05$, for two different frequencies with $\omega = 10$ and $\omega=10000$ (top and bottom, respectively). We plot the integer wavenumbers on each axis i.e.~$n_x=L_x k_x/2\pi$ and $n_y=L_y k_y/2\pi$. The flow is fully three dimensional but is dominated by a small number of large-scale modes.}
	\label{fig_ux_uy_R100_aligns}
\end{figure}

We can determine the dominant $(k_x,k_y)$ wavenumbers in the flow by evaluating the vertically integrated and temporally-averaged horizontal power spectrum of $|\hat{u}_x|^2$ and $|\hat{u}_y|^2$, which we plot in Fig.~\ref{fig_ux_uy_R100_aligns} for two different frequencies with $\omega=10$ and $\omega=10000$. We plot the integer wavenumbers on each axis i.e.~$n_x=L_x k_x/2\pi$ and $n_y=L_y k_y/2\pi$. This shows that the flow at both low and high frequencies contains multiple modes, but that the $x$ and $y$-aligned convection rolls with $(k_x,k_y)=(0,2\pi/L_y)$ or $(2\pi/L_x,0)$ are dominant. We have run several simulations with the same parameters with different random initial conditions and the flow has a similar spectrum in each case. 

\begin{figure}
	\includegraphics[width=\columnwidth]{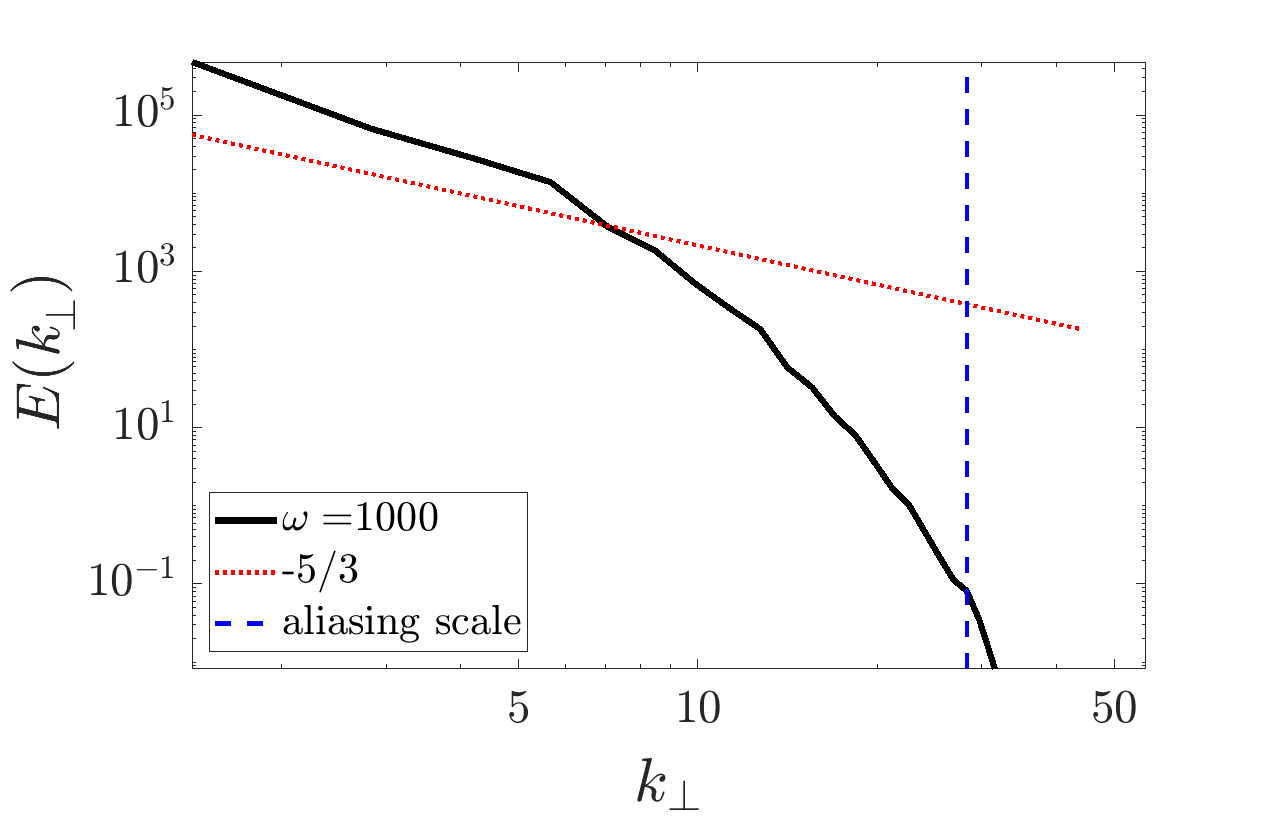}
	\caption{Temporally-averaged kinetic energy spectrum $E(k_\perp)$ as a function of $k_\perp$ in a simulation with $R=100$, $a_0 = 0.05$ and $\omega = 1000$. This is representative of all simulations with $R=100$. The red line represents the Kolmogorov scaling ($k^{-5/3}$) and the blue line shows the de-aliasing scale.}
	\label{figure_energy_spectrum_1D}
\end{figure}
In Fig.~(\ref{figure_energy_spectrum_1D}), we show the time-averaged and vertically-integrated kinetic energy spectrum $E(k_\perp)$ as a function of horizontal wavenumber $k_\perp$ (computed by summing up the mode energy in wavenumber rings of unit width in the $(k_x,k_y)$-plane, where $k_\perp=\sqrt{k_x^2+k_y^2}$). This is computed using the horizontal Fourier transform of the velocity field. This shows that the energetically-dominant scales are those on the size of the box, but that smaller scales are also present with non-negligible amplitudes. With $R=100$, the flow has a short inertial-like range, which can be seen by comparing the data (black line) with the Kolmogorov (-5/3) scaling (red line). This figure also demonstrates that our simulations are well-resolved horizontally, since the energy in wavenumbers close to the de-aliasing scale (blue dashed vertical line) is more than 6 orders of magnitude smaller than the peak.

\begin{figure}
	\includegraphics[width=\columnwidth]{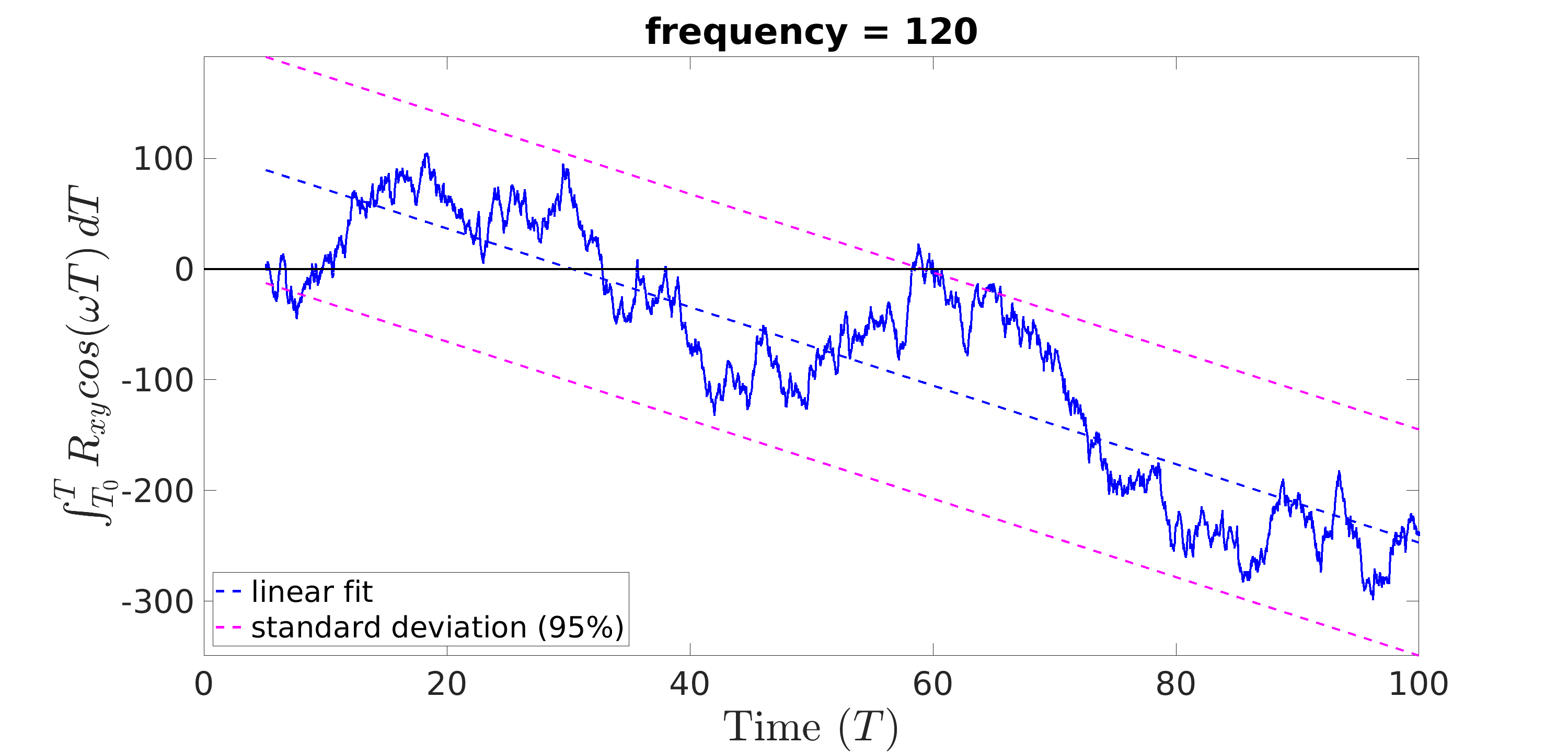}
	\includegraphics[width=\columnwidth]{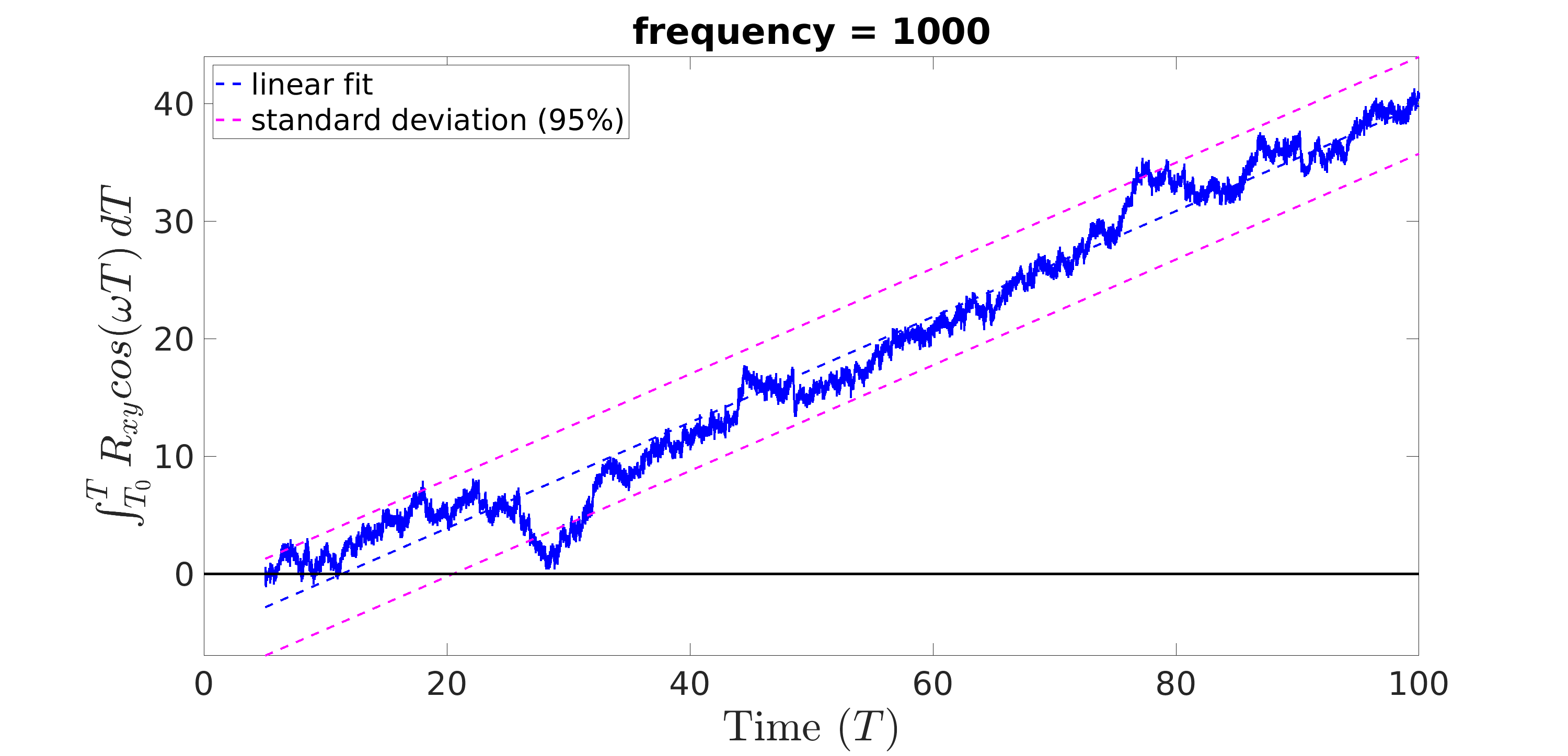}	
	\caption{Cumulative integrals used to compute $\nu_E$ based on Eq.~\ref{maths_effective_viscosity}, ignoring the factors outside the integral. Decreasing (increasing) behaviour indicates positive (negative) $\nu_E$. The blue dashed line shows the linear fit used to compute $\nu_E$, and the magenta dashed lines highlight the error, defined by $2\sigma$ from the mean trend. The cases shown here are for the low frequency, $\omega = 120$ (which has $\nu_E>0$), (top) and high frequency, $\omega =1000$ (which has $\nu_E<0$), (bottom).}
	\label{fig_cumulative_integrals_R100}
\end{figure}

Since the convective flow with $R=100$ is non-steady, accurately computing the effective viscosity (and elasticity) in this case is more challenging than for the laminar simulations. We previously outlined our approach to determine $\nu_E$ in \S~\ref{yaligned}, and in Fig.~\ref{fig_cumulative_integrals_R100} we show the results of computing the cumulative integral required to calculate $\nu_E$ as a function of the end-times $T$, omitting the constant multiplicative factors outside the integral in Eq.~\ref{maths_effective_viscosity}. This figure shows an example with low frequency ($\omega=120$) and another with high frequency ($\omega=1000$), along with the linear fit as the dashed blue line. The magenta dashed lines denote the error bars defined at 2 standard deviations from the mean. Both cases show temporal variability but exhibit a well-defined linear slope, indicating that the resulting values for $\nu_E$ are well-converged. In this figure, since we have omitted the numerical factors outside the integral in Eq.~\ref{maths_effective_viscosity}, including the sign, a trend that is decreasing (increasing) in Fig.~\ref{fig_cumulative_integrals_R100} represents a positive (negative) effective viscosity. This means that the case with $\omega=120$ has $\nu_E>0$ and $\omega=1000$ has $\nu_E<0$. The error bars are smaller for $\omega=1000$ since the simulation has been run for more tidal periods. We have analysed similar plots in all of our simulations to ensure that $\nu_E$ is always well-converged.

\begin{figure}
	\includegraphics[width=\columnwidth]{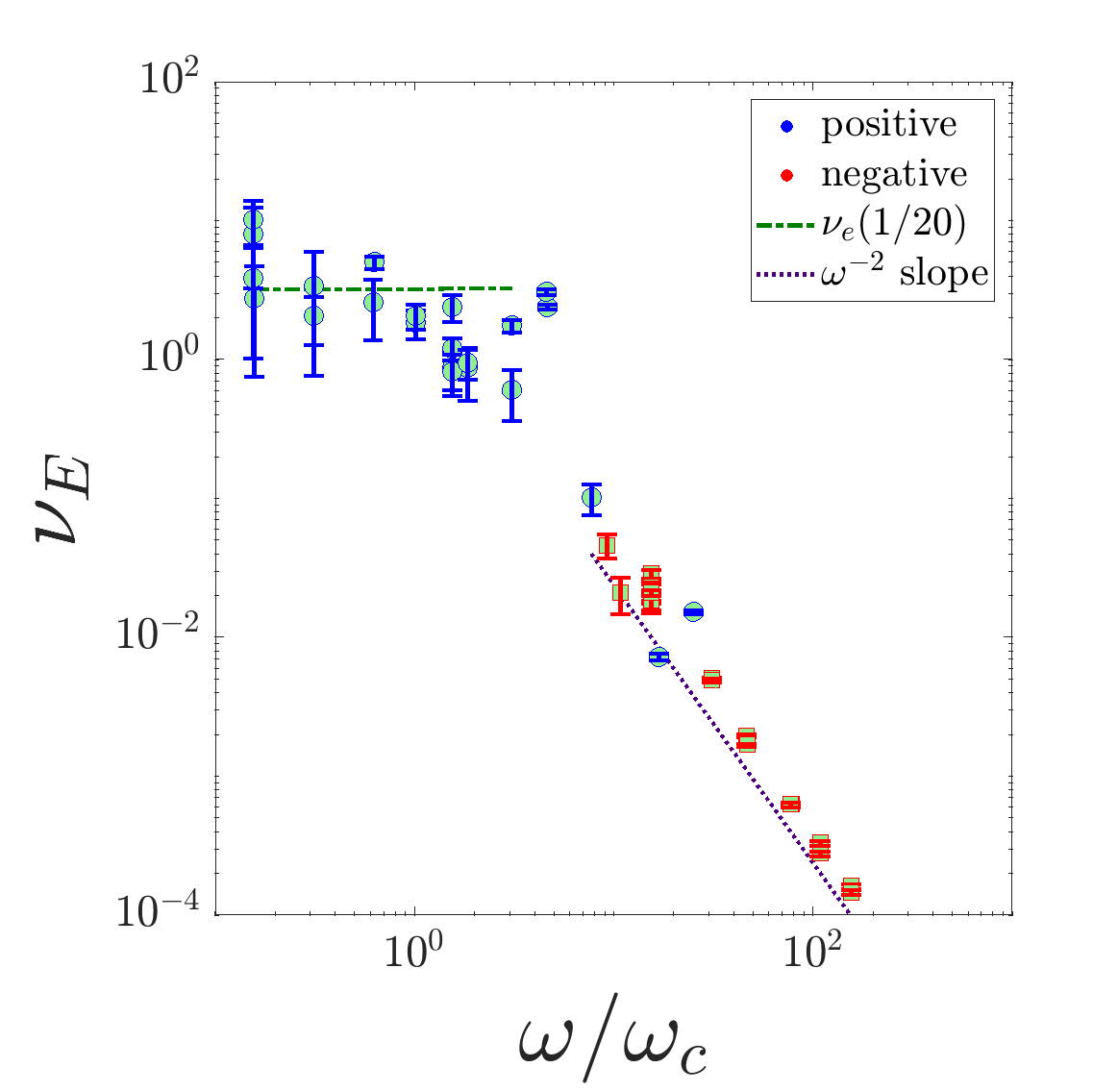}
	\caption{Plot of $\nu_E$ versus $\omega$ scaled by the convective frequency $\omega_c$ in simulations with $R=100$. Positive values are indicated by blue symbols and negative values by red. There is a clear transition from positive to negative values at $\omega/\omega_c\approx 10$. At high $\omega$ ($\omega/\omega_c\gtrsim 5$), we again find $|\nu_E|\propto \omega^{-2}$.}
	\label{figure_shearfreq_vs_VE_2sigma_mathematica_convectivescaled_R100}
\end{figure}

The results for $\nu_E$ are shown in Fig.~\ref{figure_shearfreq_vs_VE_2sigma_mathematica_convectivescaled_R100},  where we have scaled $\omega$ by the convective frequency. Similar to the laminar simulations presented in \S~\ref{LaminarR2}, this case also has a frequency-independent effective viscosity at low frequencies, with a transition to $\nu_E\propto \omega^{-2}$ at high frequencies. The transition occurs at $\omega/\omega_c\approx 5$ (where $\omega_c\approx 64$ in this case). The most surprising feature in Fig.~\ref{figure_shearfreq_vs_VE_2sigma_mathematica_convectivescaled_R100} is that the sign of $\nu_E$ changes at $\omega/\omega_c\approx 10$, with low $\omega$ having positive $\nu_E$ (plotted using blue symbols) and large $\omega$ having negative $\nu_E$ (plotted using red symbols). The negative effective viscosities at high frequencies are highly statistically significant. Indeed, the error bars -- obtained in the same way as those plotted in the bottom panel in Fig.~\ref{fig_cumulative_integrals_R100} -- on the points with $\omega/\omega_c>1$ are very small because these simulations have been run for many hundreds of tidal periods, indicating that the values of $\nu_E$ are very robust. This corroborates the tentative finding that negative effective viscosities are possible at high frequencies obtained by \cite{ogilvie_interaction_2012}. 

Also shown on Fig.~\ref{figure_shearfreq_vs_VE_2sigma_mathematica_convectivescaled_R100} is a low-frequency fit based on mixing-length theory, where we have used $\frac{1}{20}\nu_e$. The constant of proportionality is clearly smaller than the value ($1/6$) required to fit our laminar results in \S~\ref{LaminarR2}, possibly because the convective length-scales are reduced at higher $Ra$. It is possible that larger box sizes would permit larger wavelength convective modes, which can enhance $\nu_E$ and lead to a constant of proportionality that is more consistent with our laminar simulations. Further exploration of this possibility is left to future work.

\begin{figure}
	\includegraphics[width=\columnwidth]{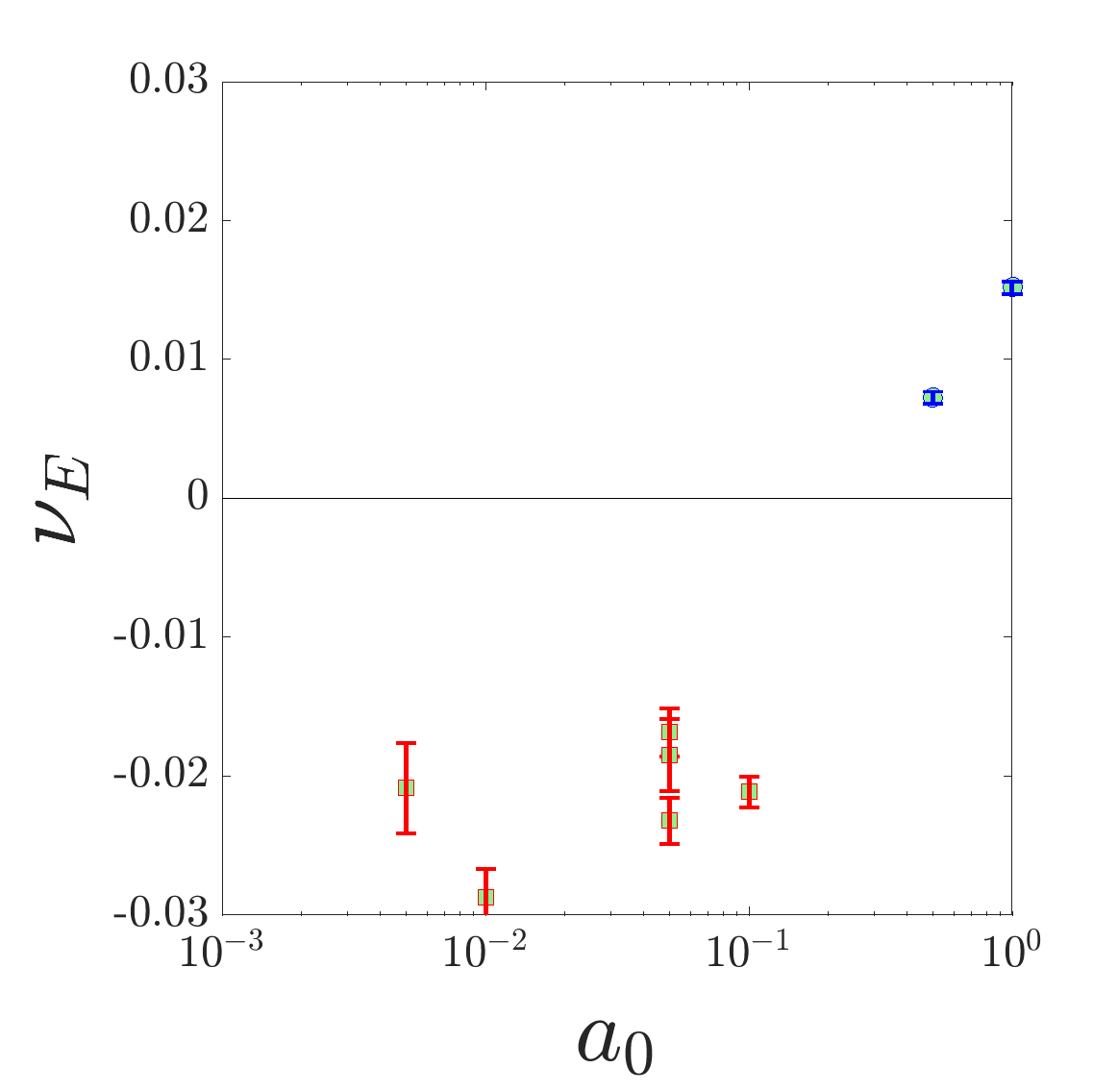}
	\caption{Plot of $\nu_E$ versus $a_0$ for $R=100$ and $\omega = 1000$ (so that $\omega/\omega_c \approx 15.6$) to explore the dependence of our results on $a_0$. Since $\nu_E$ is approximately independent of $a_0$ for $a_0\leq 0.1$, this suggests that the simulations in  Fig.~\ref{figure_shearfreq_vs_VE_2sigma_mathematica_convectivescaled_R100} are primarily exploring the regime of linear tides. For large amplitudes, the sign is observed to change, potentially indicating a departure from the linear regime.}
	\label{figure_amplitude_vs_VE_2sigma}
\end{figure}

To explore further the occurrence of negative $\nu_E$ at high frequencies, we have also performed a set of simulations in which the amplitude $a_0$ is varied. These allow us to explore whether the sign change in $\nu_E$ is related to the increasing amplitude of the tidal flow relative to the convective flow i.e.~to check whether we are still exploring the regime of linear tides at high frequencies. We focus on simulations varying $a_0$ near the transition, taking $\omega = 1000$ (so that $\omega/\omega_c\approx 15$). The resulting values for $\nu_E$ are shown in Fig.~\ref{figure_amplitude_vs_VE_2sigma} as a function of $a_0$. This shows that $\nu_E<0$ for all $a_o\leq 0.1$, but that $\nu_E>0$ when $a_0\geq 0.1$. The occurrence of negative values with very similar magnitude for all cases with $a_0\leq 0.1$, suggests that we are probing the regime of linear tides, since $\nu_E$ does not depend strongly on $a_0$. However, for larger amplitudes we obtain different results, with positive $\nu_E$, suggesting that the largest amplitude cases are no longer probing the regime of linear tides. Note that \cite{ogilvie_interaction_2012} instead fix the shear ($a_0\omega$) in their simulations, so the concurrence of our results with their observation of negative values at high frequencies also suggests that the transition to $\nu_E<0$ in Fig.~\ref{figure_shearfreq_vs_VE_2sigma_mathematica_convectivescaled_R100} is not caused by a transition out of the regime of linear tides.

Our observation of negative effective viscosities suggests the intriguing possibility of tidal anti-dissipation due to the interaction between tides and convection. This means that energy is transferred from the convective flow to the tidal flow, in opposition to the standard picture in which the tidal flow is damped by its interaction with the convection.  Unlike molecular viscosity, there is nothing in principle preventing the effective viscosity from being negative, even if this result is surprising.

\begin{figure}
	\includegraphics[width=\columnwidth]{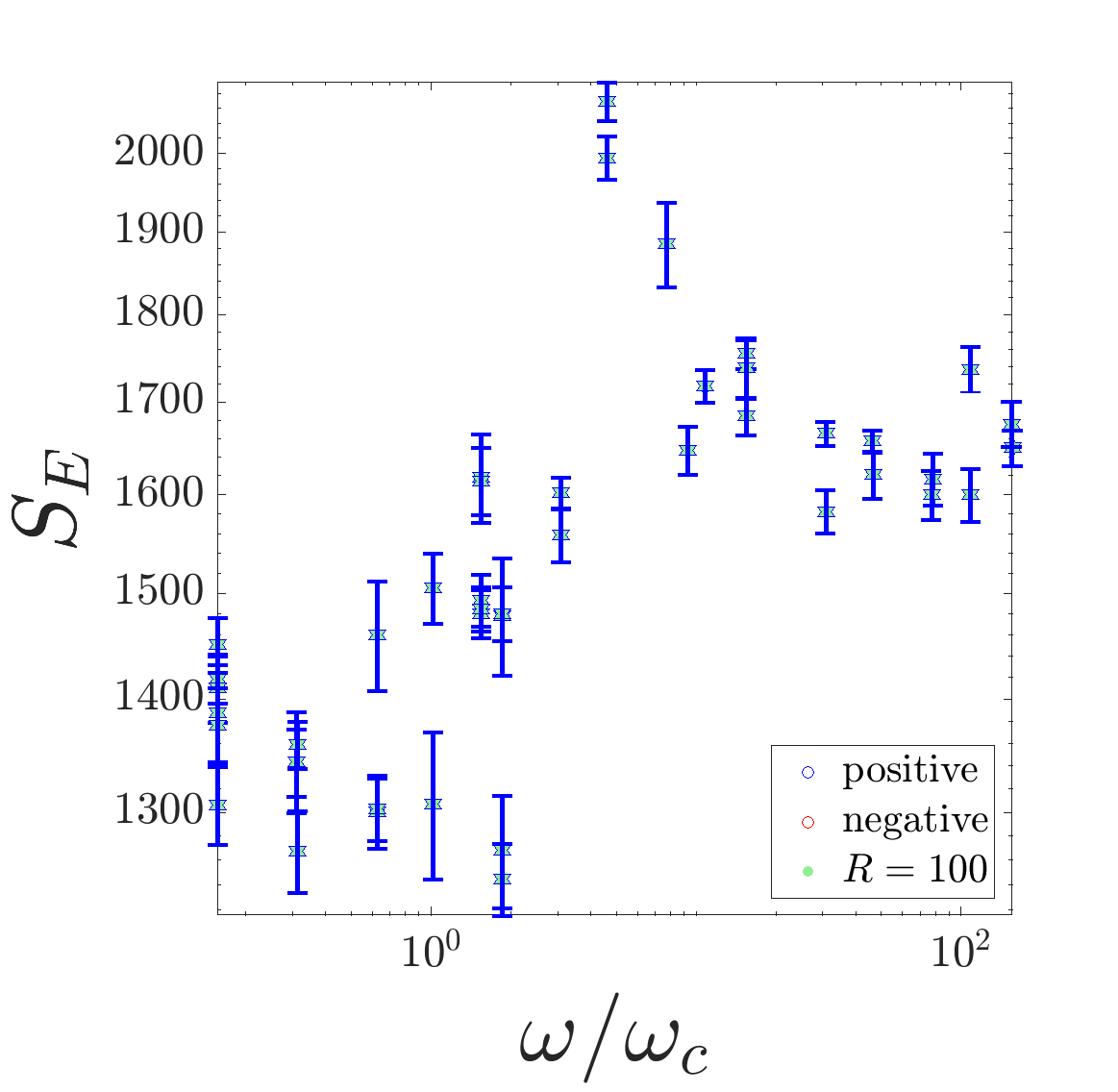}
	\caption{Plot of the frequency dependence of $S_E(\omega)$ scaled by the convective frequency $\omega_c$ for simulations with $R=100$ and $a_0 = 0.05$. For all values of $\omega$ we obtain positive values.}
	\label{figure_shearfreq_vs_SE_2sigma_mathematica_convectivescaled_R100}
\end{figure}

We plot the effective elasticity $S_E$ in Fig.~\ref{figure_shearfreq_vs_SE_2sigma_mathematica_convectivescaled_R100}. Similar to our laminar results in Fig.~\ref{figure_shearfreq_vs_SE_2sigma_mathematica_convectivescaled_yaligned}, $S_E$ increases with $\omega$ until $\omega\approx 5\omega_c$, above which there is a transition to a regime in which $S_E$ is independent of $\omega$, with a possible enhancement for $\omega\sim \omega_c$. Note that there is no change in sign and $S_E$ remains positive for all $\omega$, unlike what we have observed in our laminar simulations. 

\begin{figure}
	\includegraphics[width=\columnwidth]{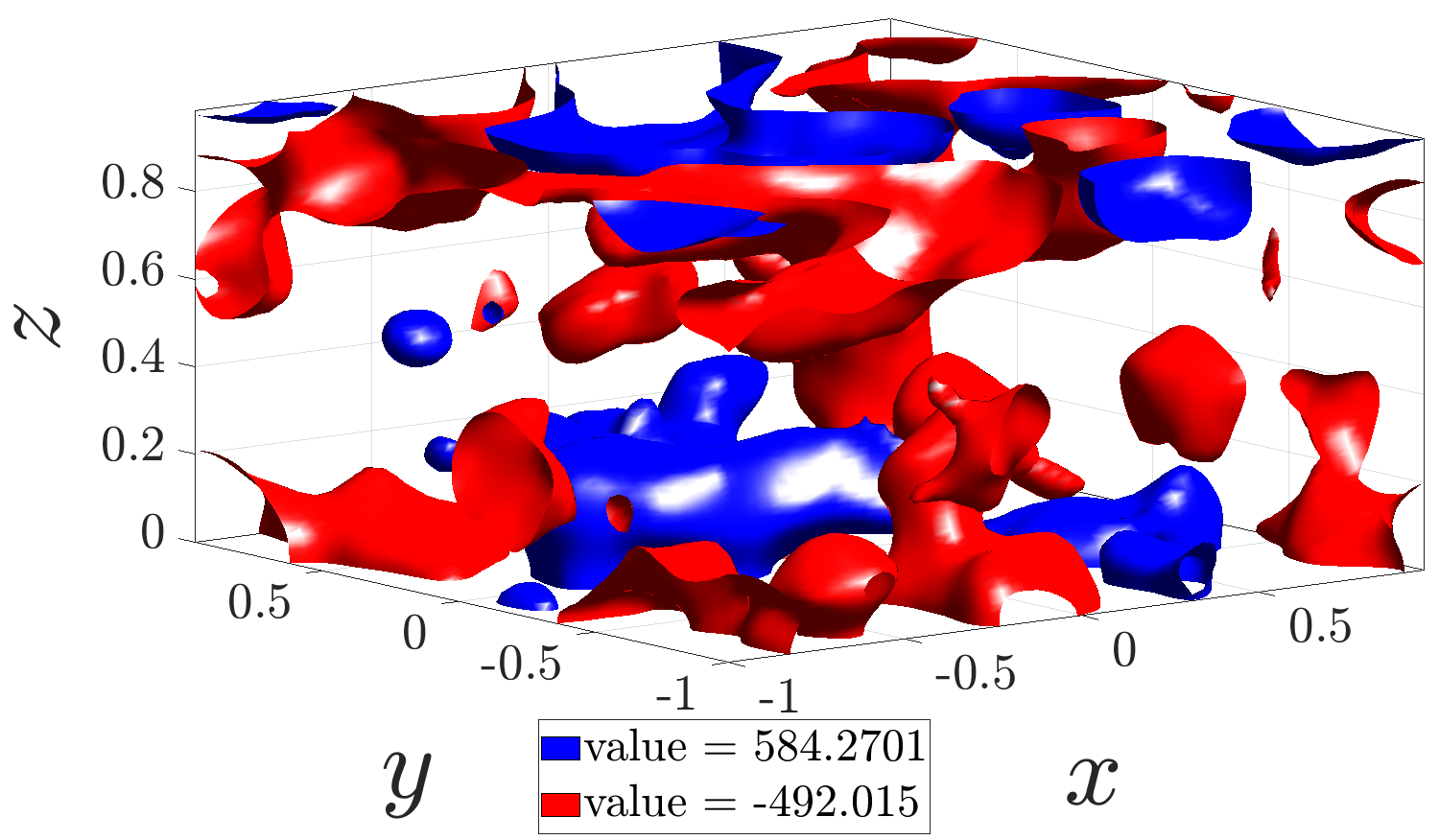}
	\includegraphics[width=\columnwidth]{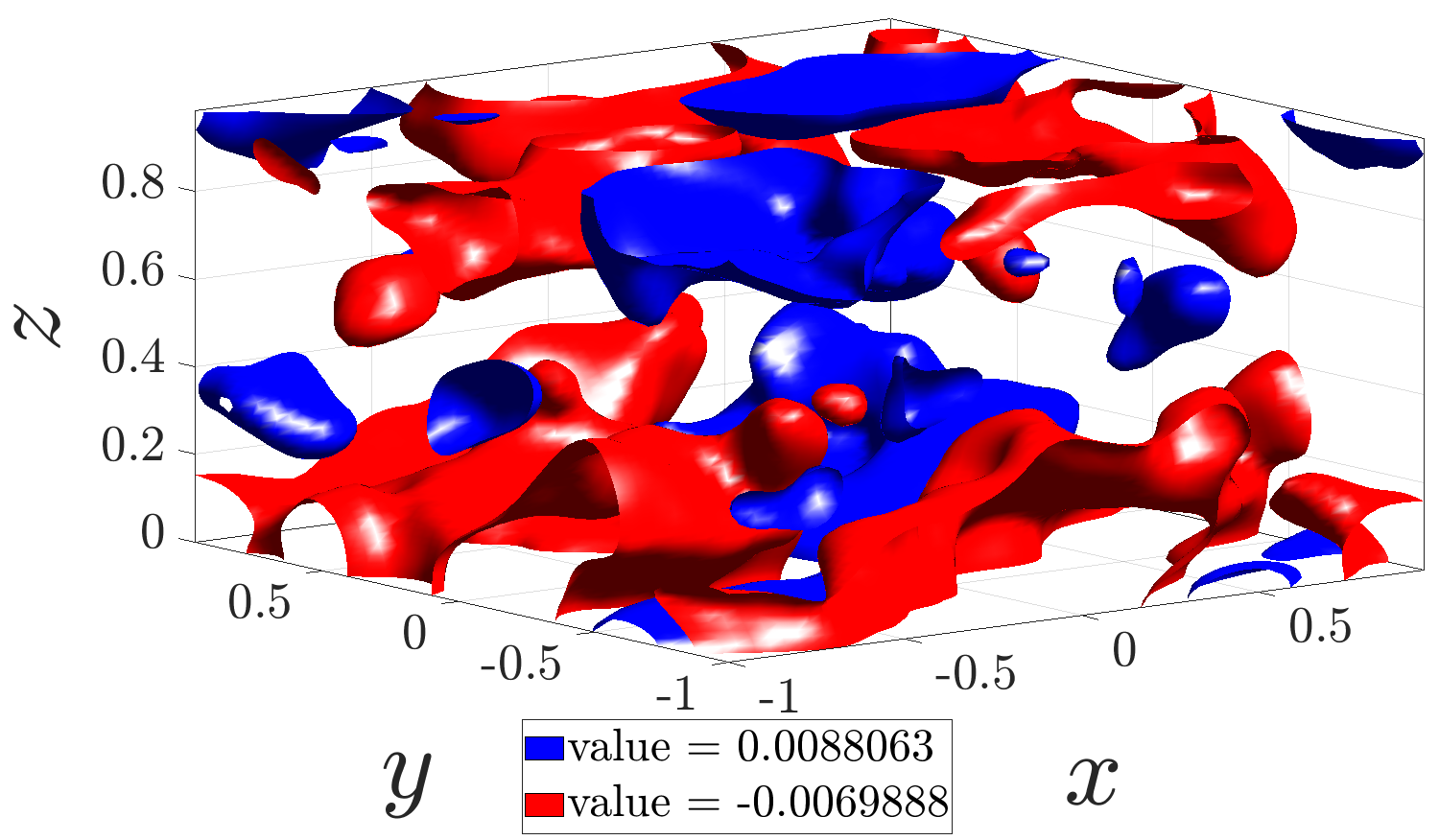}
	\caption{Iso-surfaces showing $\tilde{\nu}_E$, the spatial contribution to $\nu_E$ based on Eq.~\ref{spatialnuE}, where $80\%$ of the maximum (blue) and minimum (red) values are shown in two simulations $\omega \in\{10,10000\}$ (top to bottom respectively) with $R=100$ and $a_0 = 0.05$. The maximum values are distributed in space but are primarily localised near to the boundaries in $z$.}
	\label{figure_3D_plot_of_spatial_visc}
\end{figure}

We would like to understand where in the domain the flow contributes the most to the effective viscosity and also which scales are important. To explore this, we recorded the spatial structure using at least ten snapshots per $\omega^{-1}$ in simulations with $R=100$, $a_0 = 0.05$ and $\omega \in \{10,100,1000,10000\}$. Fig.~\ref{figure_3D_plot_of_spatial_visc} shows the result from evaluating
\begin{align}\label{spatialnuE}
	\tilde{\nu}_E(x,y,z) =  \frac{-2}{a_0\omega (T-T_0)}\int_{T_0}^{T} u_x(x,y,z,t) u_y(x,y,z,t) \cos(\omega t) \, \text{d}t  \, 
\end{align}
for the cases $\omega = 10$ and $10000$, where $T_0$ and $T$ are the start and end times of our simulation. This is effectively evaluating Eq.~\ref{maths_effective_viscosity} in a point-wise sense before volume-averaging. Note that $\nu_E=\langle\tilde{\nu}_E \rangle$.
We plot iso-surfaces representing $80\%$ of the maximum positive (blue) and negative (red) values of $\tilde{\nu}_E$ within the domain. This allows us to identify where in the domain contributes the most to both the positive and negative values of $\nu_E$.

The regions near to the boundaries in $z$ clearly provide the dominant contribution to both the positive and negative values of $\nu_E$, at both low and high frequency. This is what we might have expected based on the linear modes because we have adopted stress-free boundary conditions and the convection is only weakly turbulent. There is no obvious pattern in the spatial distribution of negative and positive contributions, nor is there an obvious change in the spatial distribution between the low and high frequency cases.

\begin{figure}
    \includegraphics[width=\columnwidth]{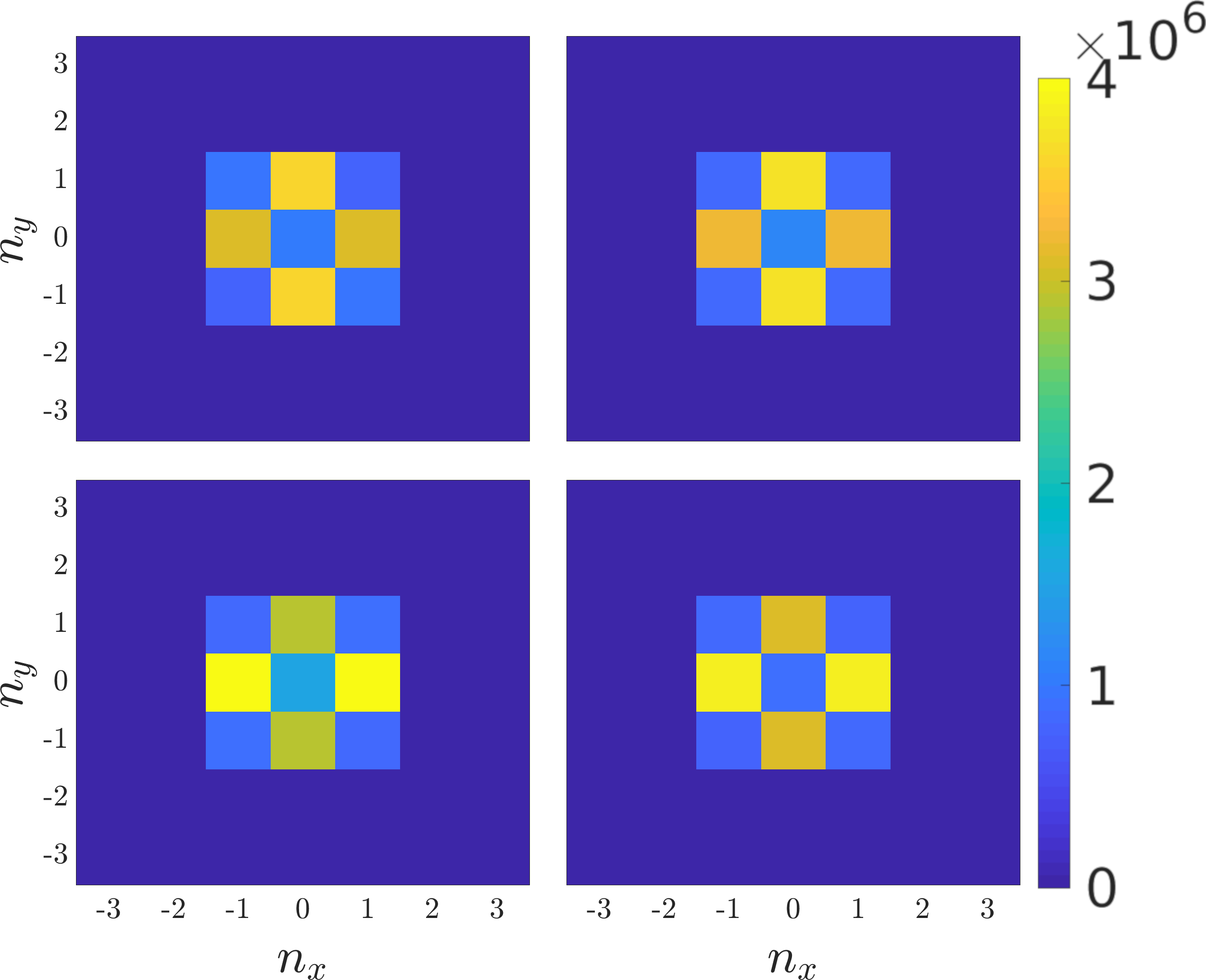}	\caption{Temporally-averaged and vertically-integrated $\hat{R}_{xy}(k_x,k_y)$ spectrum as a function of integer wavenumbers $n_x=L_x k_x/2\pi$ and $n_y=L_y k_y/2\pi$ in four simulations with $\omega = 10, 100, 1000, 10000$ (top left, top right, bottom left, bottom right, respectively), all with $R=100$ and $a_0 = 0.05$. This shows that the dominant scales for the Reynolds stress are the box-scale $x$ and $y$-aligned convection rolls, but that other modes also contribute.}
	\label{figure_kspace_kinetic_energy}
\end{figure}

In order to determine the dominant spatial scales contributing to both positive and negative values of $\nu_E$, we performed a horizontal Fourier transform of the point-wise Reynolds stress. The contribution to $\nu_E$ due to each $(k_x,k_y)$ mode is given by 
\begin{align}
	\hat{\nu}_E(k_x,k_y) =& \frac{-1}{2 a_0\omega (T-T_0)} \int_{T_0}^T  \hat{R}_{xy}\cos(\omega t) \, \text{d} t \label{maths_nu_spectrum}  \,,
\end{align}
where $T_0$ and $T$ are the start and end times of our simulation. We have defined
\begin{align}
\hat{R}_{xy}(k_x,k_y,t) =\int_0^1 \hat{u}_x(k_x,k_y,z,t)\, \hat{u}_y^{\ast}(k_x,k_y,z,t)  + \mathrm{c.c.}\,\,\text{d} z \,,
\end{align}
where c.c. denotes the complex conjugate and hats denote a horizontal Fourier transform. Note that $\nu_E$ is obtained by summing up $\hat{\nu}_E(k_x,k_y)$ over all of the modes, and we have used this to verify our method (some small differences remain due to the timestep being larger in the data used to compute the spatial structure).

First we show a temporal average of $\hat{R}_{xy}$ on the $(k_x,k_y)$-plane in Fig.~(\ref{figure_kspace_kinetic_energy}), where the axis values represent the integer wavenumbers. The largest-scale modes provide the dominant contribution to $\nu_E$, particularly the box-scale $x$ and $y$-aligned convection rolls, though smaller scales also contribute non-negligibly. 

\begin{figure}
    \includegraphics[width=\columnwidth]{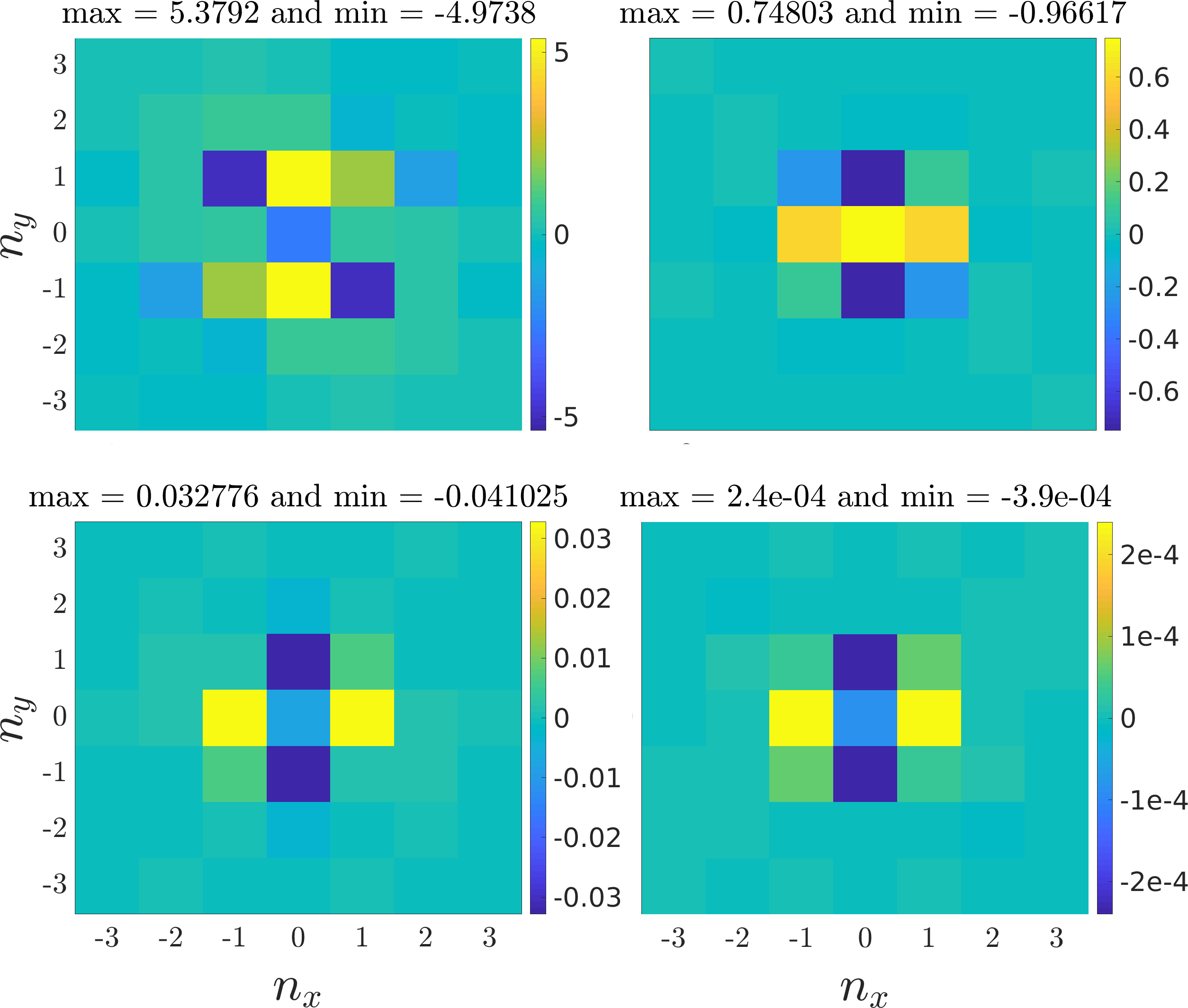}
	\caption{Spatial spectrum of $\hat{\nu}_E$ (Eq.~\ref{maths_nu_spectrum}) as a function of the integer wavenumbers $n_x=k_x L_x/2\pi$ and $n_y=k_y L_y/2\pi$, which represents the dominant contributions to $\nu_E$ due to different horizontal wavenumber modes. The cases shown are all for $R=100$ and $a_0 = 0.05$, with $\omega = 10, 100, 1000, 10000$ (top left, top right, bottom left, bottom right, respectively).}
	\label{figure_kspace_omega}
\end{figure}

We plot $\hat{\nu}_E(k_x,k_y)$ on the integer $(k_x,k_y)$-plane in Fig.~\ref{figure_kspace_omega}, which shows the contributions due to each mode to $\nu_E$. The higher frequency cases with $\omega \in \{100, 1000, 10000\}$ show that the $x$-aligned roll ($n_x=0, n_y=1$) provides a negative contribution to $\nu_E$, and the $y$-aligned roll ($n_x=1, n_y=0$) provides a positive contribution. This is in accord with our expectations based on laminar convection in \S~\ref{yaligned} and \ref{xaligned}, as well as the theory in Appendix~\ref{appendix_extension_of_ol2012}. The lowest frequency case with $\omega=10$ behaves differently however, but this is a case where the theory does not apply. In that case, the $x$-aligned roll component provides a positive contribution to $\nu_E$.

\begin{figure}
	\includegraphics[width=0.8\columnwidth]{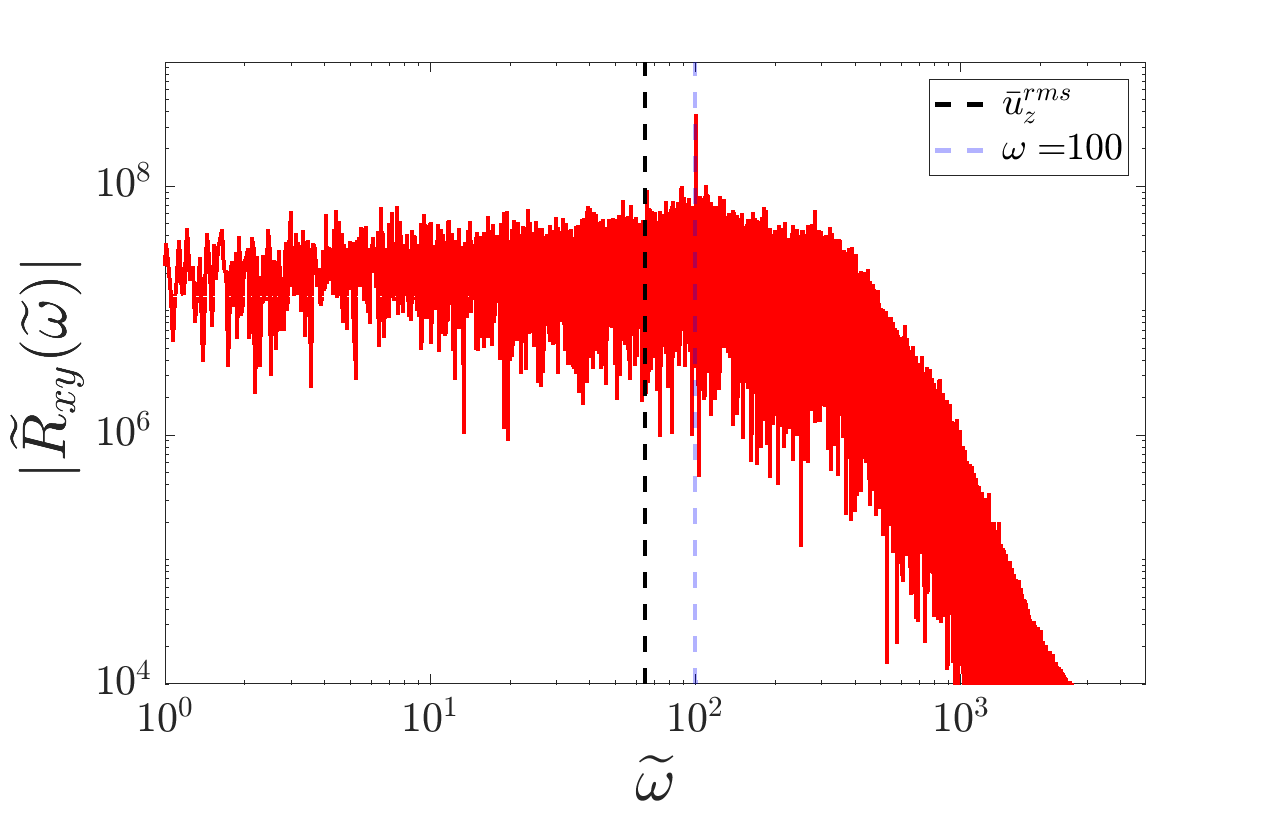}
	\includegraphics[width=0.8\columnwidth]{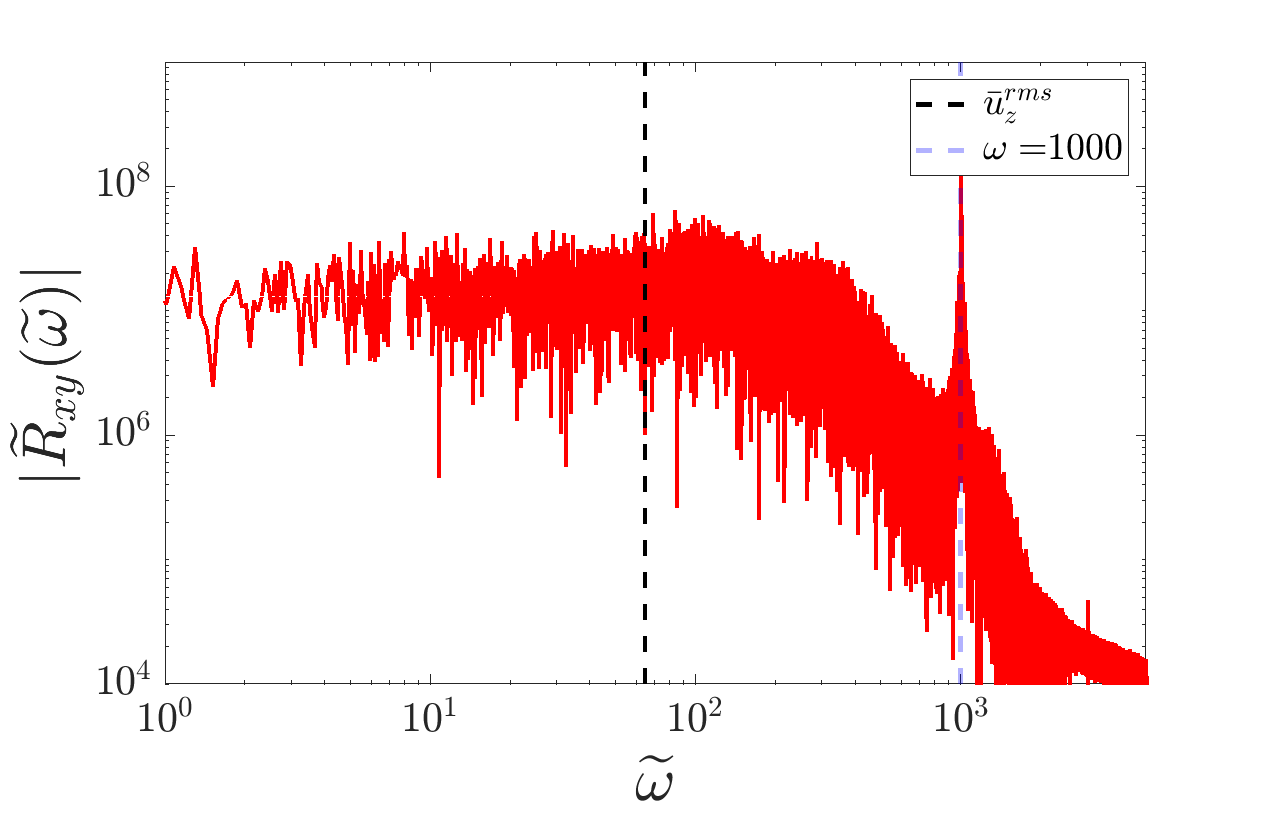}
	\includegraphics[width=0.8\columnwidth]{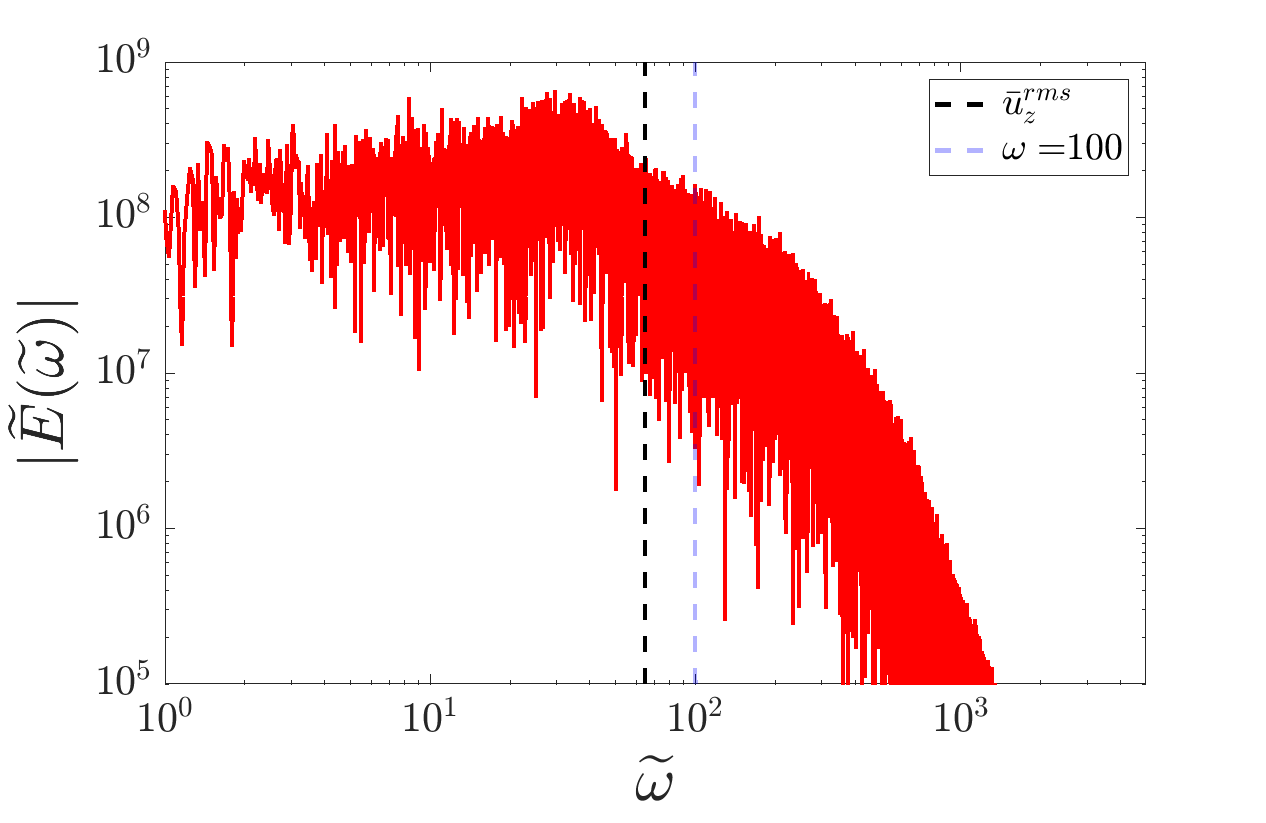}
	\includegraphics[width=0.8\columnwidth]{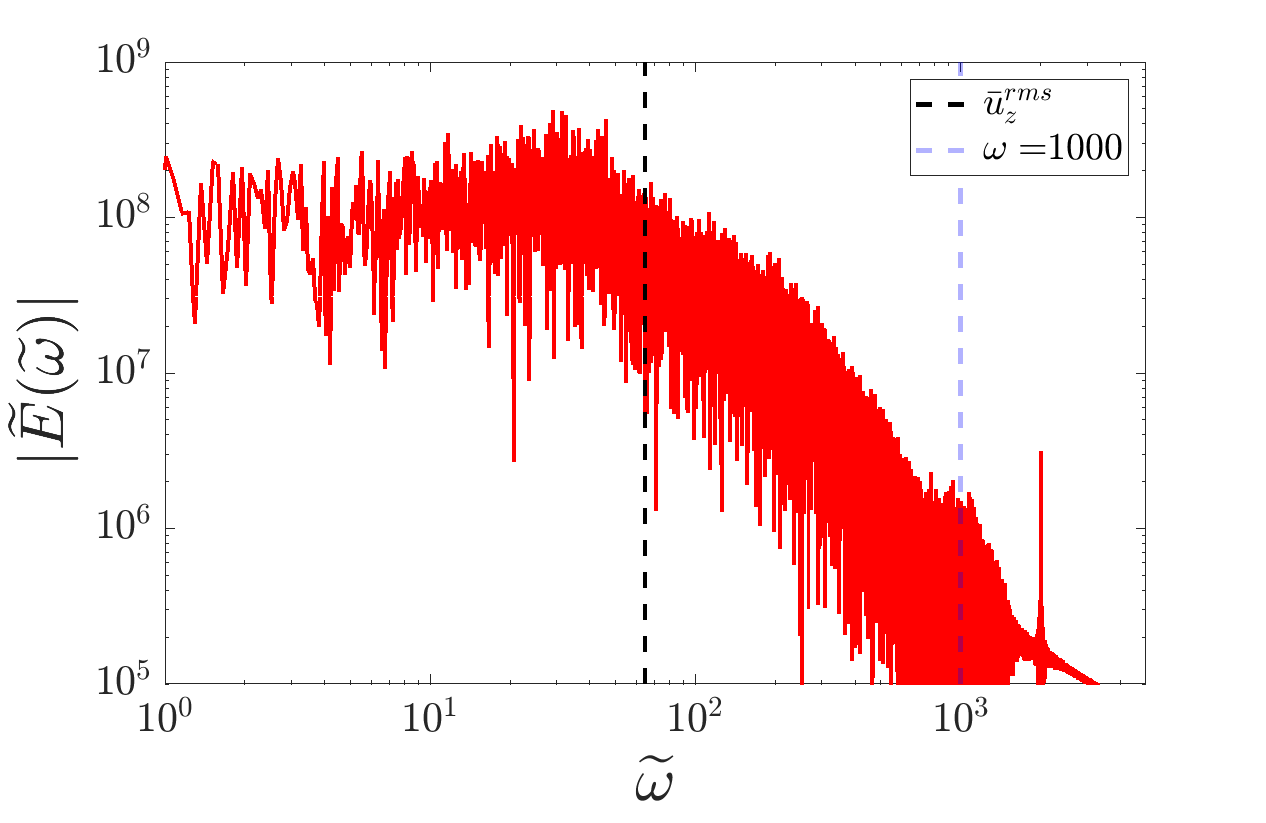}
	\caption{Temporal power spectrum of Reynolds stress $\tilde{R}_{xy}$ and kinetic energy $\tilde{E}$ as a function of frequency $\tilde{\omega}$ for one low frequency case with $\omega=100$ (first and third panel) and one high frequency case with $\omega=1000$ (second and fourth panel).}
	\label{R100FreqSpec}
\end{figure}

Finally, we show the temporal Fourier transform of the volume-averaged Reynolds stress, $\tilde{R}_{xy}$ and kinetic energy $\tilde{E}$ as a function of frequency $\tilde{\omega}$ in Fig.~\ref{R100FreqSpec}. These quantities are shown for one low frequency simulation with $\omega=100$ (with $\nu_E>0$) and one high frequency case with $\omega=1000$ (with $\nu_E<0$). The forcing frequency $\omega$ is indicated by the grey dashed vertical line. This shows that when $R=100$, the flow contains a wide range of frequencies, and that there is a peak at the forcing frequency $\omega$. We observe that $\omega=100$ coincides with a part of the $\tilde{E}$ spectrum with a shallow negative slope for a decade or so in $\tilde{\omega}$, potentially coinciding with an inertial range. On the other hand, $\omega=1000$ lies above a transition in $\tilde{E}$ to a steeper decay with $\tilde{\omega}$, potentially indicating frequencies in the dissipation range. We speculate that the sign of $\nu_E$ may be related to whether $\omega$ lies in the inertial (positive $\nu_E$) or dissipative (negative $\nu_E$) frequency range. If this is true then higher $R$ would result in an inertial range that extends to higher frequencies, hence we would require a larger value of $\omega$ to obtain negative $\nu_E$. Although our findings of negative $\nu_E$ values are robust for high frequencies, it would be worth simulating more turbulent cases with larger $R$ to explore this further.

\subsection{Effects of varying $\mathrm{Pr}$}
\label{VaryingPr}

%Pr0.1 case 005. omega = 1000. a0=0.05
%Pr10  case 005. omega = 1000. a0=0.05
% time unit 175 for all
\begin{figure}
	\includegraphics[width=\columnwidth]{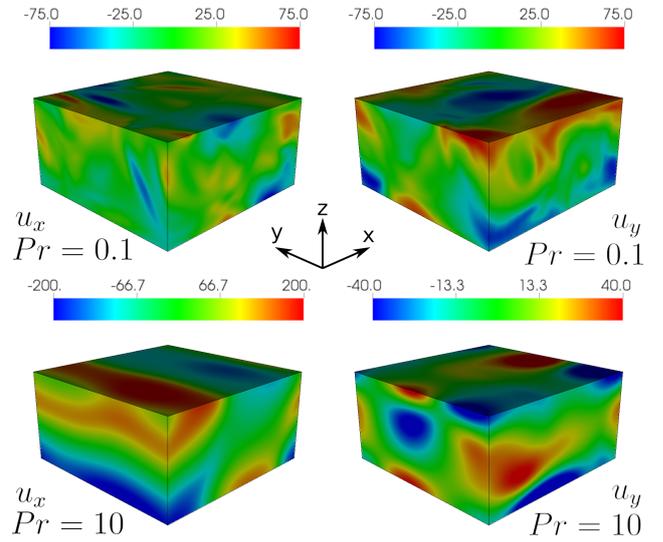}
	\caption{Snapshot of $u_x$ (left) and $u_y$ (right) for $R=100$, $a_0 = 0.05$, $\omega=1000$, at time $t=175$, with $\mathrm{Pr} = 0.1$ and $\mathrm{Pr}=10$ (top and bottom, respectively). This can be compared with Fig.~\ref{fig_flow_structure_R100} and shows the effects of varying $\mathrm{Pr}$ on the flow.}
	\label{figure_pr_flow}
\end{figure}

Our simulations until now have all adopted $\mathrm{Pr}=1$, which is the most convenient choice computationally. In stellar interiors however, $\mathrm{Pr}\ll 1$ (e.g. $10^{-6}$ at the base of the solar convection zone), and in the mantles of terrestrial planets, $\mathrm{Pr}>1$. It is therefore important to determine how changing $\mathrm{Pr}$ modifies our results. This is particularly important because the theory in Appendix~\ref{appendix_extension_of_ol2012} identifies the key role of the temperature perturbation in producing negative values for $\nu_E$. To do this, we have performed two sets of additional simulations with $R=100$ with both $\mathrm{Pr}=0.1$ and $\mathrm{Pr}=10$. We show the $u_x$ and $u_y$ velocity field in a snapshot from an illustrative simulation with each Pr in Fig.~\ref{figure_pr_flow}, which can be compared with the $\mathrm{Pr}=1$ case in Fig.~\ref{fig_flow_structure_R100}. 

\begin{figure}
	\includegraphics[width=\columnwidth]{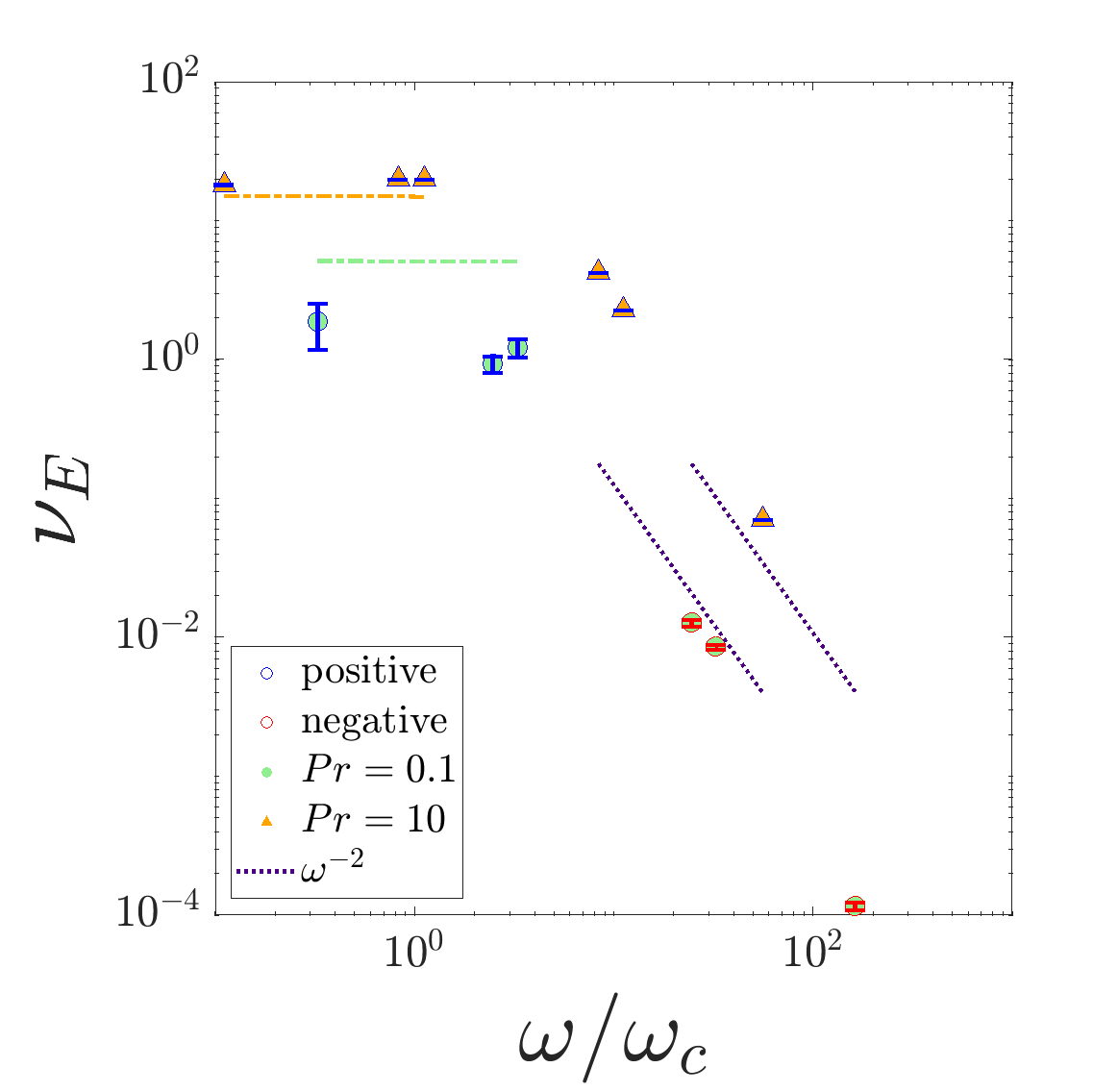}
	\caption{$\nu_E(\omega)$ as a function of frequency using the convective frequency scaling $\omega_c$ for two cases with $Pr \in\{ 0.1,10\}$ with $R=100$ and $a_0=0.05$. This shows qualitatively similar behaviour to Fig.~\ref{figure_shearfreq_vs_VE_2sigma_mathematica_convectivescaled_R100}. Positive values are indicated by blue symbols and negative quantities by red. We have used $\alpha = 1/6$ for the low frequency constant of proportionality.}
	\label{figure_nu_e_for_Pr}
\end{figure}

We show the frequency-dependence of the effective viscosity from these simulations in Fig.~\ref{figure_nu_e_for_Pr}. This figure can be compared with the simulations with $\mathrm{Pr}=1$ in Fig~\ref{figure_shearfreq_vs_VE_2sigma_mathematica_convectivescaled_R100}. To allow these simulations to be most clearly distinguished, we have not re-scaled the $y$-axis $\nu_E$ values, and our use of the thermal timescale as our unit of time in the governing equations means that $\nu_E$ is expected to move upwards as we increase $\mathrm{Pr}$ if this scales in the way predicted by mixing-length theory ($\nu_E\propto \sqrt{\mathrm{RaPr}}$). We also plot the mixing-length scaling that fits our laminar simulations ($\frac{1}{6} u_z^{rms}$) as the horizontal dashed lines in this figure.

For both $\mathrm{Pr}=$ 0.1 and 10 we also obtain a frequency-independent $\nu_E$ for $\omega/\omega_c\lesssim 5$, which transitions to $\nu_E\propto \omega^{-2}$ at higher frequencies. At low frequencies, the simulations with $\mathrm{Pr}=10$ most closely agree with the mixing-length scaling obtained in our laminar simulations, in comparison with the simulations with $\mathrm{Pr}=0.1$, which has slightly smaller values of $\nu_E$. This is presumably because the velocity field in the case with $\mathrm{Pr}=0.1$ contains more smaller scale eddies, as we can see by comparing the top and bottom panels of Fig.~\ref{figure_pr_flow}. When $\mathrm{Pr}=0.1$, $\nu_E$ transitions to negative values for $\omega\gtrsim 10\omega_c$, similar to our previously-presented cases with $\mathrm{Pr}=1$. On the other hand, our simulations with $\mathrm{Pr}=10$ do not exhibit a transition to negative values in this range of $\omega$. The occurrence of negative effective viscosities at high frequencies may provide support that such values could be possible in stellar interiors, where $\mathrm{Pr}$ is small.

\subsection{$R=1000$ with $L_x=L_y=2$}
\label{R1000}

\begin{figure}
	\includegraphics[width=\columnwidth]{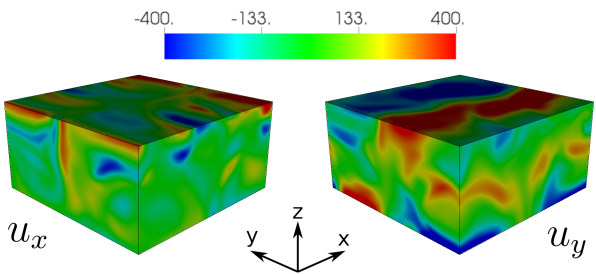}
	\caption{Plots of the $u_x$ (left) and $u_y$ (right) components of velocity for the cases $R=1000$, $a_0=0.05$ and $\omega = 50000$.}
	\label{fig_flow_structure_R1000}
\end{figure}

\begin{figure}
	\includegraphics[width=\columnwidth]{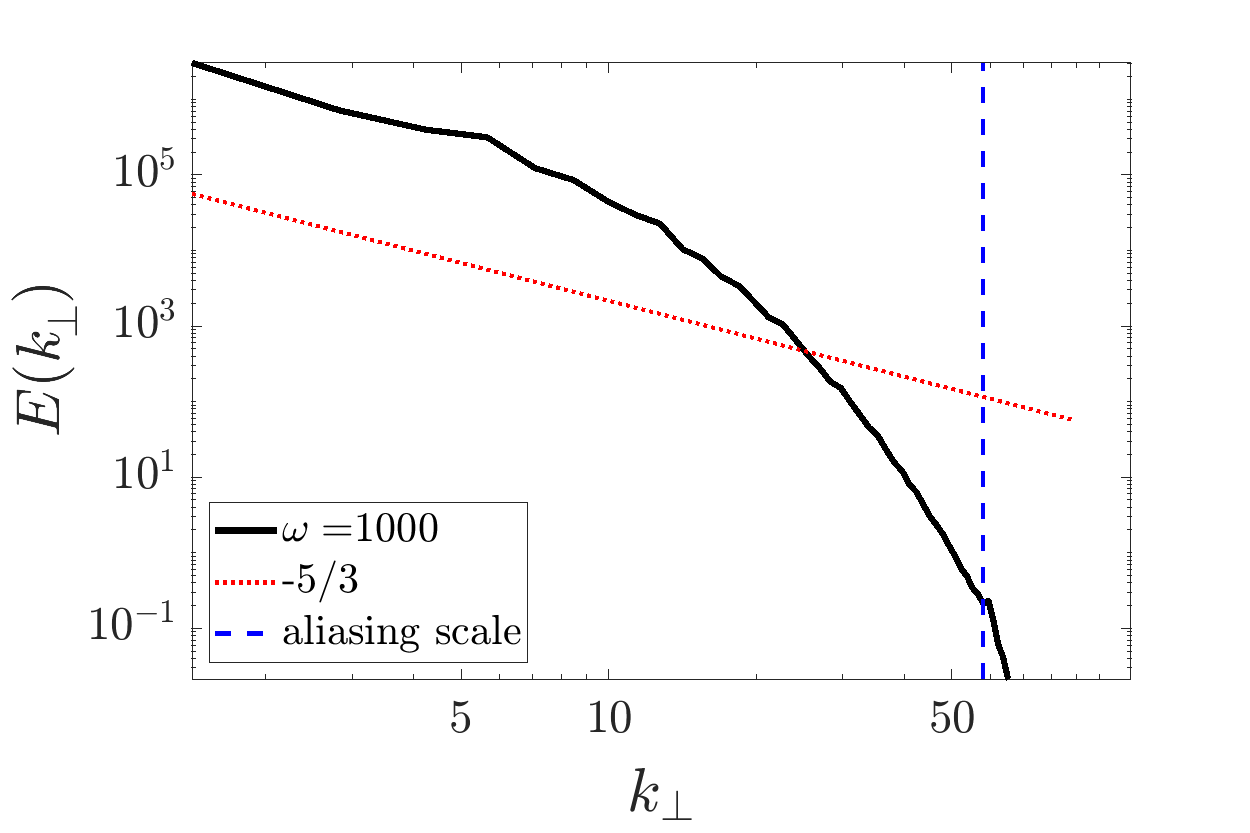}
	\caption{Kinetic energy spectrum for a simulation with $R=1000$, $a_0 = 0.05$ and $\omega = 1000$. The kolmogorov scaling ($-5/3)$ is shown as the red dotted line and the aliasing scale is shown as the blue dashed vertical line.}
	\label{R1000_spectrum}
\end{figure}

We now move on to analyse more turbulent simulations with $R=1000$ and $a_0=0.05$ with $\mathrm{Pr}=1$. We use a resolution of $(N_x,N_y,N_z) = (128,128,64)$, which was found to be sufficient for the flow to be well-resolved. In these cases, the flow is fully three dimensional and time-variable. We show the $u_x$ and $u_y$ components of the flow in a snapshot at $t=175$ in a simulation with $\omega=50000$ in Fig.~\ref{fig_flow_structure_R1000}. We also plot the horizontal power spectrum of the kinetic energy $E(k_{\perp})$ as a function of $k_{\perp}$ in Fig~\ref{R1000_spectrum}. The flow consists of many modes, though the box-scale rolls are still energetically-dominant.

\begin{figure}
	\includegraphics[width=\columnwidth]{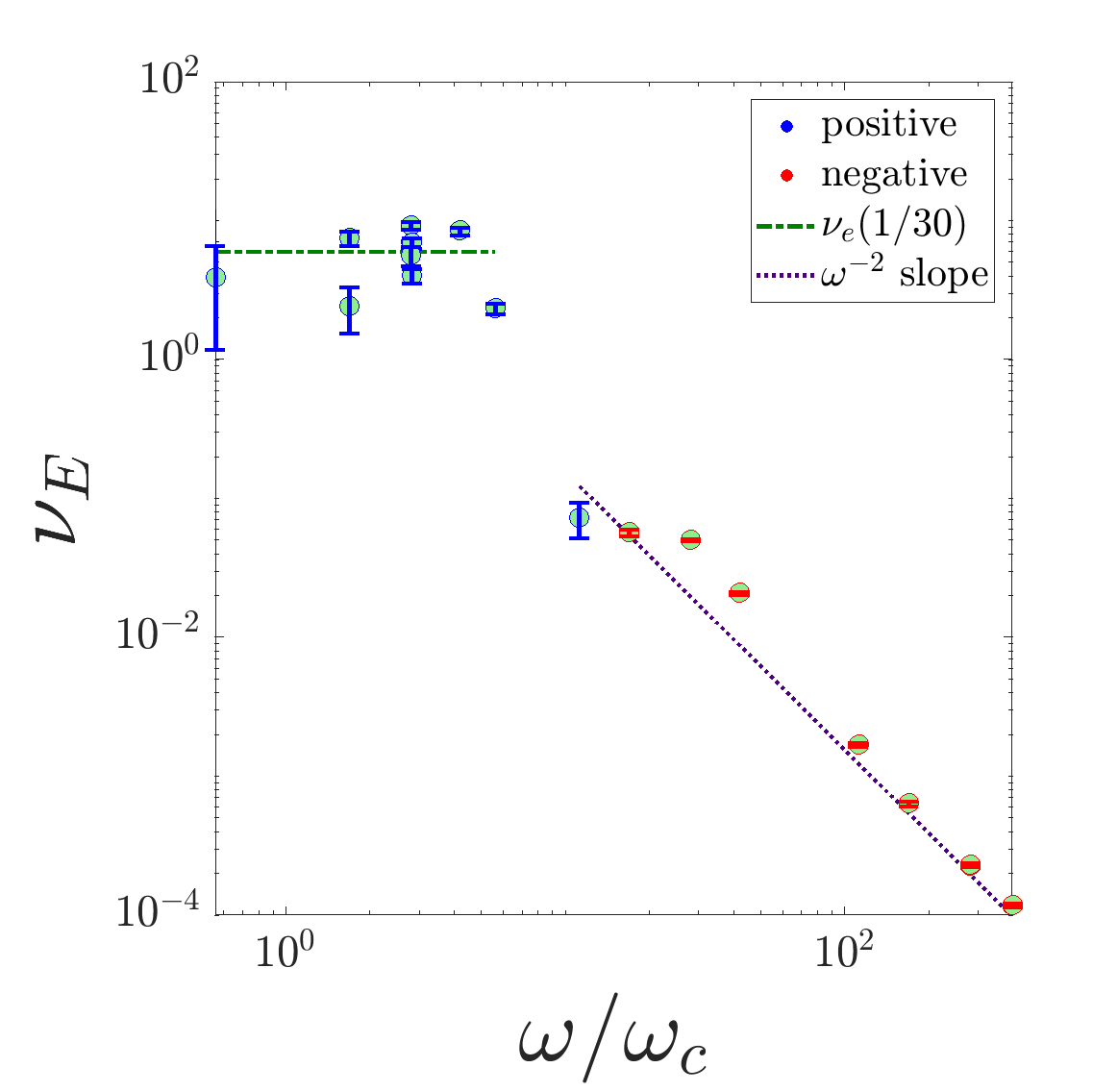}
	\caption{Plot of $\nu_E$ as a function of frequency (scaled by the convective frequency $\omega_c$) for various simulations with $R=1000$ and $a_0 = 0.05$. Positive quantities are indicated by blue symbols and negative quantities by red.}
	\label{shearfreq_vs_VE_2sigma_mathematica_convectivescaledR1000}
\end{figure}

The effective viscosity is plotted as a function of frequency in Fig.~\ref{shearfreq_vs_VE_2sigma_mathematica_convectivescaledR1000}. This shows very similar behaviour to $R=100$ (Fig.~\ref{figure_shearfreq_vs_VE_2sigma_mathematica_convectivescaled_R100}) in that $\nu_E$ remains approximately independent of $\omega$ for $\omega\lesssim 5\omega_c$. When $\omega\gtrsim 10\omega_c$, there is a transition to negative values with magnitudes falling off such that $|\nu_E|\propto\omega^{-2}$. As with the simulations in \S~\ref{R100}, we find that $\nu_E$ is smaller than we would expect based on extrapolating the mixing-length scaling that applies to our laminar simulations. Indeed, $\nu_E$ is less efficient than $(d/6) u_z^{\text{rms}}$ (where $d=1$ in our nondimensionalization), which was previously found to describe the laminar simulations. This is potentially due to the dominant length-scales being smaller than $d$. It is unclear whether larger horizontal boxes would lead to larger $\nu_E$ values, and this will be explored in further work.

\begin{figure}
	\includegraphics[width=\columnwidth]{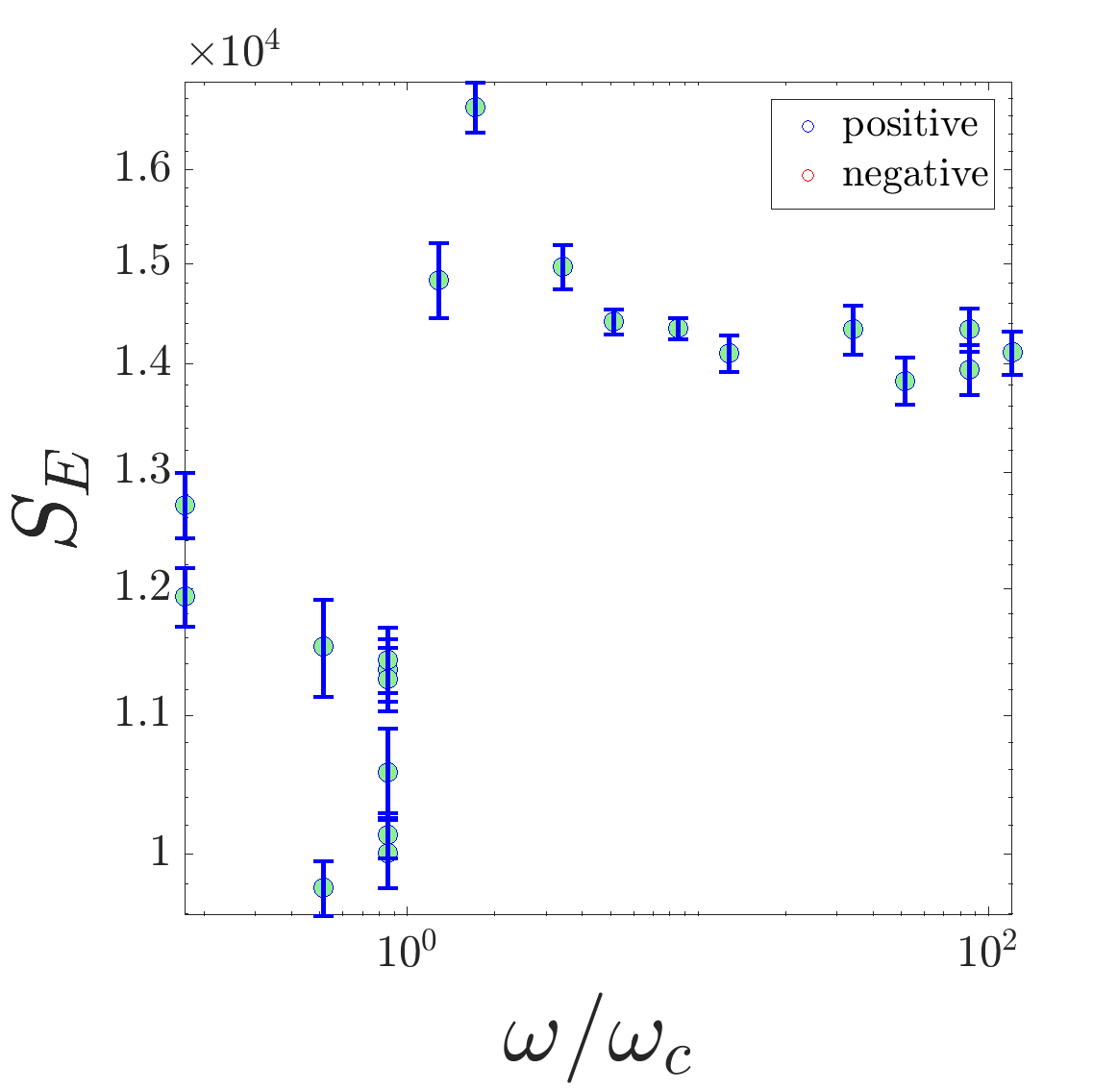}
	\caption{Plots of $S_E(\omega)$ scaled by the convective frequency $\omega_c$ for $R=1000$ and $a_0 = 0.05$. For all values of $\omega$ explored the resulting effective elasticity is positive.}
	\label{shearfreq_vs_SE_2sigma_mathematica_convectivescaledR1000}
\end{figure}

We show the effective elasticity ($S_E$) as a function of frequency in Fig.~\ref{shearfreq_vs_SE_2sigma_mathematica_convectivescaledR1000}. This shows similar behaviour to the results for $R=100$ shown in Fig.~\ref{figure_shearfreq_vs_SE_2sigma_mathematica_convectivescaled_R100}, in that for $\omega\gtrsim \omega_c$ we find that $S_E$ becomes independent of frequency. However, in the low frequency regime, when $\omega\lesssim \omega_c$, $S_E$ exhibits non-monotonic behaviour in this case. It is worth noting that the evaluation of $S_E$ for low frequencies is increasingly sensitive to error, making this regime difficult to probe.

\begin{figure}
	\includegraphics[width=\columnwidth]{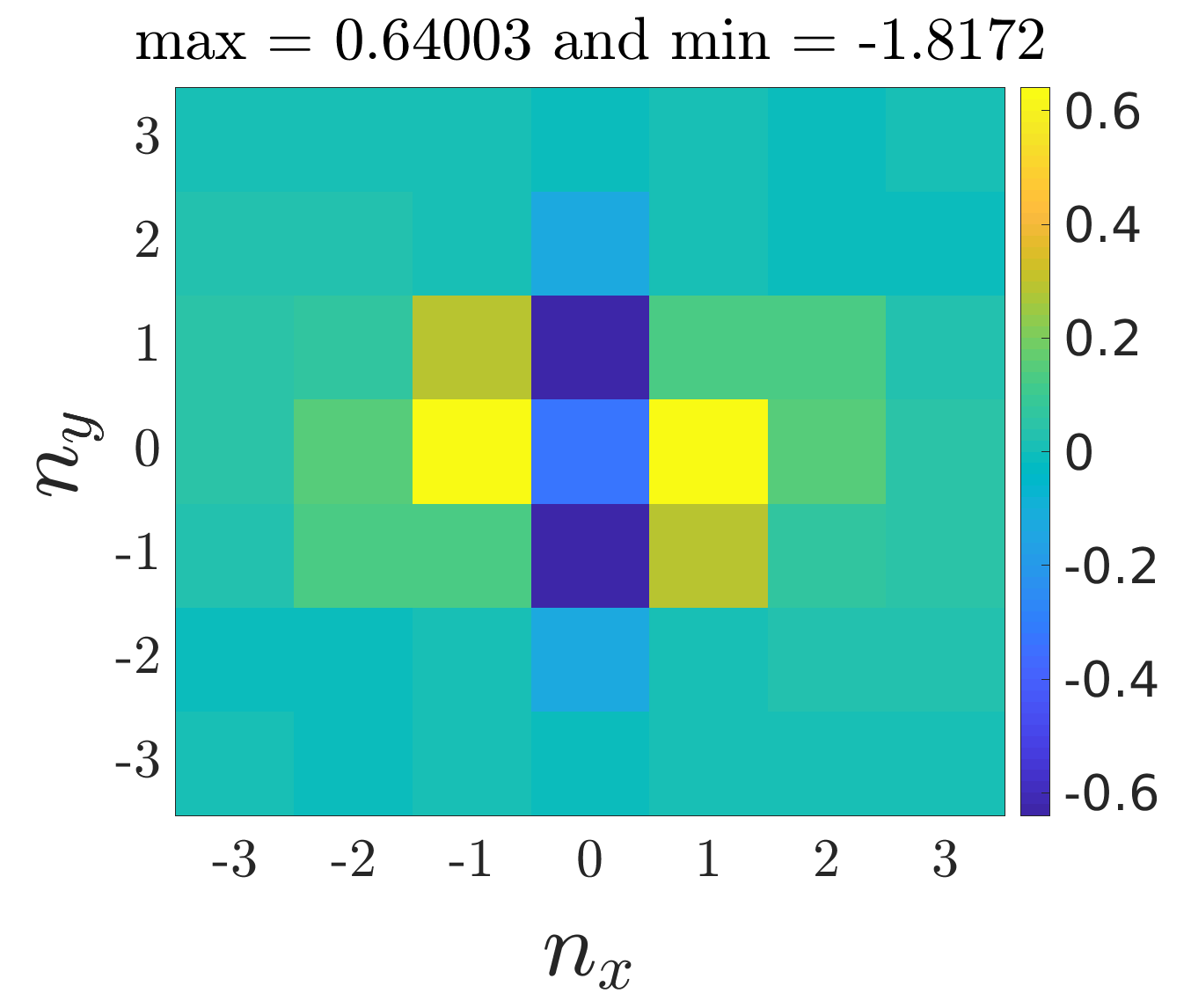}
	\caption{Spectrum of $\hat{\nu}_E$ as a function of integer $(k_x,k_y)$ for a simulation with $R=1000$, $a_0 = 0.05$ and $\omega = 1000$. The largest and smallest amplitude is indicated above the figure.}
	\label{R1000_spectrums}
\end{figure}

We show the spectrum of $\hat{\nu}_E(k_x,k_y)$ on the integer $(k_x,k_y)$-plane for the simulation with $\omega=1000$ (with $\omega_c \approx 178$) in Fig.~\ref{R1000_spectrums}, which shows the contributions to the effective viscosity due to the various modes. Multiple modes contribute to the effective viscosity in this case, but the $x$-aligned ($y$-aligned) roll continues to provide a dominant negative (positive) contribution. In this case the 3D modes also contribute appreciably.

Our simulations with $R=100$ and $R=1000$, and those with $R=100$ with smaller $\mathrm{Pr}$, indicate broadly similar results. Further work is required to explore more turbulent regimes with larger values of $R$, as well as simulations with larger boxes, to explore the robustness of these results.

%%%%%%%%%%%%%%%%%%%%%%%%%%%%%%%%%%%%%%%%%%%%%%%%%%%%%%%%%%%%%%%%%%%%%%%%%%%%%%%%%%%%%%
\section{Discussion and implications}
\label{discussion}

\begin{table}
\begin{tabular}{ r | c | c | c }
 R & $\alpha = \overline{\left(\frac{\nu_E}{\overline{u}_z^{\text{rms}}} \right)}$ & $\beta = \overline{\nu_E \omega^2} $  & $\overline{u}_z^{\text{rms}}$ \\ \hline
$2_x$       & $0.160$ & $674$   &  $5.50$\\ \hline
$2_y$       & $0.163$ & $631$   &  $5.47$\\ \hline
$2 \,\, (4\times4\times1)$ & $0.355$ & $360$ & $4.25$ \\ \hline
$5_x$       & $0.174$ & $5079$  &  $13.44$\\ \hline
$5_y$       & $0.174$ & $3691$  &  $13.44$\\ \hline
$10_x$      & $0.146$ & $13882$ &  $22.86$\\ \hline
$10_y$      & $0.150$ & $10921$ &  $22.88$\\ \hline
$100$       & $0.051$ & $17002$ &  $64.44$\\ \hline
$100\,\, (4\times4\times1)$ & $0.193$ & $20842$ & $48.87$ \\ \hline\hline
$1000$      & $0.033$ & $622001$&  $178.09$ \
\end{tabular}
\caption{Table listing values and scalings for the effective viscosity as a function of $R$ in both the low and high-frequency regimes, as well as the typical rms vertical velocity. For the laminar simulations, in the first column we use a subscript on the $R$ value to indicate whether the flow consists of $x$ or $y$-aligned convection rolls. The values of $\bar{u}_z^\mathrm{rms}$ are obtained by taking an average over all simulations with this value of $R$ (and roll orientation) for $\omega<\omega_c$.}
	\label{nuETable}
\end{table}

\begin{figure}
	\includegraphics[width=\columnwidth]{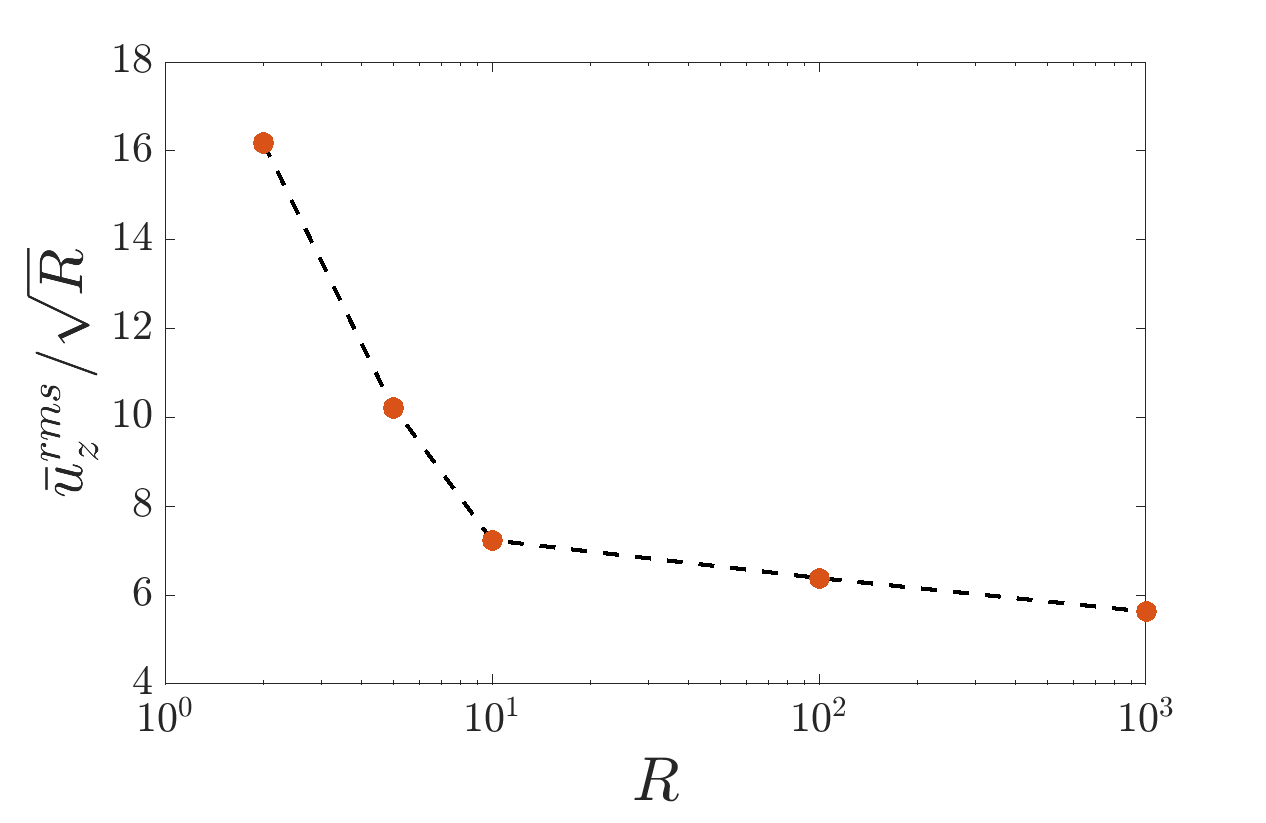}
	\caption{Comparison of the rms vertical convective velocity with the mixing-length scaling, plotting $\bar{u}_{z}^{\mathrm{rms}}/\sqrt{R}$ as a function of $R$. This indicates a trend towards $\bar{u}_{z}^{\mathrm{rms}} \propto \sqrt{R}$ for large enough $R$.}
	\label{figure_uz_R}
\end{figure}

Our simulations have demonstrated that the effective viscosity describing the damping (or otherwise) of large-scale tidal flows through their interaction with convection exhibits two (or possibly three) regimes. For low tidal frequencies such that $\omega\lesssim 5 \omega_c$, we obtain a frequency-independent $\nu_E$, which approximately agrees with the eddy viscosity from the mixing-length theory of convection (to within a constant factor e.g.~\citealt{BohmVitense1958,Zahn1989}). When $\omega\gtrsim 5 \omega_c$, we observe that $\nu_E\propto \omega^{-2}$, implying a significant reduction in the effective viscosity at high frequencies (e.g.~\citealt{goldreich_turbulent_1977}). In the high-frequency regime, we also observe $\nu_E$ to become negative, indicating the intriguing possibility of tidal anti-dissipation. Broadly, our results are consistent with \cite{ogilvie_interaction_2012} and \cite{braviner_stellar_2015}, but we used a different setup to model the convection. 

In Table~\ref{nuETable}, we list the values of $\alpha$ and $\beta$ used to fit $\nu_E=\alpha \bar{u}_z^{\text{rms}} d$ in the low-frequency regime, and $\nu_E=\beta\omega^{-2}$ in the high-frequency regime. In Fig.~\ref{figure_uz_R} we also show a comparison of the vertical convective velocity with the mixing-length scaling as a function of $R$ (taking an average over all simulations with this value of $R$). This shows that the convective velocity approaches a diffusion-free mixing-length scaling ($\bar{u}_z^{\text{rms}}\propto \sqrt{R}$) for $R\gtrsim 10$. On the other hand, we observe a departure in $\nu_E$ from the mixing-length expectation, with $\alpha$ depending on $R$, tending to decrease as $R$ is increased. We speculate that this may be due to the convection being constrained by the horizontal box size in the simulations with the largest $R$. Evidence in favour of this hypothesis includes the value of $\alpha$ increasing in simulations with $R=100$ as we increase the box size from $L_x=L_y=2$ to $L_x=L_y=4$, as listed in Table~\ref{nuETable}, in addition to the energy spectrum in e.g.~Fig.~\ref{R1000_spectrum}. Simulations with larger $L_x$ and $L_y$ are required to explore this issue further, and these will be undertaken in future work.

Our observation that $\nu_E\propto \omega^{-2}$ when $\omega\gg \omega_c$ is robust as to whether we simulate laminar convection, with only one scale, or turbulent convection with many (spatial and temporal) scales. Furthermore, our results demonstrate that the largest spatial scales (locally) are those that primarily contribute to the effective viscosity. The phenomenological arguments of \cite{goldreich_turbulent_1977}, which extended \cite{zahn_les_1966} to a turbulent flow with a Kolmogorov spectrum, assumed that only eddies with turnover times faster than the tidal period, corresponding with short spatial scales, contribute to the effective viscosity. This argument is at odds with our finding that it is the largest spatial scales that dominate this interaction. Therefore, even though we obtain a quadratic frequency reduction like \cite{goldreich_turbulent_1977}, our simulations do not support their theoretical arguments for the reason behind this scaling. It would be worth exploring further the fundamental mechanism that results in the attenuation of the effective viscosity, perhaps by building upon the theory in Appendix \ref{appendix_extension_of_ol2012}.

It is worth pointing out that in our more turbulent simulations the magnitude of the negative values of $\nu_E$ is smaller than the molecular viscosity $\nu$ (which is equal to one in each of the figures with $\mathrm{Pr}=1$) in the high-frequency regime. Whether or not the negative values would become important in reality for tidal evolution depends partly on whether $\nu_E$ increases with $R$ in the high-frequency regime so that it exceeds $\nu$. In our cases with larger $R$ the negative values of $\nu_E$ increase as $R$ is increased (see Table~\ref{nuETable}), but these values remain smaller than $\nu$. If this remains the case at larger $R$, the negative $\nu_E$ values would not be astrophysically significant for tidal evolution. In our limited exploration into the effects of changing Pr we found that decreasing Pr results in similar negative values for $\nu_E$ at high frequencies to the case with $\mathrm{Pr}=1$, but this remains to be confirmed in a more extensive parameter survey.

Our simulations adopted the Boussinesq approximation, which means that they are strictly applicable to studying the local interaction between convection and tidal flows on scales that are much smaller than a pressure scale height. In stars, the convective velocities (and length-scales) vary with radius, typically increasing (decreasing) as we approach the stellar surface, where the validity of a Boussinesq model will eventually break down. However, our results do indicate that the effective viscosity will be maximised, and therefore tidal dissipation will be dominated, by radii for which $\omega\lesssim 5 \omega_c$, which typically involve radii closer to the stellar surface than to the base of the convection zone.

Another caveat of our model is that we have followed \cite{ogilvie_interaction_2012} in only simulating one component of the tidal flow, corresponding with an oscillatory shear flow. In reality (even in a non-rotating homogeneous body with a circular companion), the tidal flow would contain additional components. We have not demonstrated in this paper whether these other components would behave in a similar manner to produce negative values for $\nu_E$, and this would be worth exploring further.

\subsection{Astrophysical implications}
\label{implications}

For just one example to illustrate the importance of considering the correct prescription for $\nu_E$ for high frequencies, we can crudely estimate the rate of orbital decay of a Jupiter-mass planet in a one day aligned circular orbit around a slowly rotating Sun-like star. Tidal dissipation in the star usually leads to planetary orbital decay (if $\nu_E>0$, but not if $\nu_E<0$), raising the question as to its long-term survival. This example is meant to represent a close-in hot Jupiter such as WASP-12 b or WASP-4 b (e.g.~\citealt{Maciejewski2016,Patra2017,Bouma2019}). We can use our results to predict the tidal orbital evolution time-scale due to the effective viscosity of the convection.

\begin{figure}
	\includegraphics[width=\columnwidth]{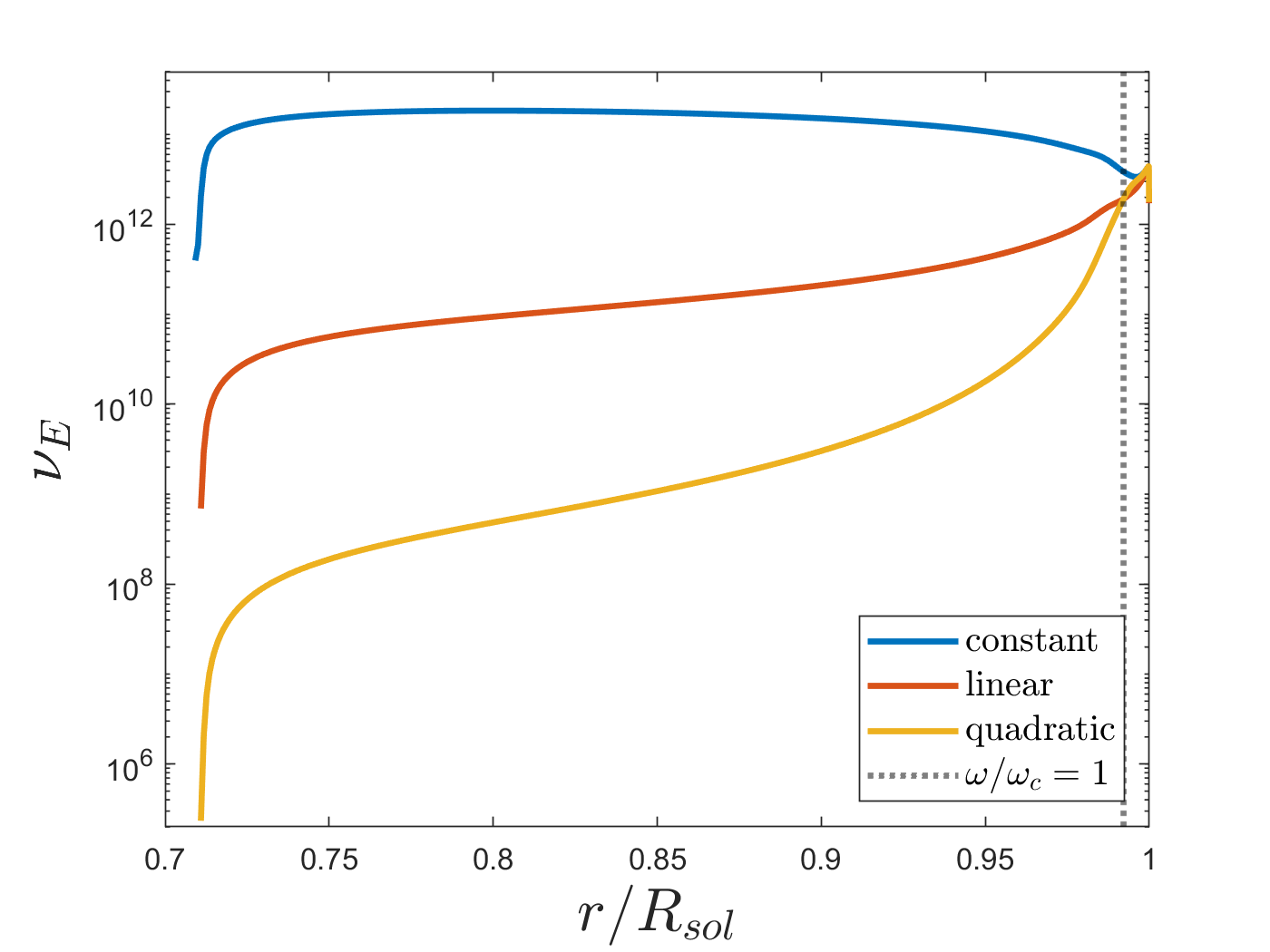}
	\caption{Effective viscosity evaluated as a function of radius in the solar convection zone according to mixing-length theory for each prescriptions for its high-frequency behaviour (Eq.~\ref{maths_nu_function_radius}). We adopt a tidal period of 0.5 d, corresponding with a hot Jupiter on a 1 d orbit. This demonstrates the importance of using the correct prescription for $\nu_E$.}
	\label{figure_nu_e_prescriptions}
\end{figure}

We show the radial dependence of the effective viscosity in a Sun-like star in Fig~\ref{figure_nu_e_prescriptions} by calculating 
\begin{equation}\label{maths_nu_function_radius}
    \nu_E(r) = \frac{1}{3} u(r) l_m(r) \left( 1 + \left( \frac{\omega}{\omega_c(r)}  \right)^\zeta \right)^{-1} \,,
\end{equation}
where $\zeta\in \{0,1,2\}$, $u$ is a convective velocity and $l_m=\alpha_M H_p$ is a mixing length, evaluated using data from a standard solar model (Model S of \citealt{ModelS1996}). We have chosen a tidal period 0.5 d, corresponding with a hot Jupiter on a 1 d orbit. We have adopted the usual mixing-length expression for $\nu_E$ (e.g.~\citealt{Zahn1989}) for this figure rather than the scalings in Table.~\ref{nuETable} since further simulations are required to conclusively determine the values of $\alpha$ and $\beta$ in turbulent convection. Here $\alpha_M\approx 2$ is the usual mixing-length parameter and $H_p$ is the local pressure scale height. The three lines represent the prediction from assuming $\zeta=0,1,2$, corresponding with no frequency-reduction, the linear frequency-reduction of \cite{zahn_les_1966}, and a quadratic reduction, respectively. The prescriptions clearly give very different predictions for the magnitudes of $\nu_E$, spanning many orders of magnitude, particularly at the base of the convection zone where $\omega\gg \omega_c$. If $\zeta=2$ (and also for $\zeta=1$), $\nu_E$ is dominated by regions close to the surface, where $\omega_c$ is largest. To obtain the total dissipation, we must integrate $\rho \nu_E r^8$ over the entire convection zone \citep{zahn_tidal_1977,Zahn1989,Remus2012}, where $\rho$ is the local density and $r$ is the radius. 

For a crude estimate, the inspiral timescale may be calculated using (e.g.~\citealt{Rasio1996})
\begin{equation} \label{maths_inspiral_time}
    \tau_a \approx f \tau_c \left(\frac{M_\star}{M_{\text{env}}}\right)\left(\frac{M_\star}{M_p}\right)\left(\frac{P}{P_{\mathrm{dyn}}}\right)^{\frac{16}{3}}.
\end{equation}
Here we assume that the planet orbits faster than the star rotates and that tidal dissipation drives inward migration. In this expression the frequency-reduction factor is accounted for by
\begin{equation}
    f=\begin{cases}
    \frac{1}{8} & \text{if}\quad\zeta=0 ,\\
    \frac{2}{13}\text{max}\left[1,\left(\frac{2\tau_c}{P}\right)\right] & \text{if}\quad\zeta=1 ,\\
    \frac{1}{5}\text{max}\left[1,\left(\frac{2\tau_c}{P}\right)^{2}\right] & \text{if}\quad\zeta=2 ,
    \end{cases}
    \label{maths_inspiral_timescale}
\end{equation}
which also contains a constant factor accounting for the different rates of migration with each prescription. We also define
$P_\mathrm{dyn}=2\pi \sqrt{R_\star^3/(G M_\star)}$ as the dynamical timescale, $G$ is the gravitational constant, $R_\star$ is the stellar radius, $M_\star$ is the stellar mass, $M_{\text{env}}$ is the mass of the convective envelope. The convective period is $\tau_c= l_m / u = \alpha H_p / u$, where we have evaluated the convective velocity using the convective luminosity. We obtain these parameters from Model S for the current Sun \citep{ModelS1996}. Following \cite{Rasio1996}, we take $\tau_c\approx 20$ d and $M_\text{env}\approx 0.0252M_\star$. We thus find
\begin{gather}
    \tau_a \approx \begin{cases}
     60 \text{Myr} & \text{if}\quad\zeta=0, \\
     2 \text{Gyr} & \text{if}\quad\zeta=1, \\
     100 \text{Gyr} & \text{if}\quad\zeta=2.
    \end{cases}
\end{gather}
Since our simulations are consistent only with $\zeta=2$, this suggests that the orbital decay timescale of the closest hot Jupiters around solar-type stars due to this mechanism would be negligible over the main-sequence lifetime of the star. This crude estimate is very simplistic, and we have not considered the integrated $\nu_E$ due to all eddies throughout the convection zone (e.g.~\citealt{Zahn1989,OgilvieLin2007,Remus2012}), but it illustrates that these prescriptions give significantly different predictions for orbital decay timescales. Note also that a modified stellar tidal quality factor for this mechanism can be defined by \citep{Zahn2008,ogilvie_tidal_2014}
\begin{equation}\label{maths_Qvalues}
    Q'_{\star} \propto \frac{1}{\omega \Delta t}\propto  \begin{cases}
    \omega^{-1} & \text{if}\quad\zeta=0, \\
    \text{const} & \text{if}\quad\zeta=1, \\
    \omega & \text{if}\quad\zeta=2,
    \end{cases}
\end{equation}
where $\Delta t\propto 1/\nu_E$ is a tidal lag time (and using Eq.~\ref{maths_inspiral_time} we find $Q'_\star$ values of $7.5\times10^4$, $3\times10^6$ and $1.6\times10^8$ for $\zeta = 0, 1, 2$ respectively). Our results therefore suggest that assuming a frequency-independent $Q'_\star$ does not correctly describe the interaction between tidal flows and convection in either regime. 

\section{Conclusions}
\label{conclusions}

 Turbulent convection is believed to act as an effective viscosity ($\nu_E$) in damping large-scale tidal flows, and this mechanism has long been thought to be one of the dominant mechanisms of tidal dissipation in stars with convective envelopes\footnote{The tidal excitation and dissipation of inertial waves, perhaps through their interaction with convection, may also be important for low tidal frequencies (see e.g.~\citealt{ogilvie_tidal_2014} and references therein).} \citep{zahn_les_1966,zahn_tidal_1977,Zahn1989,ZahnBouchet1989}. This mechanism is commonly-believed to be important for the circularisation (and synchronisation) of binary systems containing solar-type \citep{ZahnBouchet1989,Meibom2005,Meibom2006}, low-mass \citep{Triaud2017,VonBoetticherTriaud2019}, and giant stars \citep{VerbuntPhinney1995,PriceWhelan2018,Beck2019}. In principle, it could also provide an important contribution to the inferred orbital decay of some hot Jupiters (e.g.~\citealt{Maciejewski2016,Patra2017,Bouma2019}). However, there is a long-standing theoretical controversy over the efficiency of this mechanism in the regime of fast tides, when the tidal frequency ($\omega$) exceeds that of the dominant convective eddies ($\omega_c$). This regime is relevant in most applications. 

 We have presented results from an extensive set of hydrodynamical numerical simulations designed to explore the interaction between tidal flows and convection in stars (or planets) from first principles. Our simulations have adopted a local Cartesian model to represent a small patch of the convective envelope of a star (or planet), and Boussinesq convection was modelled using the well-studied Rayleigh-B\'enard setup. We have thoroughly explored the interaction between the convection in this model and an imposed oscillatory tidal shear flow, in both laminar and moderately turbulent regimes. We have also undertaken a complementary asymptotic analysis for high frequency tidal forcing by extending prior work by \cite{ogilvie_interaction_2012}. Our work significantly builds upon the previous pioneering studies of \cite{penev_dissipation_2007,penev_direct_2009} and \cite{ogilvie_interaction_2012}.
 
 Our main results are that:
 \begin{itemize}
     \item $\nu_E$ is approximately frequency-independent for $\omega\lesssim 5\omega_c$, where $\omega_c$ is evaluated using the rms vertical velocity. Our results are consistent with a scaling that follows mixing-length theory with $\nu_E\approx \alpha u_z^{\text{rms}} d$. There is uncertainty in the value of $\alpha$, but we typically find it to have an order of magnitude of $0.1$, which suggests relatively efficient tidal dissipation at low frequencies. 
     \item $\nu_E$ exhibits a transition at $\omega\approx 5 \omega_c$, above which our simulations provide strong evidence that $\nu_E\propto \omega^{-2}$. This scaling is consistent with \cite{goldreich_turbulent_1977} and \cite{goldreich_solar_1977}, though we do not find support for their arguments, but not with \cite{zahn_les_1966} and \cite{Zahn1989}. This indicates that convection is much less efficient at damping high frequency tides than is commonly believed.
     \item Statistically-significant negative effective viscosities are obtained, particularly at high frequencies, indicating the possibility of tidal anti-dissipation. In principle, this could drive the opposite tidal evolution to that which is commonly expected e.g.~excitation rather than damping of planetary eccentricities. However, it remains to be demonstrated whether this mechanism can act as tidal anti-dissipation in realistic stellar models.
     \item We have confirmed these results for laminar convection using an independent asymptotic analysis which extends prior work by \cite{ogilvie_interaction_2012}. We identify an additional thermal contribution which is essential to obtain negative effective viscosities.
 \end{itemize}
 
Our observation that $\nu_E\propto \omega^{-2}$ in the high-frequency regime is consistent with the previous simulations of homogeneous convection by \cite{ogilvie_interaction_2012} as well as the simulations of the convection-like ABC flow by \cite{braviner_stellar_2015}. We also corroborate a tentative finding in \cite{ogilvie_interaction_2012} that negative effective viscosities are possible for high frequency tidal forcing. We have confirmed this result with simulations of much longer duration to obtain statistically-significant results. We have also shown using both simulations and asymptotic analysis that negative values are possible in laminar convection. This indicates that the appearance of negative effective viscosities is a robust feature in this model.

On the other hand, our results disagree with the high-frequency scaling law of \cite{zahn_les_1966} and \cite {Zahn1989}, and apparently also with the prior simulations of \cite{penev_dissipation_2007} and  \cite{penev_direct_2009}. The reason for this discrepancy has not yet been elucidated, but it may be related to differences in the turbulent temporal power spectra between these simulations. Another possibility is that compressible convection responds differently to tidal forcing in an important way compared with the Boussinesq convection that we have modelled. The simulations in \cite{penev_dissipation_2009} adopted the anelastic approximation, allowing them to simulate multiple density scale heights, whereas our model is effectively restricted to single scale height. However, our simulations have been run for a much longer duration, over a much wider range of frequencies, and they have also been confirmed for laminar convection with an independent asymptotic analysis. Further work should explore the effects of vertically-varying density on the interactions between tides and convection. A further difference between our models is that their simulations adopted a body force to drive the tidal flow rather than a background flow in a shearing box.

Our results confirm that tidal dissipation in convective regions due to this mechanism does not behave like a frequency-independent tidal quality factor. For low frequency tidal forcing, our results indicate that this mechanism behaves in a similar way to the commonly-adopted constant tidal lag-time model, at least in the simplest cases of either an aligned circular orbit or an aligned, spin-synchronised, weakly eccentric orbit, where there is only one tidal frequency \citep{Darwin1880,Alexander1973,Mignard1980,Hut1981,Eggleton1998,Ivanov2004}. However, our result that $\nu_E\propto \omega^{-2}$ when $\omega\gtrsim 5\omega_c$, indicates in the general case that if any component of the tide has a frequency that is comparable with or larger than the dominant convective frequency, then different components of the tide will be damped at different rates i.e.~they will have different lag times \citep{Lai2012}, and these will no longer be constant (see also \citealt{Ivanov2004}). The consequences of this should be explored in future work.

One implication of our results is that the inferred orbital decay of the shortest-period hot Jupiters with observed transit timing variations (e.g.~\citealt{Maciejewski2016,Patra2017,Maciejewski2018,Bouma2019}) is unlikely to result from the interaction between tides and convection. Instead, these observations are probably explained by the tidal excitation and dissipation of internal gravity waves in radiative regions \citep{GoodmanDickson1998,OgilvieLin2007,BO2010,B2011,Weinberg2012,Essick2016,Chernov2017}. This mechanism is able to provide the required level of dissipation to explain the decaying orbit of WASP-12 b (e.g.~\citealt{B2011,Chernov2017,Weinberg2017,Bailey2019}), if we assume that these waves are fully dissipated. However, uncertainties remain regarding the structure of the star (whether or not it has a radiative core), and whether these waves should in fact be fully damped.

Another implication of our results is that the strong tidal dissipation required to explain the outward migration of the satellites of Jupiter and Saturn \citep{Lainey2009,Lainey2012,Lainey2017} is unlikely to be produced by the convective damping of large-scale tidal flows. This supports prior theoretical arguments by \cite{goldreich_turbulent_1977}. However, these planets rotate sufficiently rapidly that convection is likely to be strongly rotationally constrained, motivating further simulations like those in this paper, but in the presence of rapid rotation (e.g.~\citealt{BDL14}, by building upon the phenomenology of e.g.~\citealt{mathis_impact_2016}).

Much further work is required to explore the interaction between tidal flows and convection, to study: more turbulent regimes of convection, the incorporation of rotation, modelling significant density variation (e.g.~by simulating convection using the anelastic approximation), and the extension of this problem to idealised (and, ultimately, more realistic) global models of stars or planets.

%%%%%%%%%%%%%%%%%%%%%%%%%%%%%%%%%%%%%%%%%%%%%%%%%%%%%%%%%%%%%%%%%%%%%%%%%%%%%%%%%%%%%%%%%%%%%%%%%%%%%%%%%%%%%%%%%%%%%%%%%%%%%%%%%%%%%%%%%%%%%%%%%%%%%%%%%%%%%%%%%%%%%%%%%%
%%%%%%%%%%%%%%%%%%%%%%%%%%%%%%%%%%%%%%%%%%%%%%%%%%%%%%%%%%%%%%%%%%%%%%%%%%%%%%%%%%%%%%%%%%%%%%%%%%%%%%%%%%%%%%%%%%%%%%%%%%%%%%%%%%%%%%%%%%%%%%%%%%%%%%%%%%%%%%%%%%%%%%%%%%
%%%%%%%%%%%%%%%%%%%%%%%%%%%%%%%%%%%%%%%%%%%%%%%%%%%%%%%%%%%%%%%%%%%%%%%%%%%%%%%%%%%%%%%%%%%%%%%%%%%%%%%%%%%%%%%%%%%%%%%%%%%%%%%%%%%%%%%%%%%%%%%%%%%%%%%%%%%%%%%%%%%%%%%%%%

\section*{Acknowledgements}
CDD was supported by EPSRC CDT in Fluid Dynamics EP/L01615X/1.  
AJB was supported by STFC grants ST/R00059X/1 and ST/S000275/1, and initially by the Leverhulme Trust through the award of an Early Career Fellowship. CAJ was supported by STFC grant ST/N000765/1. We would like to thank the referee for their helpful report and for constructive comments and Gordon Ogilvie for sending us his comments on a draft version of the paper.
This work was undertaken on ARC1, ARC2 and ARC3, part of the High Performance Computing facilities at the University of Leeds, UK. Some simulations were also  performed using the UKMHD1 allocation on the DiRAC Data Intensive service at Leicester, operated by the University of Leicester IT Services, which forms part of the STFC DiRAC HPC Facility (www.dirac.ac.uk). The equipment was funded by BEIS capital funding via STFC capital grants ST/K000373/1 and ST/R002363/1 and STFC DiRAC Operations grant ST/R001014/1. DiRAC is part of the National e-Infrastructure.

%%%%%%%%%%%%%%%%%%%%%%%%%%%%%%%%%%%%%%%%%%%%%%%%%%
%%%%%%%%%%%%%%%%%%%% REFERENCES %%%%%%%%%%%%%%%%%%

\bibliographystyle{mnras}
\bibliography{mybib}

%%%%%%%%%%%%%%%%%%%%%%%%%%%%%%%%%%%%%%%%%%%%%%%%%%
%%%%%%%%%%%%%%%%% APPENDICES %%%%%%%%%%%%%%%%%%%%%

\appendix

\section{Asymptotic linear analysis for high frequency tidal flows}\label{appendix_extension_of_ol2012}

In this appendix, we extend the linear asymptotic analysis of \cite{ogilvie_interaction_2012} for high-frequency oscillatory shear. We set out to provide a complementary analysis to explore further the response at high frequency, and to validate the results of our simulations of laminar convection, particularly the surprising appearance of negative effective viscosities. Our primary extension is to incorporate buoyancy forces and to consider perturbations to the temperature field. This was found to be necessary since \cite{ogilvie_interaction_2012} considered a flow driven by an imposed body force that is divergence-free, but buoyancy forces in Rayleigh-B\'{e}nard convection do not satisfy this property.

This section builds upon section 3 in \cite{ogilvie_interaction_2012} (see also \citealt{braviner_stellar_2015} who corrects some typos), and we adopt their notation and use the dimensional equations for ease of comparison with their work. The momentum equation is
\begin{equation}
(\partial_t + u_i \partial_i ) u_j = -\partial_j p + \theta \delta_{j3} + \nu\Delta u_j \,,
\end{equation}
where $\Delta\equiv\nabla^2$. The velocity field satisfies $\partial_i u_i=0$,
and we incorporate the heat equation 
\begin{equation}
\label{heateqnappendix}
	(\partial_t + u_i \partial_i ) \theta = N^2 u_z + \kappa \Delta \theta \,.
\end{equation}
We define horizontally-sheared coordinates
\begin{equation}
 x' =x, \quad y'=y-a(t)x, \quad t'=t,
\end{equation}
so that partial derivatives transform according to
\begin{equation}
	\partial_x = \partial_x^{\prime} - a \partial _y^{\prime}\,, \quad \partial_y = \partial_y^{\prime}\,, \quad \partial_z = \partial_z^{\prime}\,, \quad \partial_{t} = \partial_t^{\prime} - \dot{a} x \partial_y^{\prime}\,, \nonumber
\end{equation}
and define the velocity components
\begin{equation}
	u_x = v_x\,, \quad u_y = v_y + \dot{a} x \,, \quad u_z = v_z \nonumber.
\end{equation}		
Equation \ref{heateqnappendix} gives us an extra equation to be used with equations 7 and 8 of \cite{ogilvie_interaction_2012}. In sheared coordinates we obtain the system
\begin{gather}
	\left(\partial_t^{\prime} +v_j (\partial_j^{\prime} - a \delta_{j1} \partial_y^{\prime}) \right) v_i
	+\dot{a}v_x \delta_{i2} = - \left(\partial_i^{\prime} - a \delta_{i1} \partial_y^{\prime} \right) P \nonumber \\
	\hspace{1cm}+\nu (\partial_j^{\prime} - a \delta_{j1}\partial_y^{\prime})(\partial_j^{\prime} - a \delta_{j1}\partial_y^{\prime})v_i + \theta \delta_{i3},  \\	
	\left( \partial_t^{\prime} +v_j (\partial_j^{\prime} - a \delta_{j1} \partial_y^{\prime}) \right) \theta
	= N^2 u_z + \kappa (\partial_j^{\prime} - a \delta_{j1}\partial_y^{\prime})(\partial_j^{\prime} - a \delta_{j1}\partial_y^{\prime}) \theta,	 \\	
	\left( \partial_t^{\prime} - a \delta_{i1}\partial_y^{\prime} \right) v_i = 0 \label{maths_RBC_in_sheared_coords}\,.
\end{gather}
Since we are interested in small amplitude shear, we linearise the above equations in the shear amplitude. The basic convective flow in the absence of the shear satisfies
\begin{gather}
	(\partial_t^{\prime} + v_j \partial_j^{\prime} ) v_i = - \partial_i^{\prime} P + \nu \Delta^{\prime} v_i + \theta \delta_{i3}, 
	\label{maths_linearised_inshear_momentum} \\
	( \partial_t^{\prime} + v_i \partial_i^{\prime} ) \theta = N^2 u_z + \kappa \Delta^{\prime} \theta,
	\label{maths_linearised_inshear_temperature} \\
	\partial_i^{\prime} v_i = 0 \,
	\label{maths_linearised_inshear_continuity},
\end{gather}
and the pressure satisfies
\begin{gather}
	\Delta^{\prime} P = \partial_z^{\prime} \theta - (\partial_i^{\prime} v_j) (\partial_j^{\prime} v_i),
\end{gather}
where the linear operator $\Delta^{\prime}$ is defined by
\begin{equation}
    \Delta^{\prime} = \Delta + 2a \partial_x^{\prime} \partial_y^{\prime} - a^2 \partial^{\prime 2}_y \,. \nonumber
\end{equation}
The presence of buoyancy forces here is the key difference compared with \cite{ogilvie_interaction_2012}.

We use the method of multiple scales, introducing a fast time variable to represent the rapid shear, $T' =t'/\epsilon$, where $\epsilon \ll 1$ is a small parameter that characterises the ratio of convective to shear frequencies. We pose the asymptotic expansions 
\begin{gather}
	\delta v_i = \delta v_{i0} + \epsilon \delta v_{i1} + \dots, \\
	\delta P_i = \frac{1}{\epsilon}\delta P_{0} +  \delta P_{1} + \dots, \\
	\delta \theta = \delta \theta_0 + \epsilon \delta \theta_1 + \dots.
\end{gather}
At leading order we obtain equations 24 and 25 in \cite{ogilvie_interaction_2012}, so we do not repeat those here. Buoyancy forces, and their modifications to the pressure field, do not enter at this order. The linearised shear stress $\delta R_{xy0} = \langle v_x\delta v_{y0} + \delta v_{x0} v_y\rangle$, therefore satisfies
\begin{align}
    \partial_T^{\prime}(-\delta R_{xy0})=-\dot{a}\langle v_x^2-2(v_x\partial_y^{\prime}+v_y\partial_x^\prime)\partial_y^{\prime}\Delta^{\prime-1}v_x\rangle,
    \label{1storder}
\end{align}
where angled brackets denote a volume average. To obtain the perturbed shear stress at the next order, which is necessary to obtain the effective viscosity, we start from equation 33.1 in \cite{ogilvie_interaction_2012} (the decimal place indicating counting from the last labelled equation):
\begin{align}
	\partial_T^{\prime}(-\delta R_{xy1}) =& 
	\langle v_x (\partial_t^{\prime} + v_j \partial_j^{\prime}) \delta v_{y0} + v_y (\partial_t^{\prime} + v_j \partial_j^{\prime} ) \delta v_{x0} \nonumber \\
	& + v_x(\delta v_{j0} \partial_j^{\prime} - a v_x \partial_y^{\prime} ) v_y + v_y ( \delta v_{j0} \partial_j^{\prime}- av_x \partial_y^{\prime})v_x \nonumber\\
	&+(v_x \partial_y^{\prime} + v_y \partial_x^{\prime}) \delta p_1 \underbrace{- a v_y \partial_y^{\prime} p}\nonumber\\
	&- \nu v_x (\Delta^{\prime} \delta v_{y0} - 2 a \partial_x^{\prime} \partial_y^{\prime} v_y) \nonumber \\
	&- \nu v_y (\Delta^{\prime} \delta v_{x0} - 2 a \partial_x^{\prime} \partial_y^{\prime} v_x) 
	\label{dTRxy1},
\end{align}	
where $\delta R_{xy1} = \langle v_x\delta v_{y1} + \delta v_{x1} v_y\rangle$, and
where we have highlighted the only term that requires modification. Perturbations to the temperature field do not enter at this order either. The pressure is determined by
\begin{gather}
	 P = \Delta^{\prime -1}\partial_z^{\prime} \theta - \Delta^{\prime -1}(\partial_j^{\prime} v_j) (\partial_j^{\prime} v_i) \nonumber \,.
\end{gather}
which we can substitute into Eq.~\ref{dTRxy1}. The new contribution to the highlighted term is $\Delta^{\prime -1}\partial_z^{\prime} \theta$, which becomes
$-a v_y \Delta^{\prime -1}\partial_y^{\prime}\partial_z^{\prime} \theta$ in Eq.~\ref{dTRxy1}.
A second term arises from equation 33.2 in \cite{ogilvie_interaction_2012} that has the form
\begin{gather}
	- \Delta^{\prime -1} (\partial_y^{\prime} v_x - \partial_x^{\prime} v_y) (a \partial_x^{\prime}\partial_y^{\prime} P ) \nonumber
\end{gather}
which on considering only the additional contribution due to buoyancy forces we obtain
\begin{gather}
	-a \Delta^{\prime -1} (\partial_y^{\prime} v_x - \partial_x^{\prime} v_y) (  \Delta^{\prime -1}\partial_x^{\prime} \partial_y^{\prime} \partial_z^{\prime} \theta)\nonumber
\end{gather}	
thus equation 33.3 in \cite{ogilvie_interaction_2012} will have additional terms of 
\begin{gather}
	-a v_y \Delta^{\prime -1}\partial_y^{\prime}\partial_z^{\prime} \theta -a \Delta^{\prime -1} (\partial_y^{\prime} v_x - \partial_x^{\prime} v_y) (  \Delta^{\prime -1}\partial_x^{\prime} \partial_y^{\prime} \partial_z^{\prime} \theta)\nonumber
\end{gather}	
which need to be followed until equation 33.6 in \cite{ogilvie_interaction_2012}. The first step is to apply $\partial_T^{\prime}$ which gives
\begin{gather}
	-\dot{a}\left( v_y \Delta^{\prime -1}\partial_y^{\prime}\partial_z^{\prime} \theta + \Delta^{\prime -1} (\partial_y^{\prime} v_x - \partial_x^{\prime} v_y) (  \Delta^{\prime -1}\partial_x^{\prime} \partial_y^{\prime} \partial_z^{\prime} \theta)   \right)  \nonumber \\
	 \equiv -\dot{a} \mathcal{G}_1^{\theta}\,.\label{maths_extra_term_for_OL12}
\end{gather}

We can write equation \ref{1storder} and \cite{ogilvie_interaction_2012} equation 33.6 with the new term in equation \ref{maths_extra_term_for_OL12} in the form
\begin{gather}
	\partial_T' (-\delta R_{xy0}) = \dot{a} \mathcal{G}_0, \nonumber \\
	\partial_T^{\prime 2}(-\delta R_{xy1}) = -\dot{a} \left(\mathcal{G}_1 +\mathcal{G}_1^{\theta}\right) , \label{maths_G0_G1_expressions_modified}
\end{gather}
where $\mathcal{G}_0$ and $\mathcal{G}_1$ are the same as those in \cite{ogilvie_interaction_2012} equations 40 and 41. These are the final results required to obtain the effective elasticity and viscosity of the flow. These are straightforward to evaluate for laminar flows since the $\mathcal{G}$ quantities on the right hand side depend only on the basic convective flow. Note that $S_E=\mathcal{G}_0$ and $\nu_E=(\mathcal{G}_1+\mathcal{G}_1^\theta)/\omega^2$ to this order. These equations therefore indicate that for high frequency shear, the leading order response is a frequency-independent effective elasticity, and an effective viscosity that scales quadratically with frequency. In principle, the $\mathcal{G}$ coefficents can take either sign, depending on the basic flow.

To evaluate these quantities for a given convective flow for comparison with simulations, we assume a single convective mode with
\begin{gather}
    u_z=\Re \Big(\hat{u}_z\mathrm{e}^{i k_x x + i k_y y}\sin (\pi z)\Big),
\end{gather}
and similarly for $u_x, u_y$ and $\theta$. The amplitude of each component is obtained by comparison with the amplitude of the corresponding mode in a simulation. This then allows us to calculate all of the $\mathcal{G}$ coefficients and hence $S_E$ and $\nu_E$ for a given flow. This approach is only expected to work for steady laminar convection near onset. However, the asymptotic theory in principle applies to more complex flows if they can be adequately represented.

We have found that the new term $\mathcal{G}_1^\theta$ is essential to obtain negative effective viscosities for laminar convection. Note that it vanishes for $y$-aligned rolls, since $\partial_y^{\prime}=0$. On the other hand, for $x$-aligned rolls, $\partial_x^{\prime}=0$, so only the first term in $\mathcal{G}_1^\theta$ contributes. For a 3D convective mode, all terms in $\mathcal{G}_1$ could be important in principle.

%%%%%%%%%%%%%%%%%%%%%%%%%%%%%%%%%%%%%%%%%%%%%%%%%%%%%%
%%%%%%%%%%%%%%%%%%%%%%%%%%%%%%%%%%%%%%%%%%%%%%%%%%%%%%

\section{Simple derivation of $\nu_E$ scaling for $y$-aligned convection rolls}\label{appendix_simple_model}
Here we provide a simpler explanation (than the formal analysis in Appendix~\ref{appendix_extension_of_ol2012}) for why $\nu_E\propto \omega^{-2}$ for high-frequency shear in the specific case of $y$-aligned laminar convection rolls (as described in \S~\ref{yaligned}). We consider the $\boldsymbol{e}_y$ component of the momentum equation (using the dimensionless variables in the body of paper),
\begin{equation}\label{maths_balance_of_terms}
	\partial_t u_y =	 - S u_x \cos(\omega t)+ \nabla^2 u_y -\boldsymbol{u}\cdot\nabla u_y-\boldsymbol{u}_0\cdot\nabla u_y -\partial_y p\,.
\end{equation}	
For a steady $y$-aligned roll, $u_x$ is approximately time-independent, $u_y$ is small in the absence of shear, and $\partial_y\approx 0$. For high-frequencies and laminar flows, the dominant balance in Eq.~\ref{maths_balance_of_terms} is
\begin{align}
	\partial_t u_y \approx - S u_x \cos(\omega t),
\end{align}
which implies $u_y\approx -a_0u_x\sin\omega t$, such that the response is primarily elastic (out of phase with the shear). We have confirmed that the amplitude of $u_y$ is indeed approximately independent of $\omega$ in our simulations in this high-frequency regime.
The resulting effective viscosity (Eq.\ref{maths_effective_viscosity})
\begin{align}
\nu_E \propto \frac{1}{\omega} \int\langle u_x u_y\rangle \cos(\omega t) \, \mathrm{d}t,
\end{align}
therefore vanishes to leading order. 

To obtain the frequency scaling for the effective viscosity, we must therefore consider the next largest terms in Eq.~\ref{maths_balance_of_terms}. We may write $u_y=u_{y,0}+u_{y,1}+\dots$, where $u_{y,0}$ is the dominant contribution just obtained, and $u_{y,1}$ is the leading order correction. For laminar convection, we suppose that $u_{y,1}$ is dominated by the viscous term acting on the $u_{y,0}$ component, and hence
\begin{align}
	\partial_t u_{y,1} \approx \nabla^2 u_{y,0}, 
\end{align}
implying $u_{y,1} \propto \frac{1}{\omega}\cos\omega t$. We thus obtain 
\begin{align}
\nu_E \propto \frac{1}{\omega} \int\langle u_x u_y\rangle \cos(\omega t) \, \mathrm{d}t \propto \frac{1}{\omega^{2}}
\end{align}
for the largest nonzero contribution.
This scaling would also be expected if the nonlinear and pressure terms (involving $u_{y,0}$) were instead dominant.

%%%%%%%%%%%%%%%%%%%%%%%%%%%%%%%%%%%%%%%%%%%%%%%%%%%%%%
%%%%%%%%%%%%%%%%%%%%%%%%%%%%%%%%%%%%%%%%%%%%%%%%%%%%%%

% Don't change these lines
\bsp	% typesetting comment
\label{lastpage}
\end{document}